\documentclass[singlespace,tocchapterhead]{ccw_chithesis}
\usepackage{astroextras}
\usepackage{natbib}
\usepackage{graphicx}
\usepackage{amsmath}
\newcommand{\p}[1]{(\ref{#1})}

\def\l{ \Lambda}

\def\k{ \mathrm{K}}

\def\fb{ \bar{f}}
\def\bd{ \bar{d}}
\def\bl{ {\bar{\lambda}} }

\def\ma{\mbox{$\mathcal{A}$}}
\def\fl{\lq\lq flame\rq\rq}
\def\e{\epsilon}
\def\stur{S_{t}}
\def\slam{S_{l}}
\def\num{\mathrm{num}}
\def\gcm{ \mathrm{g\:cm^{-3}} }
\def\kms{ \mathrm{km\:s^{-1}} }
\def\mch{ M_\mathrm{Ch}}
\def\rof{ \rho_\mathrm{fuel}}
\def\lcr{ \lambda_\mathrm{cr}}
\def\Le{ \mathrm{Le}}
\def\st{\mathrm{st}}
\def\num{\mathrm{num}}
\def\ie{\emph{i.e.}}
\def\eg{\emph{e.g.}}
\newcommand{\bnabla}{\mbox{\boldmath$\nabla$}}
\newcommand{\R}[2]{\mbox{$ #1 \bigr|_{#2}$}}
\begin{document}

\title{Analysis of reaction-diffusion systems for Flame Capturing in Type Ia Supernova 
simulations}
\author{Andrey V. Zhiglo}
\department{Physics}
\division{Physical Sciences}
\degree{Doctor of Philosophy}
\date{June 2009}
\maketitle
\makecopyright


\begin{abstract}
We present a study of numerical behavior of a thickened flame used in Flame Capturing (FC, Khokhlov~(1995))
for tracking thin physical flames in deflagration simulations. This technique, used extensively
in astrophysics, utilizes artificial flame variable to evolve flame region, width of which is resolved 
in simulations, with physically motivated propagation speed. 
We develop a steady-state procedure for calibrating flame model used in FC, and test it against analytical
results. Original flame model is properly calibrated with taking matter expansion
into consideration and keeping artificial flame width at predetermined value regardless of expansion. 

We observe numerical noises
generated by original realization of the technique. Alternative artificial burning rates are 
discussed, which produce 
acceptably quiet flames (relative dispersion in propagation speed within 0.1\% at physically 
interesting ratios of fuel and ash densities). 

Two new quiet models are calibrated to yield required \lq\lq flame\rq\rq\ 
speed and width, 
and further studied in 2D and 3D setting. Landau-Darrieus type instabilities of the flames are observed. 
One model also shows significantly anisotropic propagation speed on the grid, both effects increasingly 
pronounced at larger matter expansion as a result of burning; these 2D/3D effects make that model 
unacceptable for use in type Ia supernova simulations at fuel densities below 
about 100 tons per cubic centimeter
. Another model, first presented here, looks promising
for use in flame capturing at fuel to ash density ratio of order 3 and below, the interval
of most interest for astrophysical applications. No model was found
to significantly inhibit LD instability development at larger expansions without increasing flame width.
The model we propose, \lq\lq Model B\rq\rq, yields flames 
completely localized within a region 6 cells wide at any expansion. 

We study Markstein effect (speed of the flame dependence on its curvature) 
for flame models described, through direct numerical simulations and through 
quasi-steady technique developed. By comparing results obtained by the 2 approaches we demonstrate that 
Markstein effect dominates instability effects on curved flame speed for Model B in 2D simulations 
for fuel to ash density ratio of about 2.5 and below. We find critical wavelength of LD instability
by direct simulations of perturbed nearly planar flames; these agree with analytical predictions
with Markstein number values found in this work. We conclude that the behavior of model B is well understood,
and optimal for FC applications among all flame models proposed to date.

\end{abstract}

\acknowledgments{
I am grateful to Alexei Khokhlov for proposing original problem of correct calibration of his flame model,
which over time evolved into extensive study of various aspects of behavior of flames. His encouraging comments,
stimulating discussions on various astrophysical issues, references suggested --- especially at the beginning of 
the project --- are greatly appreciated. Most numerical simulations presented here were performed with code {\tt{ALLA}},
in creating which Alexei was a main contributor.

Thanks to the Committee members: Juan Collar, Robert Rosner and Thomas Witten, whose comments contributed to shaping 
the final form of the thesis.

Thanks to FLASH Center for numerous talks and meetings organized over the years, which broadened my horizons in astronomy, 
astrophysics and computational physics; and for supporting my graduate study. 
Alan Calder, Tridivesh Jena, Dean Townsley helped me with basics of {\tt{FLASH}} code operation, which is appreciated. 

Thanks to Alexei Poludnenko for numerous inspiring conversations on variety of scientific and general topics, 
and for great job in maintaining computer cluster. 

Numerous seminars, presentations, colloquia at Physics, Math, 
Astronomy and Astrophysics Departments of the University of Chicago were encouraging and helped to keep motivation.

Thanks to my friends at the University of Chicago and beyond. In particular, I am indebted to Lenya Malyshkin and Sasha Paramonov for their help
during my last days in Chicago.

Thanks to my family.
}

\tableofcontents
\listoffigures
\listoftables


\mainmatter

\chapter{Introduction}
\label{cha:intro}
In the thesis we study features of propagating flame solutions in reaction-diffusion (RD) systems.
Such flames are ubiquitously observed in chemical combustion in terrestrial experiments 
and in industry, as well as in certain astrophysical phenomena, with fast composition
transformation taking place in nuclear deflagration fronts propagating through stellar 
material. 
For most of the study we concentrate on a few specific
RD models used in, or proposed here for use in simulations of deflagration phase
in type Ia supernova (SN~Ia; plural SNe~Ia) phenomenon. An important applied goal of the study is
to propose an optimal flame model to be used in the simulations, the one that reliably 
reproduces expected physical propagation of a deflagration front in SN~Ia problem, without
introducing unphysical numerical artifacts, such as numerical noises, excessive
front instabilities and anisotropic behavior related to computation grid. 

We start with a brief overview of SN~Ia phenomenon, providing motivation for the 
study, and discussing general physics of flames in this example system. We then proceed 
to describing techniques currently used for numerical simulations of SN~Ia. The 
chapter is concluded with the list of problems the current simulations face that we
address in this work, and with more detail on organization of the thesis.

\section{Supernovae: observations}

Extreme cases of variable stars, which are now classified as novae and supernovae,
attracted human attention since ancient times. The first accurate record of 
such an event, with documented dates (185 AD) and position (close to $\alpha$ Centauri)
is due to Chinese astronomers (\cite{sn185}). Crab Nebula M1 and the famous short-period pulsar 
within are remnants of another supernova, observed by Chinese, Japanese and Arab astronomers 
in 1054; for 23 days it was bright enough to be visible in daylight, and at nights it was visible
with naked eye for 653 days. 
In 1572 Tycho Brahe measured parallax of another star of 
this class (to find no detectable parallax); he also recorded its dimming 
with time. That was the first well-documented supernova in western 
world; Brahe's book, \lq\lq De Stella Nova\rq\rq, gave rise to the name we currently use for 
a different explosive phenomenon, much more frequent and dim than supernovae. 

There were more than a hundred classical novae (now understood as a
\ result
of explosive burning of accreted hydrogen-helium fuel on a surface of a White Dwarf star
in binary systems, when certain critical mass of the fuel is accumulated) studied by 1919,
when Lundmark suggested that the distance to M31, \lq\lq Andromeda Nebula\rq\rq, 
was $7\times 10^5$ light years (at that time
 most astronomers believed that everything observed on the sky belonged to our Galaxy). 
Even with this 
significantly underestimated distance it followed that S Andromedae explosion of 1885 was
extragalactic, and hence was more luminous than classical novae studied by about 3 orders of 
magnitude (\cite{lundmark}). With improved distance measurements \cite{baade} suggested to 
differentiate between novae and supernovae, in the way as used nowadays.

Typical supernova brightness rises rapidly for a few weeks, reaches \lq\lq maximum light\rq\rq,
 then drops within months, approximately exponentially after a few weeks of less regular transient
dimming
.
At maximum light SNe are among the brightest objects in the Universe, often shining brighter than 
the whole host galaxy. \cite{zwicky38} proposed collapse of a star made of ordinary matter into a 
neutron star as a mechanism for SN phenomenon; currently accepted scenario of certain types of 
core-collapse
supernovae is based on that idea. All supernovae are understood now to be a result of cataclysmic
event happening to some stars in late stages of their evolution, after almost all hydrogen and 
helium are transformed into carbon and heavier elements. As observed energy emitted  
within a few weeks following such an event (around maximum light), 
is comparable to energy produced by the Sun over its 
multi-billion year history,
it is clear that SN event must involve drastic transformation of the whole star structure. In fact,
the transformation of most of the star composition 
takes just a few seconds based on current 
understanding of the physics of SNe.

Historically supernovae are categorized into a few types based on their spectral features. 
Supernovae of type II show hydrogen absorption lines in the spectrum near maximum luminosity, 
those of type I show no hydrogen. Supernovae of type~I 
showing strong Si II lines belong to class Ia; those with no silicon lines are further 
subdivided into types Ib and Ic if they show or lack helium lines, respectively. 

Observationally SNe~Ia are special in that vast majority of them show close peak luminosity 
and light curve decline features, whereas variety within other types is significant.
Typical SN~Ia light curve (brightness vs time) shows initial rise interval about 20 days
long, followed by about as fast decline, about 3 magnitudes within a few weeks,
then dimming more slowly by about a magnitude per month. 
For most spectroscopically normal SNe~Ia the absolute bolometric magnitude at maximum light 
lie in narrow interval from $-19.4^m$ to $-18.7^m$. \cite{phillips} observed that
this scatter in peak luminosities is correlated with light curve width (characterized by 
$\Delta m_{15}(\mathrm{B})$, star magnitude drop within 15 days after maximum light);
this Phillips relation makes it possible to accurately predict SN~Ia absolute magnitude
based on observed dimming rate.
Coupled with extreme brightness of supernovae this makes SN~Ia excellent standard candles
for probing cosmological distances. Peculiarities in observed SN~Ia magnitude as a function of 
their redshift was a first indication that the Universe is expanded with (positive) acceleration 
(see \cite{leib} for a review).

SNe~Ia also differ from SNe of different types in their environment: Ia is the only supernova
type observed in elliptical galaxies, although they still occur about twice more frequently in 
younger stellar populations in spiral and irregular galaxies (\cite{cappellaro}). 

These features, together with low luminosities of SN~Ia in radio, X-ray and gamma bands suggest 
less massive progenitor of SNe~Ia than those of other types of supernovae, and no collapse happening
in the process of explosion. All other supernova types are believed to involve core collapse to a 
neutron star or a black hole as a main source of energy powering disruption of and energy deposition 
into the envelope.

\section{SNe~Ia: physical models}
Below we briefly overview physical models of SN~Ia, mostly paying attention
to physical issues relevant for our study. See, \eg, \cite{hillniem00mech} for more detail 
on explosion models.

\subsection{Overview}
SNe~Ia are believed to be a result of thermonuclear explosion of a White Dwarf (\cite{HoFo60}, 
\cite{ar69}), a late stage of evolution of light stars, with main sequence mass below about 8 solar 
masses. All models involve heavy White Dwarfs (WDs), with masses close to or exceeding solar mass $M_\odot$. 
Such WDs are formed from contracted central parts of stars after hydrogen and helium burning there
is complete and H/He-rich shell is dissipated in space. They are composed of 
${}^{12}\mathrm{C}$ and ${}^{16}\mathrm{O}$ mostly, and are supported by the cold pressure of 
degenerate electrons. WDs with smaller amounts of carbon, but rich in heavier elements, 
${}^{20}\mathrm{Ne}$ and ${}^{24}\mathrm{Mg}$, are also candidates to undergo a thermonuclear 
explosion, however some calculations (\cite{gutierrez96}) indicate that core collapse is a 
probable outcome of near-central ignition in such WDs. 

For all specific models (briefly discussed 
below) not overtly contradicting observations successful result of the explosion is stellar material
transformed mostly into ${}^{56}\mathrm{Ni}$ with considerable amount of intermediate-mass elements
(Si, S, Ca) within about 3~s of proper explosion; enough energy is released for total disruption
of the WD. Explosion itself is practically invisible due to opacity of dense hot ashes of the 
explosion; significant part of the released energy is spent to overcome gravitational potential of
the compact star,
with excess transformed into kinetic energy of expanding material. A few days after explosion,
expanding envelope starts looking bright to an external observer, though it remains optically 
thick until 
much later times. Double $\beta$-decay of ${}^{56}\mathrm{Ni}$ to ${}^{56}\mathrm{Co}$ 
($t_{1/2}=6.08\,\mathrm{d}$) and then to stable ${}^{56}\mathrm{Fe}$ ($t_{1/2}=77.2\,\mathrm{d}$)
powers the light curve (\cite{truran67}, \cite{colgate69}), that is it heats the expanding
gas, which captures emitted in the decay positrons and $\gamma-$quanta and eventually reemits light
outside, mostly in optical band. Both the initial light-curve decline rate (governed by slow diffusion
of light out of the gas nebula) and peak luminosity (\cite{arnett82}) turn out to be proportional to
amount of ${}^{56}\mathrm{Ni}$ produced, thus providing an explanation for Phillips relation.

Specific models proposed differ in particulars of progenitor, ignition and burning mechanism. 
The most natural model to account for uniformity of most SN~Ia events, single-degenerate
scenario with nearly Chandrasekhar mass $M_\mathrm{Ch}$ 
(the maximal mass of the WD that degenerate electron 
pressure can support, \cite{chandra}; about $1.4\,M_\odot$), involves ignition near the center 
of a heavy WD as a 
result of thermonuclear runaway, readily happening in highly-degenerate fuel (C+O) in central part
of a WD with near-critical mass. The progenitor WD is believed to be formed with lower mass, 
and to have its mass increased by accretion from a companion giant or main-sequence star. 
Parameters of the close binary system must be rather special so that accreting hydrogen or helium 
steadily burned into C/O on the surface of a WD, without nova outbursts (which do not on average 
lead to increase in a WD mass); cf. \cite{nomoto00} for study of the accretion process.

The amount of nickel produced, and thus the maximal luminosity (if
explosion does unbind the WD) is likely to depend on WD composition, rotation, and on
specifics of explosion process, thus leading to variation in nickel mass produced. In the model of 
\cite{ar69} explosion starts as detonation, and it is in this detonation front sweeping 
through entire star where all \lq\lq fuel\rq\rq, C+O, is transformed into ash; the latter consists
almost completely of nickel after detonation at original high density of WD material. This 
contradicts observations, which show a few tenths of solar mass of intermediate mass elements
produced by all SNe~Ia. Pure detonation model (in near-Chandrasekhar mass WD) is thus ruled out by 
observations (\cite{ar74}, \cite{os74}). It is believed that \lq\lq explosion\rq\rq\ in a single 
degenerate near-$M_\mathrm{Ch}$ model
is started as deflagration, possibly turning into detonation at a later stage. We will discuss this
in more detail in the next section.

Other class of models involve WDs with masses substantially below $M_\mathrm{Ch}$. While such 
lighter WDs (still massive enough to consist mostly of carbon/oxygen) will not self-ignite, 
detonation front may be sustained once initiated, according to simulations. 

\cite{woosley94}, \cite{livnearnett95}
present 1D and 2D simulations, in which detonation is initiated on the surface of C/O 
subchandrasekhar WD, by self-detonating (through runaway at the bottom) accreted helium shell of 
$\sim 0.2M_\odot$. Detonations at lower densities in such lighter WDs can produce significant
amounts of intermediate-mass elements, though some features observed in simulations 
(like large amounts of ${}^{44}\mathrm{Ti}$ produced) are not supported by observations. 
Models of this 
type are attractive as binary systems containing lower mass WDs are more abundant; however 
accumulation of significant mass of He is still required for the mechanism to work
; also, due to a range of possible progenitor masses there is no natural explanation for uniformity
of SN~Ia peak luminosities. Finally, simulations performed in 2D with postulated axial symmetry
are just not reliable, as 3D simulations of the last decade suggest.

In double degenerate scenario (\cite{webbink84}, \cite{IbenTutukov84}) ignition is 
initiated by collision of 2 sub-$M_\mathrm{Ch}$ WDs.
This scenario requires very close binary system of WDs orbiting around each other: their separation 
must not exceed $\sim 10^6\,\mathrm{km}$ so that the merging happened on cosmological 
timescales, assuming gravitational radiation as the main damping mechanism. Like the model in the 
previous paragraph there is no special mass for merging system, thus significant dispersion
in maximum luminosities is to be expected. Such non-standard mechanisms may be an explanation
for a small number of SNe~Ia with peculiar spectra and substantially high or low peak luminosity.

\subsection{Deflagration and detonation in a White Dwarf: physics}
\label{cha:intro_physics}
As reviewed in the previous section, nuclear explosion of a near-$M_\mathrm{Ch}$ WD is currently 
considered the best explanation for SN~Ia. What is usually loosely referred to as \lq\lq  
explosion\rq\rq\ physically corresponds to one of 2 possible modes of fast fuel transformation 
(through a network of thermonuclear reactions) into ash, in thin front
propagating 
through the system (WD). Reaction rates show sharp dependence on 
temperature, of Arrhenius type. Temperature in the reaction zone (within the front) must be raised
to approximately reaction activation energy for the reaction to progress with substantial rate. 

Detonation mode is characterized by supersonic front speed; adiabatic compression by  
significant pressure jump across the front is the mechanism raising the temperature high enough
for reactions to proceed. Heat released in the reactions sustains the shock wave propagation. Due
to fast detonation front velocity, if detonation is initiated near WD center as the start of the
explosion, the whole WD will be swept by the front at nearly original high density --- density 
distribution has just no time to change. Reaction goes all the way through nuclear statistical 
equilibrium (NSE) composition at densities above about $5\times 10^7\gcm$; and this composition is 
mostly iron-group elements (having the highest binding energies per nucleon) at high fuel 
densities. This is why too little lighter elements is produced in pure detonation model.

In deflagration mode a flame (thin zone with large gradients of temperature, density, composition; 
it contains reaction zone) propagates into the fuel by diffusion-reaction mechanism. Temperature is raised 
in front of the flame due to heat conduction from the hot reaction zone. Temperature distribution has
characteristic for diffusion processes exponential tail deep into cold fuel; right in front of the 
reaction zone (where still the fuel is not depleted in the reaction) the temperature continuously approaches 
high flame temperature, leading to high reaction rate; flame this way propagates forward. 
Thermal energy released in reaction zone prevents temperature profile from smearing out by heat 
conduction; steady solutions (of the form of a kink wave) typically exist describing propagating flame with 
constant in time distribution of physical quantities within, in the flame rest-frame. 
See Chapter 2 for details; Fig.~\ref{stepx} shows ash fraction distribution in one particular flame model.

Species diffusion, and their production/depletion in the reaction zone contribute to distribution 
of composition and temperature near the flame (see Sec.~\ref{markstein:intro} below for details). 
In terrestrial combustion in nearly ideal gases burning is affected to approximately the 
same degree by diffusion as it is by heat conductivity, as Lewis number (ratio of heat diffusivity to 
matter composition diffusivity) $\Le\approx 1$. For WD 
matter thermal conductivity by far dominates other transfer phenomena, $\Le\sim 10^7$, thus
species diffusion plays negligible role in flame propagation.

Laminar flame speed as estimated in \cite{TimWoo92} depends on fuel composition (that is local WD 
composition in front of the flame) and density; it is about $80\:\kms$ in central region of 0.5C+0.5O
near-$\mch$ WD (central density $\rof\approx 2\times 10^9\:\gcm$) and lower at smaller densities in outer 
layers. This velocity can be roughly estimated by equating characteristic energy release timescale, 
\[t_\mathrm{reac}\approx 1/R\]
with heat diffusion timescale (assuming comparable or smaller contribution to flame dynamics from species 
diffusion, \ie\ $\Le\approx 1$ or greater)
\[t_\mathrm{diff}\approx \frac{W^2}\kappa . \]
$R$ here stands for some characteristic value of reaction rate in flame zone, such that the average 
energy release rate in the flame $\langle dQ/dt\rangle = Rq$, $q$ being total heat release of the reaction; 
$\kappa=\lambda/(c_V\rho)$ is temperature diffusivity coefficient ($\lambda$ meaning heat conductivity, 
$\rho$ density and $c_V$ specific heat). This balance of timescales determines flame width $W$, which 
further provides rough estimate for laminar flame speed:
\[ W\approx \sqrt{\kappa /R},\quad \slam \approx W/t_\mathrm{reac}\approx \sqrt{\kappa R}\, . \]
See \cite{zelbook,williams} for more accurate estimates, and for more on physics of deflagration.

\subsection{Pure deflagration models}
Whether deflagration alone can account for observable features of SNe~Ia is not straightforward to answer.
Laminar flame speed is far less than sound speed anywhere in the WD, thus flame propagation is nearly 
isobaric; star structure changes as heat is released in the flame, WD expands, this hydrodynamic
evolution must be computed simultaneously, coupled to flame propagation calculations. Early 1D simulations 
suggested that fast WD expansion after deflagration sets in quenches burning too early, so that not 
enough energy is released to unbind the star; WD contracts back after burning stops. It was understood, 
however, that in reality the flame will never propagate as an ideal spherical flame with its laminar 
burning speed, due to instabilities discussed below. 

It is known that deflagration fronts are subject to a number of instabilities, which 
can drastically affect flame propagation. As soon as transverse flame dimensions exceed characteristic 
instability scales $\lcr$ flame surface starts developing various features (bubbles, wrinkles --- the 
geometry depends on specific instability involved) increasing its surface compared to a smooth surface 
stable laminar flame would have had; this effectively increases integral burning rate. In flamelet regime, 
that is when characteristic spatial scale of flame surface perturbations by flame instabilities and background
turbulence significantly exceeds laminar flame width, kinetics of the burning (and hence distribution of 
physical quantities) within the flame is not affected significantly; integral burning rate is increased 
approximately in proportion to flame surface increase (compared to stable laminar burning). 
When a flame is perturbed on scales comparable to or 
less than its laminar width burning description based on laminar flame physics becomes not applicable, 
the flame is generally torn apart; such regime is called distributed burning (\cite{Damkoehler40}). 

For nuclear deflagration in degenerate WD matter, where kinematic viscosity and species diffusivity are
negligible compared to temperature diffusivity $\kappa$, hydrodynamic instability  
of the type (abbreviated LD below) studied by \cite{lan}, \cite{dar} 
is important. Besides, the flame is subject to
Rayleigh-Taylor (RT) instability (basically a buoyancy instability --- occurring when lower density  
reaction products support denser cold fuel above 
in WD gravitational field, directed towards its center), and to Kelvin-Helmholtz (KH) instability, 
perturbing shear flow interfaces, like on sides of rising RT bubbles. LD instability is most important
on scales comparable with flame width. Such scales are never resolved in full star simulations 
(flame width is submillimeter for most of deflagration phase \cite{TimWoo92}). 
On the largest scales RT instability dominates, driving hydrodynamics on the scale of the star. See
\cite {X94, X95} for features of RT-driven burning in cylinders with uniform gravitational field, and 
in 3D SN~Ia simulations. One result of studies in cylindrical geometry important in applications 
(see next section) is that after certain transient time steady (in 
average) burning sets in 
uniform gravity driven RT burning, with integral burning velocity (\lq\lq turbulent flame speed\rq\rq, 
$\stur$) depending only on geometry of the cylinder cross-section, 
and independent of laminar flame speed. 
\begin{equation}\label{St}
\stur\simeq 0.5\sqrt{\ma g \Delta}
\end{equation}
is the value confirmed in \cite{X95}, \cite{Zhang06} as reasonably accurate. 
$\Delta$ here stands for linear scale of
cylinder cross-section (in the cited works horizontal cross-section was a square with side $\Delta$);
\ma\ denotes Atwood number for fuel--ash, 
\begin{equation}\label{At}
\ma=(\rho_\mathrm{fuel}-\rho_\mathrm{ash})/(\rho_\mathrm{fuel}+\rho_\mathrm{ash}),
\end{equation} 
$g$ stands for gravitational acceleration.
 
The smaller the laminar 
speed the more intricate surface of the flame develops, its area in steady regime scaling inversely 
proportional to $\slam$, laminar speed of the flame. If $\slam$ is increased, on the other hand, the 
surface becomes smoother; when $\slam>\stur$ (as given in \p{St}) flame is essentially flat and 
propagates with its 
laminar speed; that is for any geometry of the system flames with large enough $\slam$ are stable to RT.
See \cite{X95} for more on self-regulation mechanism behind this behavior; qualitatively, for flames
with larger laminar speed laminar burning overruns slower developing smaller RT bubbles, thus only 
larger-scale bubbles (which rise faster) develop --- the surface is polished by burning on lower scales.

In non-steady 3D setting in deflagrating WD characteristic large scale burning speed $\stur$ on large scales
increases as the flame gets farther from the WD center, as Atwood number, gravity and available tangential 
dimension increase; in addition, the instability had been allowed more time to develop
before the flame got to such larger radius. Laminar flame speed, on the 
other hand, rapidly decreases, as fuel density decreases. Hence, early 1D simulations grossly underestimated 
total burning rate, except at the very center of the WD. 1D models are used for certain purposes till now, 
however the flame speed is prescribed based on estimates of $\stur$ from 3D analysis. Often this burning
rate is being left as a (time dependent) free parameter, tuned to get desired features on the output
of the simulation, close to observations; some calculations of this type are notably successful in yielding
results closely resembling observed SN~Ia features (\cite{w7}).

Real 3D simulations are needed to judge about viability of a pure deflagration model. These still
use artificial techniques to propagate the flame with right velocity (improving one of such techniques, 
proposed in \cite{X94}, is a subject of this thesis) as the scales physically governing flame propagation
are not resolved in any WD-scale 3D simulations. However the techniques aim to get the flame propagating
with correct, physically motivated effective speed to reproduce, on resolved scales, speed of real thin physical 
flame; no tuning of flame speed with the purpose of getting particular output results is used. We will
discuss the techniques used in the next section; more detail on implementation and the results may be 
found in \cite{X00}, \cite{X03}, \cite{X05}, \cite{Rein02}, \cite{Reinlanl02}. 

No consensus exists right now whether deflagration
alone may explain a normal type Ia supernova. With currently used subgrid models, prescribing flame
speed so that to get total burning rate as the physical one (as the latter is currently understood), of 
submillimeter-thick nuclear flame wrinkled on unresolved scales, different groups get enough energy 
to unbind the WD, for near-central single-spot ignition. The amount of energy released in such deflagration
models is still lower than in normal SNe~Ia, as is amount of iron-group elements produced. Significant
amount of unburned carbon and oxygen remains close to the center of the star, a result of sinking RT 
fingers of cold fuel. After homologous expansion of unbound WD sets in (about 10 seconds after original 
ignition) these fingers are observed as large amounts of unburned material expanding with low
velocities; on the other hand in real SNe~Ia light elements are observed almost exclusively in outer layers 
of expanding nebula, that is having largest expansion speeds. 

Attempts were undertaken (\cite{Roepke05}) to fix some of these shortcomings with extending burning 
simulation at lower densities, in distributed burning regime (starting at densities below about 
$10^7\:\gcm$, when flame preheating zone becomes wider than Gibson scale.\footnote{%
This is the scale on which average turbulent velocity fluctuations are equal to the laminar flame speed. 
On lower scales the turbulent fluctuations are smaller, therefore flame propagation is insensitive to turbulence 
on such lower scales. Above Gibson scale flame surface is bent and stretched by background turbulent flows.
The turbulence on all scales is mostly generated (through Kolmogorov cascade) by large-scale RT motions, 
augmented with vorticity generated by KH instability.}

Another (effective) approach (\cite{Travaglio04, Roepke07}) is based on multipoint ignition. Igniting at several
points distributed in the central part of a WD simultaneously reduces the amount of unburnt fuel near the 
center and increases the amount of nickel produced. Say, simultaneous ignition in 1600 spherical spots 
in \cite{Roepke07} resulted in asymptotic kinetic energy of ejecta of $8.1\times 10^{50}$~erg, with 
$0.606\:M_\odot$ of iron group elements produced, including $0.33\:M_\odot$ of ${}^{56}\mathrm{Ni}$. 
Corresponding results in \cite{Rein02b} with single spot central ignition (albeit smaller resolution) are 
$4.5\times 10^{50}$~erg, $0.5\:M_\odot$ and $0.3\:M_\odot$. Simulation in \cite{Travaglio04}, in one octant 
of a full WD (octahedral symmetry assumed) with 30 bubbles (larger than in \cite{Roepke07}, 10~km radius 
vs 2.6~km; and located closer to WD center) ignited at $t=0$ resulted in $6.5\times 10^{50}$~erg asymptotic 
kinetic energy of ejecta and $0.418\:M_\odot$ of ${}^{56}\mathrm{Ni}$ produced. Kinetic energy of ejecta of 
a typical observed SN~Ia is slightly above $10^{51}$~erg. Exact pre-explosion conditions (smoldering phase)
at the center of a WD are not known well, so it is not obvious if single point ignition is the best
choice of initial conditions. Yet significant dependence on this choice of initial conditions is noteworthy,
even though dispersion in Ni masses quoted above would lead to dispersion in brightness within $0.8^m$,
comparable to dispersion among observed Ia\rq s.

\subsection{Delayed detonation models}
The best fit with observations is obtained in hybrid models, where \lq\lq explosion\rq\rq\ starts as 
deflagration, and later on detonation is triggered in some spot of the WD preexpanded during the deflagration
phase. Detonation happening at lower densities under such a scenario produces sufficient amount of 
intermediate mass elements (in outer layers of preexpanded WD), transforms RT fingers of fuel near
WD center into iron-group elements (together with the whole dense central part: detonation front is smooth,
on scales characteristic for RT instability, 
in contrast with intricately corrugated deflagration front surface), leaves less fuel unburnt and produces 
more energy. Exact mechanism, what triggers detonation, is not yet agreed upon. When originally proposed, 
\cite{X91} assumed deflagration to detonation transition (DDT) at smaller densities (a few times 
$10^6\:\gcm$, characteristic for distributed burning regime) as the most probable mechanism. DDT is
observed in terrestrial chemical flames, however it is more problematic to explain it in unconstrained WD 
setting. 

Different indirect mechanisms for triggering detonation were proposed over time, involving creation of 
extended regions with large temperature gradients by colliding masses of WD material with different
velocities. Pulsating detonation scenario involves recollapse of the WD after deflagration phase in which 
not enough energy was released to unbind the star. The original implementation (see \cite{Hoeflich96})
seems not plausible now, as 
accurately modeled 3D deflagration with central ignition does ordinarily produce enough energy to unbind 
a C+O WD, as reviewed in the previous subsection. 

Asymmetric variations of this scenario, with slightly offcenter
ignition, Gra\-vi\-ta\-ti\-o\-nal\-ly-Confined Detonation (GCD, \cite{gcd}), Detonating Failed Deflagration Model
(DFD, \cite{Plewa07}) do trigger detonations in 3D simulations. In these only a small fraction of fuel is burnt
before the bubble driven by buoyancy breaks through WD surface, creating surface wave intensive enough to 
trigger detonation after colliding at diametrically opposite point of the WD surface. The main problem of
these models is to get enough preexpansion before detonation so that significant amount of intermediate 
mass elements be produced, to agree with observations. Preexpansion, depending on the mass burnt before 
the bubble breakthrough, strongly depends on the position of original ignition site. The problem is thus
to propose some robust mechanism to rule out significantly offcenter ignitions. These models tend to 
overproduce iron-group elements in all simulations to date.

\section{Simulations of deflagration phase}
Here, after a brief overview of standard hydrodynamics, we describe flame-capturing (FC) technique 
(\cite{X95}) used in SN~Ia simulations for tracking flame position and prescribing heat release 
in artificially thickened (to be resolved in simulations) flame zone with intent to reproduce,
on resolved scales, results of real, non-discretized, hydrodynamics. 
We then describe some problems of the original implementation
of FC found in this study, and outline topics presented in the thesis, 
mostly related to improving the FC model.

\subsection{Hydrodynamics involved in simulations}
Deflagration is described by a standard set of hydrodynamic equations with reaction source terms.
These include a mass continuity equation; Navier-Stokes equation for velocity of the matter subject to gravity,
pressure gradients and viscosity; transport equation for internal energy, with a source term 
describing heat release in reactions taking place in the system; and finally equations describing 
diffusion and reactions for species involved in reactions. Equations of state, expressing pressure
and internal energy as functions of density and temperature make the system of equations closed.
For SN~Ia problem the system is usually simplified by neglecting viscosity and diffusion
of species (as thermal conductivity by several orders of magnitude dominates other transfer phenomena:
Prandtl number is small, Lewis number is large).

For large scale simulations the system is modified further: detailed reaction networks are not used, 
instead only a few species are taken into account, with model reaction rates for these, found separately 
in such a way that the reduced system imitated kinetics of the full system with acceptable accuracy.
Such a simplification is used in chemical combustion simulations as well: this makes it possible to save
computational resources, to be able to use higher resolution for example, when it is more important to
resolve some critical hydrodynamic scales then to get exact distributions of species. This is usually 
the case in engineering applications, when, for example, knowing pressure or temperature distribution in 
combustion chamber is more important than knowing detailed chemistry of burning process\footnote{%
The latter is studied separately, for calibrating the reduced scheme in part, in simplified (usually steady
1D) setting, allowing to concentrate on kinetics rather than (trivial in such setup) aerodynamics.}.

In SN~Ia problem using detailed reaction network in large-scale hydrodynamic simulations 
makes little sense in principle, as physical reaction zone (across which at least some species 
concentrations change significantly) 
is never resolved in such simulations, and this situation will not change in any 
foreseeable future. As the total simulation box size must be of the WD size, that is a few thousand 
kilometers, grid cells are of order 1 km size for most detailed of current simulations. Flame width
is submillimeter for most of the deflagration (at density in front of 
the flame $\rho_\mathrm{fuel}\sim 10^8\:\gcm$ and above, \cite{TimWoo92}). Because of this disparity, 
all full star simulations evolve flame surface indirectly, without resolving flame width\footnote{%
The flame propagates directly on its own when the set of hydrodynamic equations described above is 
solved exactly. Then, as described in Sec.~\ref{cha:intro_physics}, 
interplay between heat release in the flame and heat 
conduction into cold material in front of the flame leads to flame propagating forward, with correct (by 
 definition) speed. In simulations described in this paragraph, however, derivatives are modeled via 
finite differences
of quantities in adjacent grid cells. This modeling is designed to be accurate enough for fluid motions
on scales exceeding a few grid spacings, yet obviously such differences cannot model derivatives 
of quantities changing significantly on subgrid scales, which is the case for reaction species abundances.
}.

One approach (\emph{level set method}, LSM; used by Munich group in part, 
\cite{Rein02,Roepke07}), 
motivated by 
tiny width of real flames, is to (formally) describe the flame as infinitely thin surface, on which matter 
density, velocity and other physical variables have hydrodynamically prescribed discontinuities. 
One then propagates this discontinuity surface with the speed prescribed so that 
to mimic, in average, real flame region evolution.
This surface is represented as the zero-level of certain \lq\lq level function\rq\rq\ $G$: flame
manifold$|_t=\lbrace\mathbf{r}(t):G(\mathbf{r},t)=0\rbrace$. 
Time evolution of $G$ (advection plus normal 
propagation into the fuel with prescribed \fl\ speed) used strives to ensure the needed evolution of
the flame surface. Various numerical tricks are employed to take care of the tendency of $G$ to develop 
unphysical peculiarities.
 Hydrodynamical solvers (based on piecewise-parabolic reconstruction, \cite{cowo})
used in conjunction with this formally
infinitely thin flame model spread discontinuities over several grid spacings, exact fractions of fuel/ash 
in grid cells intersected by the \lq\lq flame surface\rq\rq\ have more of numerical significance for the 
scheme than physical meaning. 

We will stick to another flame-tracking prescription (described below), where the deflagration front
 is represented with 
artificial thick \lq\lq flame\rq\rq . Quotation marks are used below to distinguish the 
artificial numerical flame wherever confusion with real physical flame seems possible. We also refer to 
this numerical construction as \lq\lq flame zone\rq\rq\ as it is intended to be evolved in such a way so 
that to contain the physical flame within. The \fl\ is governed by the same reaction-diffusion equations
as physical flames, which makes it clear what behavior of the \fl\ to expect (instabilities,
interaction with background turbulence and sound waves), no new physics is introduced. This seems a clear 
advantage over LSM.

\subsection{Flame capturing}
Flame 
Capturing (FC) technique (\cite{X95})  employs 
artificial scalar quantity $f(\mathbf{r},t)\in [0;1]$, \lq\lq reaction progress 
variable\rq\rq, to track flame evolution: $f=0$ in the unburnt material, and changes to 1 
behind the \lq\lq flame\rq\rq, when the burning is complete. The \lq\lq flame\rq\rq\ in the 
sense of this numerical technique is a region where $f$ is neither 0 nor 1 (but strictly between);
it is made a few 
grid sizes thick by appropriately choosing parameters governing $f$ evolution (see below\footnote{%
This in essence is a particular realization of artificial viscosity approach, \cite{vn}.
}); the value of $1-f$ is intended to mimic real physical fuel fraction in fuel/product mixture 
within the grid cell in the flame region. 

$f$ is evolved via a diffusion-type
equation, 
\begin{equation}\label{I1}
    \frac{\partial f}{\partial t}+\mathbf{u}\cdot\bnabla f
    =\bnabla(K\bnabla f)+\Phi(f).
\end{equation}
Physical quantities (pressure, temperature, composition, matter velocity) are simultaneously
evolved through a standard system of Euler equation \p{euler}, diffusion-type equations with 
reaction terms for internal energy density (heat conduction equation) and species concentrations
(if tracked separately; in standard FC realization they are just set to be a linear function of
$f$), and equations of state (in code {\tt{ALLA}} that we use for most non-steady simulations these are 
implemented as functions taking matter density and internal energy density as input, and returning
pressure and temperature)
. $f$ is coupled with physical variables through advection (by local matter
velocity $\mathbf{u}$ in (\ref{I1})); and through heat-release term in hydrodynamic 
equations, which is governed directly by $f$: $\delta Q=q\,df$, that is heat is released
linearly with $f$ increasing (reaction progressing) up to $q$, heat release of complete 
nuclear burning, per unit mass. (This $q$ depends on local fuel composition and density.) 
Artificial diffusivity $K$ and artificial \lq\lq burning rate\rq\rq\ $\Phi(f)$ 
were prescribed (\cite{X95}\footnote{%
Only constant $K$ was used before \cite{zh07}. Eq.~\ref{I1} here is written in the form we
use below with more general flame models, in which $K$ is a function of $f$.%
}) 
so that to make the \lq\lq flame\rq\rq\ $\sim 4\Delta$  
thick, $\Delta$ denoting grid zone size, and propagating with prescribed velocity. That original
prescription in effect led to systematic error in flame velocity, due to matter expansion
(in the process of burning) neglected when estimating front speed in a system \p{I1}. This
was corrected in \cite{zh07} (and will be described in the next chapter).

Scales larger than the flame width, which govern flame instabilities development (LD and RT), are
not resolved in SN~Ia simulations as well --- for most of the simulation, at large densities; 
a portion of a flame within any grid cell is thus not smooth, but corrugated 
by instabilities. Real flame surface therefore exceeds the surface 
implied from numerical observations on grid scale.
This is an extra complication, not encountered in shock-tracking techniques 
used similarly to FC in certain simulations with shocks.
To address this issue all simulations include some 
subgrid model for prescribing renormalized, corrected for missing scales velocity, 
larger than physical laminar flame speed. 

Observations of stationary RT-accelerated burning in 
\cite{X95} led to
using \lq\lq turbulent flame speed\rq\rq\ $\stur$, defined through \p{St}
where for $\Delta$ the grid spacing is used, as the value for artificial
\lq\lq flame\rq\rq\ propagation speed  
ensuring correct integral burning rate. When the laminar flame speed 
$\slam$ exceeds $\stur$, before developed RT-driven burning regime sets in,
the \fl\ is required to propagate with laminar speed $\slam$. See 
\cite{Zhang06} for further verification of this flame speed prescription, and 
for proposed technique modification in transient, non-stationary setting. 
We will not touch below on the best
prescription for the \fl\ speed, but concentrate on the very model
propagating the \fl\ with a given speed.

\subsection{Goals and scope of the thesis}
To hope to get meaningful physical results in simulations it is critically important to understand
the properties of the model used for tracking the flame region 
and for estimating heat release there. It is important that flame tracking model itself, 
in part, did not introduce unphysical 
artifacts in uncontrolled way into simulations. One needs to study model behavior in 
density range between $\sim 2\times 10^9\:\mathrm{g\:cm^{-3}}$ (central WD 
density) and a few $\times 10^6\:\mathrm{g\:cm^{-3}}$ (when distributed burning sets in, 
and hypothetical deflagration to detonation transition could take place), and favor
models not introducing much noise into simulations, and not demonstrating significant 
instabilities of their own, unless there is a hope that the latter mimic instabilities of
actual physical flame zone. 

When this work was started no analysis of features of the flame model used for FC existed, 
apart from initial study of isotropy of \fl\ propagation on the grid in \cite{X95}, together 
with analytical estimates that led to prescription for parameters in \p{I1}; the
parameter values proposed there were used essentially 
verbatim by several groups thereafter (\cite{X03,gcd,brown05}). By now some results on numerical
noises generated by flame models, including description of a new burning rate proposed for use in FC,
were presented by our colleagues (\cite{Jordan08}). 

Numerical simulations always pose a question how reliable the results are, what in the results
is due to actual physics of the system studied, and what is an artifact of the numerical 
scheme/approximation/model used. This question is particularly important for systems
so complex that simulations are the only way to get an estimate of the results. With these
concerns results are always tested against other simulations, with different resolution, utilizing 
different discretization algorithm or based on analytically different solver. A suite of
problems with known analytical behavior is used to compare these known results with results
of simulations, albeit on simple systems. With all this care every so often discrepancies between 
results of different groups appear. \cite{X95} reports that 3-cell wide cylindrical flame 
propagates isotropically on a square grid. Same FC model implemented by Niemeyer required 
8-cell wide flame to make anisotropies acceptable. GCD model reach robust detonation conditions
in 3D in FLASH group simulations (\cite{Jordan08}), the same model never detonates in simulations by
\cite{Roepke07}. SN~Ia simulations in octant with central ignition show clear tendency of the flame 
to preferentially develop features along grid directions from the very beginning of the simulation,
which is a clear numerical artifact (this further facilitates fast RT instability development).
Concerns like this were a part of motivation to study flame model features in detail.

In Chap.~2 we present results obtained mostly without using 
actual non-steady hydrodynamical solvers. We describe a method we developed for finding flame speed
by solving eigenvalue steady-state problem. We use that method to calibrate the model flame used in
\cite{X95} for flame capturing. In original calibration in \cite{X95} matter expansion (we characterize 
it with expansion parameter 
$\Lambda={\rho_\mathrm{ash}}/({\rho_\mathrm{fuel}-\rho_\mathrm{ash}})\:$) in a result
of burning was neglected; we correct for this and propose additional improvement, allowing 
to keep flame width independent of matter expansion (this expansion increases with fuel density 
decreasing, due to decreasing degeneracy of the electron gas). We present analytical solution for 
the original model flame profiles, which allows us
to directly check the eigenvalue numerical scheme. Flame profile (distribution of $f$ in space for 
steadily propagating flame solution of \p{I1}) has an exponential tail; we propose another class of
models, with
$f$-dependent diffusivity coefficients, that produce flames localized in space (having finite total 
width, with no tails). We argue about the advantages of models having such finite flames, and find 
certain representative model of that class with nearly linear flame profiles (similar in that to real 
fuel distribution in shaped with RT-instability flame zone in simulations of SN~Ia), which are 
furthermore insensitive to $\l$. For this model we present simple fits of the the parameters
entering \p{I1} as functions of $\l$, yielding very simple implementation of the new model proposed 
in flame capturing numerical scheme. We conclude Chap.~2 by presenting results of actual 
implementation in FC technique, observed flame speeds, widths and profiles. Discrepancy with 
steady-state results are clearly connected to discretization effects by varying resolution,
number of cells within the flame width.

Chap.~3 is devoted to studying noises produced by FC scheme in 1D simulations. Two new quiet models
are introduced, that have finite flame widths and unique eigenvalues for flame speeds. 
They are calibrated to yield correct \fl\ speeds and widths using the methods of Chap.~2.

In Chap.~4 we study 2D and 3D behavior of model flames. By modeling circular flames in 2D we
observe that anisotropic propagation speed is a generic feature of reaction-diffusion flames, more
prominent at higher expansions.  
This is a purely numerical effect, and is cured by increasing flame width. For one of the new 
quiet flame models, sKPP, this effect is especially pronounced: for flame width of about 8 cells 
between reaction progress variable values $f=0.01$ and $f=0.99$ the flame propagates along
grid axes by $>5\%$ faster than at 45 degrees to them, for expansions corresponding
to deflagration in WD at relevant densities of $10^8\:\gcm$ and below. 
A free parameter of the second of the quiet models of Chap.~3 is used to eliminate propagation
anisotropy at all relevant expansions.  
We observe LD instability, particularly
severe for sKPP model of Chap.~3 at larger expansions. These 2D effects are a strong
argument to avoid using sKPP model at densities of $\lesssim 10^8\:\gcm$. Second of the quiet models, 
model B, shows significantly larger critical LD wavelength, and slower growth of this instability.
At all densities above $3\times 10^7\:\gcm$ in SN~Ia problem maximal combined 2D asphericity due to both LD and
numerical effects stays within $1.2\%$ for simulation times exceeding those used in SN~Ia simulations. 
At lower densities, and in 3D simulations, combined asphericity is larger yet tolerable for the \fl\ width used, 
about 5.5--6 cells between $f=0.01$ and $f=0.99$, which is thinner than in original model of \cite{X95}\footnote{%
The thinner the \fl\ is the better for resolving its small scale features, ultimately for better accuracy of the
simulation.}. We present simulations of perturbed planar flames in 2D; observed critical wavelengths for 
perturbations to grow agree with theoretical $\lcr$ for LD instability (in Markstein approximation of 
infinitely thin flame with curvature-dependent propagation speed).

We develop quasi-steady technique for estimating Markstein numbers $M$ (quantifying flame speed dependence
on flame curvature for curved multidimensional flames) in Chap.~5, and calculate these for models described
in the thesis. $\lcr$ for LD instability depends on $M$. Comparison with $\lcr$ estimated directly in 
Chap.~4, as well as comparisons of exact $M$ found here with $M$\rq s estimated numerically provides 
evidence that Model B is well-understood, and in its observed behavior physical effects dominate over 
numerical ones. 

We conclude that model B proposed here is best suited for use for Flame Capturing.

\chapter{Steady state analysis of 1D flames}\label{steady}
In the present chapter we decouple equation for reaction progress variable $f$ in \p{I1} from the rest
of hydrodynamic equations, and study its flame-like solutions in steady 1D setting. We write
a master equation for steady flame profile $f(x)$ in dimensionless form, and find dimensionless
velocities $d$ and widths $w$ for a few artificial burning rates $\Phi(f)$ used in combustion literature,
as well as, in Sec.~\ref{stepK}, for nonconstant diffusivity $K(f)$ proposed during this work.
Accuracy of the numerical scheme proposed for finding $d$ and $w$ is tested against analytic results
obtained for one model (original one, \cite{X95}) and qualitative results for another one, KPP.
Restoring normalization of the master equation \p{artvisfront0} leads to a simple prescription
for choosing scale-factors $R$ and $\tilde{K}$ of reaction rate and diffusivity, such that the flame with
coefficients \p{dimless} will propagate with prescribed speed $D$ and have prescribed width $W$, for any 
expansion $\l$. These scalings \p{KR} are determined based on dimensionless $d(\l)$, $w(\l)$ found
in the first sections of this chapter and fitted in Sec.~\ref{1dsuggest} for efficient numerical 
implementation. In the last section we check our results in practice, in {\tt{ALLA}} code. 

Methods for calibrating flame models presented in this chapter are used for different models
in the rest of the thesis; qualitative results for various possible burning rates are used
for constructing a new flame model (Model B) in the next chapter, which we ultimately suggest for use
in FC based on properties of the \lq\lq flames\rq\rq\ it produces; eigenvalue problem used in this
chapter for finding $d$ and getting flame profiles is generalized in Chap.~\ref{markstein} for
spherical flames in higher-dimensional problems, to study effects of curvature on flames.

\section{Physical formulation and numerical method}\label{steadyK1}
Here we study time evolution of the reaction progress
variable profile $f(\mathrm{r})$, which determines heat release in hydrodynamic
simulations. In 1D it is possible to decouple equation \p{I1} for $f$ from the rest
of hydrodynamic equations for any equations of state --- provided the burning rate
can be considered depending on $f$ only. This is the case for 1 stage reactions when  
$\Le=1$ (when temperature and reactant concentration distributions are similar, 
\cite{z-fk}; relevant for terrestrial chemical flames), and for FC progress variable $f$,
by construction. Here we demonstrate this decoupling, formulate the steady-state problem
and describe numerical procedure we use for its solution.

\subsection{Boundary problem for flame speed}
\label{bndryd}
All models studied in this chapter are of form
\p{I1}.
One particular type of solutions of \p{I1} coupled to hydrodynamic equations (as well as of real combustion systems) 
in homogeneous medium is steady 1D flame front, propagating with constant speed $D_f$ 
into the fuel ($D_f$ is defined here as the speed of the flame in the reference frame where the fuel rests). 
Here we study these particular 1D solutions, with physical quantities (and $f$) being functions of $x-D_ft$ only. 
In such a steady solution it is convenient to study the system in a flame rest-frame, 
in which all physical quantities depend on $x$ only (after Galilean transformation).
Matter velocity (in this 1D steady setting) is determined by continuity equation: 
\begin{equation}\label{front}
u(x)\rho(x)=\mathrm{const}=-D_f\rho_0.
\end{equation}
This further determines $f(x)$ from \p{I1},
which after this substitution for $u(x)$ reads  
\begin{equation}\label{artvisfront0}
   -D_f\frac{\rho_0}{\rho(x)}\frac{df}{dx}=\frac{d}{dx}\left( \frac{Kdf}{dx}\right)+\Phi(f).
\end{equation}

To close the system one needs to express $\rho(x)$ via heat release distribution, $dQ/dx=q\, df/dx$. 
For this, in isobaric burning,
one determines $\rho(x)$ from enthalpy conservation, 
\[ H(p_0,\rho_0) + qf=H(p_0,\rho), \]
which in particular provides $\rho_0/\rho(x)$
as a function of $f$ (depending on the particular equation of state). This makes \p{artvisfront0} a second order equation for
$f$ only. 

In certain physically interesting situations, \eg\ near the center of a WD (main contribution to pressure from ultrarelativistic
degenerate electron gas), or for ideal gas (terrestrial flames) $\rho\propto 1/H$ at constant pressure\footnote{%
this generally is the case for \lq\lq $\gamma$ equation of state\rq\rq, 
$H=\frac\gamma{\gamma -1}\frac p\rho$ (for ideal gases this is the case when specific heat is independent 
of temperature.)
}, 
thus \p{artvisfront0} assumes the following form, 
\begin{equation}\label{artvislam}
    -D_f\left(1+\frac{f}{\Lambda}\right)\frac{df}{dx}=\frac{d}{dx}\left( \frac{Kdf}{dx}\right)+\Phi(f),
\end{equation}
which we use for steady-state flame speed and width estimates. 
\begin{equation}\label{lam}
    \Lambda=\frac{\rho_\mathrm{ash}}{\rho_\mathrm{fuel}-\rho_\mathrm{ash}}
\end{equation}
quantifies how the matter expands as a result of burning; with $\Lambda$ defined this way eigenvalues found from \p{artvislam}
are close to true ones even for situations with $H\rho$ slightly nonconstant across the flame. 
$\Lambda=\frac\gamma{\gamma-1}\,\frac{P_0}{q\rho_0}$for ideal gas equation of state, $\gamma$ denoting adiabatic parameter
($\gamma=c_p/c_V$ for ideal gas).
 
For solution $f(x)$ to describe a physically meaningful flame profile it needs to satisfy physical boundary conditions, 
\begin{eqnarray}\label{bndl}
f(-\infty)&=&1  \\ \label{bndr}
f(+\infty)&=&0.
\end{eqnarray}
These may be satisfied only for certain values of $D_f$ (a parameter in \p{artvisfront0}); these eigenvalues of $D_f$ are by definition
possible flame propagation speeds in 1D. Corresponding eigenfunctions, $f(x)$ satisfying (\ref{artvislam}--\ref{bndr}), 
are steady flame profiles; in flame capturing these will determine \fl\ thickness and fuel distribution within the \fl\ 
--- approximately, as long as the \fl\ segment propagation may be treated as 1D steady-state, and up to discrepancies due to 
numerical (discretization) effects.

\subsection{Numerical procedure}\label{steadynum}
First, by representing burning rate and diffusivity as products of constant dimensionful scale-factors
 and ($f$-dependent in general) dimensionless form-factors, 
\begin{equation}\label{dimless} 
  \Phi(f)=R\Phi_0(f),\: K(f)=\tilde{K} K_0(f)
\end{equation}
we rewrite \p{artvislam} in dimensionless form:
\begin{equation}\label{artvislamdim}
    -d\left(1+\frac{f}{\Lambda}\right)\frac{df}{d\chi}=\frac{d}{d\chi}\left( \frac{K_0(f)df}{d\chi}\right)+\Phi_0(f).
\end{equation}
To accomplish this we have introduced dimensionless flame speed, 
\begin{equation}\label{dimlessd}
d=D_f/\sqrt{\tilde{K}R} ,
\end{equation} 
and dimensionless coordinate
\[ \chi=x\sqrt{R/\tilde{K}}\] along flame propagation.
Eigenvalue $d$ of boundary value problem \p{artvislamdim} with boundary conditions following from (\ref{bndl}--\ref{bndr})
may be found numerically following the procedure described next (see \cite{zh07} for more detail).

As (\ref{artvislamdim}) is invariant under translations in $\chi$ it can generically
be reduced to a first order equation by rewriting it in terms
of
\[ p=-\frac{df}{d\chi} \] considered a function of $f$:
\begin{equation}\label{master1}
\frac{d}{d f}\bigl( K_0(f)p\bigr)-d\Bigl(1+\frac f\Lambda\Bigr)+\frac{\Phi_0(f)}p=0
\end{equation}
This form is used below for qualitative and numerical analysis of
the problem. Corresponding boundary conditions are $p(0)=p(1)=0$.
Eigenfunctions $p(f)$ are nonnegative.

In this section we demonstrate the technique for systems with constant diffusivity: 
the original model of \cite{X95} ($f_0=0.3$ used there)
with step-function rate
\begin{equation}\label{Phistep}
\Phi=\left\lbrace
\begin{array}{ll}
  R &\: \mathrm{for\ } f_0<f<1\\
  0 &\: \mathrm{otherwise}
\end{array} \right.
\end{equation}
and KPP (\cite{KPP})
\begin{equation}\label{Phikolm}
\Phi=R\, f(1-f)\quad (\mathrm{for\ }0<f<1).
\end{equation}

Equation \p{master1} is integrated by the fourth order Runge-Kutta
algorithm starting from $f=1,\: p=0$.\footnote{%
It is imperative to
start from $f=1$ for parabolic $\Phi_0(f)$ (KPP) as a general solution
for $f(x)$ near $x\to+\infty$ (where $f\to 0$) is a superposition of
two decaying exponentials, and the faster decaying one is lost when integrating
$dp/df$, thus making it impossible to satisfy $p|_{f=1}=0$.}
We use constant grid spacing
($\Delta f=10^{-5}$ for most of the runs), except near zeroes of $p$ (the
starting point $p(f=1)=0$, and at most one more), where it was refined further.

Integration is actually started at $f_\delta=1-\delta$, with $\delta$ being the smallest
refined integration step, as $p(f=1)=0$ appears in the denominator of \p{master1}.
Analytically found asymptotic expansion is used as initial value for $p(f_\delta)$ (see Chap.~5
for more details in more complicated setting in $D>1$ dimensions. Using $\alpha=D-1=0$  
in Eq.~\ref{asymp}, and similar equations below that for $p$ for different flame models, 
yields asymptotics valid for 1D flames.)

The $d$ eigenvalue is then found {for step-function $\Phi_0(f)$} by
requiring $p|_{f=1}=0$. Namely, Newton-Raphson algorithm (see, \eg\
\cite{numfor}) is applied, $\frac{\partial
\left(\vphantom{p^2}p(f=1)\right)}{\partial d}$ is found
simultaneously with $p(f)$, ensuring fast convergence.
$d(\Lambda=\infty)$ (found beforehand by solving \p{Lambdainf}) is
used as a seed at each new $f_0$ value for the first $\Lambda$ in a
row; for subsequent $\Lambda$\rq s the previous one provides seed value
for $d$; 4--13 iterations were enough to get $d$ with $10^{-8}$
precision. Results are summarized in Sec.~\ref{markstein:intro}. For the case
of KPP burning rate the spectrum of steady flame speeds is actually continuous,
as for the studied (\cite{KPP}) case with no matter expansion. We explain
qualitatively this feature of the spectrum in the next section.

\section{Spectrum of flame speeds and flame profiles. Qualitative and numerical analysis}
Here we show qualitatively why an eigenproblem with step-function burning rate has 
unique solution for $D_f$ but the problem based on KPP has continuous spectrum
of eigenvalues. We describe asymptotic behavior of flame profiles near $f=0$ and 1,
to better understand physical versus numerical features of 1D non-steady simulated 
\fl\ in Sec.~\ref{nsteady1D}, and, with this understanding, to propose new flame
models in Sec.~\ref{noises:models}. Qualitative consideration is confirmed by numerically
solving the eigenproblem for $d$, and by numerical integration of \p{master1} for 
several $d$\rq s. 

\subsection{Step-function burning rate}\label{step:qual}
\begin{figure}[htbp] \begin{centering}
\includegraphics[width=5.1in]{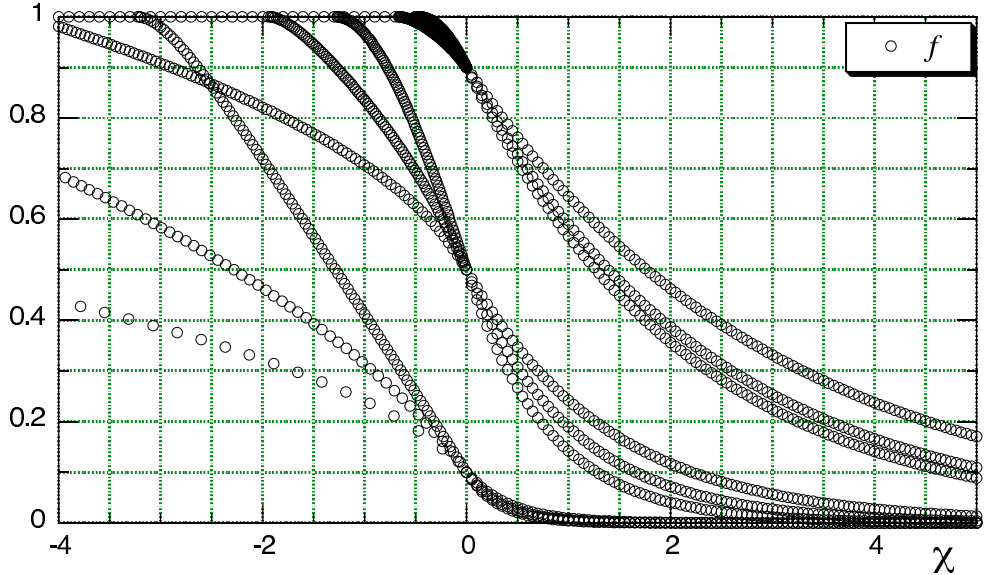}
\caption[Flame profiles, step-function burning rate]{Flame profiles 
(\ie\ numerically found eigenfunctions). For each $f_0$
(the ordinate of the curve intersection with $\chi=0$) 3 pairs of profiles
are depicted, for $1/\Lambda=0.05,\,4,$ and 20, larger $1/\Lambda$
corresponding to the curves intersecting $\chi=0$ axis at larger
angles.
\label{stepx}}
\end{centering}\end{figure}

We start with the model described by (\ref{artvislamdim}) with
step-function burning rate \p{Phistep}. 
One can observe that an eigenfunction $f(\chi)$ is monotonically
decreasing, in accord with physical expectations (see typical flame profiles in 
Fig.~\ref{stepx}). More than that, $f(x)$ is convex
at $f<f_0$ and concave elsewhere. Really, the solution of \p{master1}
at $f<f_0$ is $p=df(1+f/2\Lambda)$, positive and increasing. It is thus enough to show
that $p$ is monotonically decreasing in $(f_0;1)$ (then $p$ is automatically positive as
it goes to 0 at $f\to 1$ as dictated by the boundary conditions). If $p$ were not decreasing
there would have existed $f_r\in(f_0;1)$ such that $d p/d f(f_r)=0$. By \p{master1}
$p(f_r)=(d(1+\frac{f_r}{\Lambda}))^{-1}$, $d^2p/df^2(f_r)=d/\Lambda>0$, thus $p$ 
would have been increasing 
in some right neighborhood of $f_r$, and then \p{master1} would require that $p$ grew on entire
$(f_0;1)$, making $p(f=1)=0$ impossible.

Any self-similar solution satisfying physical boundary conditions must behave as follows 
(for certain $\chi_1$):
\begin{equation}\label{leftb}
f=1 \;\forall
\chi<\chi_1,\,\;df/d\chi\,(\chi_1)=0,\;0<f<1\mbox{ at }\chi>\chi_1,\;\lim\limits_{\chi\to\infty}f(\chi)=
0.
\end{equation}
Really, any solution
of (\ref{master1}) equal to 1 at any point with non-zero $df/d\chi$
would necessarily exceed 1 nearby. While we could allow systems with profiles $f$ 
taking values outside $[0;1]$, for our particular model with rate $\Phi(f)=0$ at $f\ge 1$
any solution exceeding $f=1$ at some $\chi$
monotonically goes to $+\infty$ to the left of such a $\chi$, as $\chi\to -\infty$ 
(so the boundary conditions are not satisfied).
Thus $f$ must either become identically 1 on some half-line $(-\infty;\chi_1]$
or approach $f=1$ asymptotically from below as $\chi\to -\infty$. 
Behavior of the solution of (\ref{master1}) in the region where $|1-f|\ll 1$ 
coincides with that of the linearized equation,
\begin{equation}\label{f1step}
f=\bar{c}_1-\chi/\bar{d}+\bar{c}_2\exp(-\bar{d}\chi),
\end{equation}
where
\begin{equation}\label{dbar}
\bar{d}=d\left(1+1/\Lambda\right).
\end{equation}
The latter does not remain in the vicinity of 1 as $\chi\to -\infty$ (in fact is
unbounded) for any values of integration constants $\bar{c}_{1,2}$,
unless it is differentiably glued to the $f=1$ solution to the left of some $\chi_1$.
This completes the proof for the behavior near $f=1$. 

Similar analysis in the region where $|f|\ll 1$ yields a general solution of
the linearized equation
\begin{equation}\label{f0step}
 f=c_1+c_2\exp(-d\chi) ,
\end{equation}
which tends to 0 at $\chi\to -\infty$ if and only if $c_1=0$. From this it follows
that any solution of
\p{master1} such that $f(\infty)=0$ cannot equal
zero at any finite point. It has an infinite tail,
decaying exponentially. 

Summing up, the desired eigenfunction asymptotically behaves like (\ref{f0step})
with $c_1=0$ as $\chi\to +\infty$, and like (\ref{f1step}) as $\chi\to \chi_1+$
($\chi_1=x_1\sqrt{R/K}$), with $\bar{c}_{1,2}$ such that
\begin{equation}\label{lbdry}
 f(\chi_1)=1,\; df/d\chi(\chi_1)=0.
\end{equation}

For any fixed $\chi_1$ a solution $f_{\chi_1}$ satisfying (\ref{lbdry}) is unique, and
any solution with $f(-\infty)=1$ is a translation of such. In order for this $f$ to 
vanish at $+\infty$ it must belong to another one-parameter subset of solutions. 
For this to happen there must be one functional dependence among the parameters in \p{master1}.
As it is shown below, this actually leads to a unique value for $d$ for any fixed
values of the other parameters ($\Lambda$ and $f_0$ here), for which
solutions of the boundary problem exist.

In the $q=0$ ($\Rightarrow\rho(x)=\rho_0=\mbox{const}$) case 
(\ref{master1}) actually \emph{is} piecewise linear, thus
the above restrictions immediately yield
$d(\Lambda=\infty,f_0)$ as the solution of equation 
\begin{equation}\label{Lambdainf}
f_0d^2=1-e^{-d^2}.
\end{equation}
(Cf. \p{xi1}; $f(\chi_1)$ is then expressed in elementary functions as $\Lambda=\infty$).
One can write the solution as expansion in $f_0$ asymptotically as
$d^2=\frac 1{f_0}-\frac {e^{-1/f_0}}{f_0}-\frac
{e^{-2/f_0}}{f_0^2}-\left(\frac 3{2f^3_0}+\frac 2{f^2_0}+
\frac 2{f_0}\right)e^{- 3/f_0}+O\left(
(f_0 e^{1/f_0})^{-4}\right)$; at $f_0=0.3$ this yields $d$ 2\% smaller
than the value in \cite{X95} (where in $q=0$ approximation $f(x)$ was found incorrectly);
at $\Lambda$\rq s of interest the difference will be
more significant. In the limit $f_0\to 1$ $d$ vanishes
as $d^2=2(1-f_0)+\frac 2 3 (1-f_0)^2+\frac 7 9
(1-f_0)^3+\ldots\;$(leading term agreeing with an estimate in
\cite{zelbook}, p.~266).

Flame profiles 
found numerically using the procedure described in the previous section are presented 
in Fig.~\ref{stepx}. They demonstrate asymptotic behavior as found above,
and show how the profile width depends on expansion parameter $\l$.

\subsection{KPP burning rate}

It is known (see \cite{zelbook} for detailed discussion) that the spectrum
of eigenvalues of KPP problem is continuous, comprising all reals
above some $d_1$. In this section we show that the same holds if one
includes the term arising from gas expansion, find similar spectrum for a wide
range of $\Phi$\rq s and verify these conclusions numerically.

One can qualitatively analyze the spectrum for burning rate \p{Phikolm}
along the same lines as for \p{Phistep} described in the previous section.
Upon linearization in the region where $1-f\ll 1$ ($\chi\to -\infty$)
\p{artvislamdim} yields
\begin{equation}\label{kleftasym}
f=1-\bar{b}_+e^{-\bar{\lambda}_+\chi}-\bar{b}_-e^{-\bar{\lambda}_-\chi},\quad
\bl_\pm=\frac\bd 2\pm\sqrt{\frac{\bd^2}4+1},
\end{equation}
$\bar{d}=d(1+1/\Lambda)$ as before; thus there is a one parameter subset of physical solutions, those behaving
asymptotically like \p{kleftasym} with $\bar{b}_+=0$.

By linearizing \p{artvislamdim} in the region where $f\ll 1$ one gets asymptotic
behavior of a solution
\begin{equation}\label{krightasym}
f\approx {b}_+e^{-{\lambda}_+\chi}+{b}_-e^{-{\lambda}_-\chi},\quad
\lambda_\pm=\frac d 2\pm\sqrt{\frac{d^2}4-1}.
\end{equation}\\
There are a priori three different situations for drawing further conclusions:\\
\indent 1) $d<2$: any solution getting to a neighborhood of 0 necessarily
becomes negative. \\
\indent 2) $d>2$: any $b_+,b_-\ge 0$ describe physically acceptable
behavior, as does a certain subset of $b_+<0,\;b_->0$. Thus one may
conjecture that \emph{for all} $d>2$ there is an
eigenfunction: it belongs to the described above 1-parameter subset
of solutions asymptotically approaching 1 as $\chi\to -\infty$
and on becoming small
at larger $\chi$ it behaves asymptotically like \p{krightasym},
exponentially approaching zero as $\chi\to+\infty$. The resulting
1-parameter set of physical solutions corresponds to one of them translated
arbitrarily along $\chi$, \ie\ there is a unique flame profile for
any $d$ (the term \lq\lq unique\rq\rq\ as related to profiles is used 
below in this sense, \ie\ up to translations).

By analyzing \p{artvislamdim} one expects the $f$ to decrease monotonically,
and it is easy to see that a solution cannot asymptotically approach any value in $(0;1)$
as $x\to +\infty$ (say, by linearizing \p{artvislamdim} near such a value). Thus a solution will
eventually get to a neighborhood of $f=0$, where it will behave as \p{krightasym}; unless
it becomes negative (\ie\ have unphysical $b_\pm$, $b_-<0$ in part; numerical results below
show that this does not happen, as well as confirm the
form of the flame profile) it will asymptotically
go to zero, hence being physical. It is the presence of 2-parameter
set of solutions decaying to zero at $\chi\to +\infty$ which accounts for the continuous
spectrum of $d$ here, in contrast with the situation with step-function burning rate. 

To
compare side-by-side one
may start with some physical (satisfying $f(-\infty)=1$)
$f$ on the left and check if it can satisfy $f(+\infty)=0$ as well.
For step-function there is a unique solution (modulo translations) which goes to zero
as $\chi\to+\infty$. Other solutions asymptotically approach non-zero values; a solution
asymptotically approaching some $c_1<f_0$ reads 
\begin{equation}\label{fc1}
f(\chi>0)=c_1+2\Lambda\left(\Bigl(1+\frac{2\Lambda}{f_0-c_1}\Bigr)e^{d(1+\frac{c_1}\Lambda)\chi}-
1\right)^{-1}
\end{equation}
(up to translations in $\chi$). As $c_1$ increases
from 0 to $f_0$ derivative $\left.\frac{df}{d\chi}\right|_{f=f_0}$ increases monotonically from
$-f_0d\left(\frac{f_0}{2\Lambda}+1\right)$ to 0. If for a given $d$ the
(unique up to translations) $f$ going to 1 as $\chi\to-\infty$ has
$\left.\frac{df}{d\chi}\right|_{f=f_0}\in[-f_0d(\frac{f_0}{2\Lambda}+1);0]$
it will go asymptotically to the
corresponding $c_1$ as $\chi\to +\infty$, namely it will be exactly
\p{fc1} where $f<f_0$; if its derivative is more negative, it will be
described by \p{fc1} with negative $c_1$ until it intersects $f=0$ at
some finite $\chi$ (and this is not a solution we are interested in).
Easily established monotonousness properties (how the slope 
$\left.\frac{df}{d\chi}\right|_{f=f_0}$ varies with $d$) prove this
way the claim that there is a unique $d$ for which the physical
solution (according to its behavior near $f=1$) goes asymptotically
to zero as $\chi\to-\infty$. 
For the $\Phi(f)$ \p{Phikolm}, on the other hand, there are \emph{no} solutions
going asymptotically to any constant but 0 as $\chi\to+\infty$, but those going
to 0 represent a two-parameter set of solutions, and depending on the $b_\pm$ in
their asymptote the $\left.\frac{df}{d\chi}\right|_{f=f_0}$ may take different
values, thus making it possible to match this slope of the solution physical near $f=1$ for
a range of $d$\rq s. (In this case $f_0$ denotes some intermediate value
of $f$, between 0 and 1.)

\indent 3) $d=2$: asymptotic solution \p{krightasym} must be rewritten as
$f(\chi)=(b+b_0\chi)e^{-\frac d 2 \chi}$, and there is again a 2D domain of
physical $(b,b_0)$, yielding $f>0$, $f\to 0$ as $\chi\to+\infty$. In this case a
meaningful burning profile exists as well.\smallskip

\begin{figure}[htbp] \begin{centering}
\includegraphics[width=5.1in]{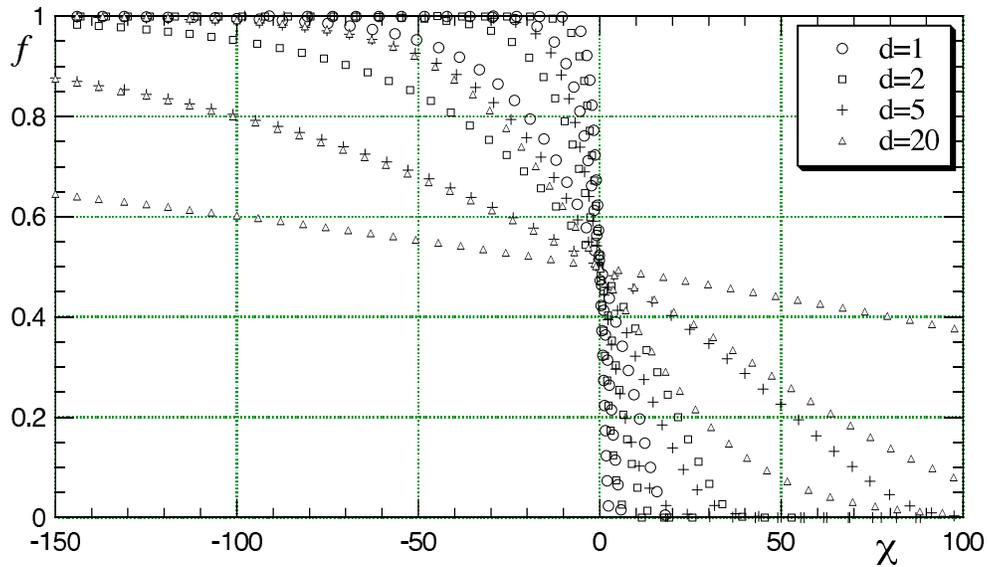}
\caption[KPP flame profiles at different $\Lambda$ and $d$]
{KPP flame profiles at different $\Lambda$ and $d$.
For each $d$ 3 curves are depicted, for $1/\Lambda=0.05,\,5,$ and 20,
larger $1/\Lambda$ corresponding to the curves
intersecting $\chi=0$ axis (where $f$ was set to 0.5) at larger
angles, and having larger widths.
\label{xkpp}}
\end{centering}\end{figure}

\begin{figure}[htbp] \begin{centering}
\includegraphics[width=5.1in]{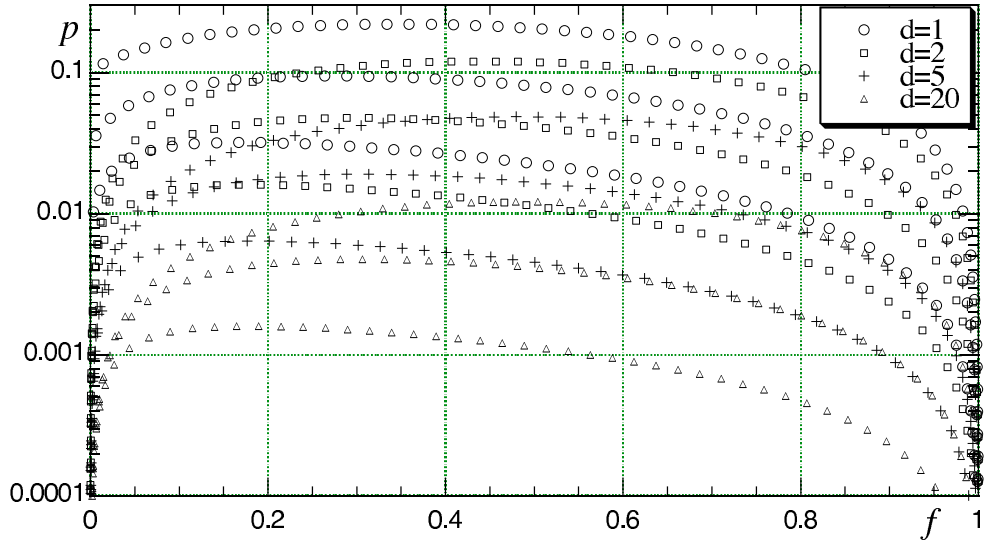}
\caption[Slope $p(f)$ of the profile of KPP flame]{Slope $p(f)$ of the profile of KPP flame, integrated from $f=1,\, p=0$ at $\Lambda$ and $d$ as in Fig.~\ref{xkpp};
larger $1/\Lambda$ correspond to smaller $p$.
Note that at $d=1$ $\,p(f=0)$ is non-zero, in contrast with the $d\ge 2$ curves.
\label{pkpp}}
\end{centering}\end{figure}

Flame profiles $f(\chi)$ found numerically for $\Phi_0(f)=f(1-f)$
are shown in Fig.~\ref{xkpp} for four values of $d$. These seem to satisfy the boundary
conditions for $d\ge 2$, whereas the profiles for $d=1$ (each integrated
from $p(1)\equiv -df/d\chi|_{f=1}=0$) intersect $f=0$ at
finite $\chi$ with non-zero $df/d\chi$. Corresponding integral curves
$p(f)$ at the same values of $d$ and $\Lambda$ are presented in Fig.~\ref{pkpp}.
Note that for $d=1$ \ $p|_{f=0}\ne 0$, whereas for $d\ge 2$ it was
checked that by refining the grid $p(0)$ became correspondingly closer to 0,
down to $p(0)\approx 10^{-10}$ at the bulk grid spacing
of $10^{-7}$ (at finer grid rounding errors in double precision reals dominated).
This numerically confirms our qualitative conclusions about continuous spectrum $[2;\infty)$
for $d$ in KPP model.

One can observe that the flame width grows fast with $d$. As in the original
model (\cite{KPP}) 
one may conclude that only propagation with the smallest velocity is stable
(only asymptotes near $f=0$ are essential for the
argument to hold, and these have the same exponential form regardless of $\l$). 
Of course, if the $f(x)$ at some time corresponds to some
eigenfunction above, such a profile will propagate with
corresponding $d$. But if one considers a process of setting up the
steady-state propagation, with initial $f(x)$ corresponding to pure
fuel on a half-line (or $f$ decaying with $x$ faster than the fastest
exponent $\lambda_-(d=2)=1$ in \p{krightasym}) the resulting self-similar 
front will be the the
eigenfunction with the smallest velocity. More generally, by
considering the evolution of some superposition $\Psi$ of the
steady-state profiles found above one concludes that amplitudes of
all of them but one will (asymptotically) decrease in time in favor of the one with
the smallest $d$ in the spectrum of $\Psi$ (they interact due to
nonlinearity of the system). As a generic
perturbation is a superposition of all eigenfunctions, however
small it is it will eventually reshuffle the profile into that for
smallest $d$.

\subsection{General burning rate}\label{qual:gen}
Observations above may be extended for more general burning rate function as follows:\\
\indent i) If $\Phi_0(f)$ goes to some positive constant as $f\to 0$ there
are no solutions going to 0 as $x\to+\infty$. From a physical viewpoint a
system reacting with finite rate in the initial state is unstable and
self-similar solutions cannot exist. Problems with
$\Phi_0(0+)=\nu<0$ do have eigenvalues (discrete spectrum). Corresponding eigenfunctions are 
identically zero on the right of some $\chi_2$, $f=-\nu(\chi-\chi_2)^2/2+O((\chi-\chi_2)^3)$ 
as $\chi\to\chi_2-$. sKPP model studied in the next chapter is of this type.\\
\indent ii) For $\Phi_0(f)=\mu f +o(f)$ at $f\to 0$ non-negative solutions going to 0 as
$x\to +\infty$ exist iff $d\ge 2\sqrt{\mu}$, $\mu>0$ for $\Phi_0(f)$ to be positive (what
is usually assumed in  literature. See below for $\mu<0$ case). 
General solution getting to a vicinity of  $f=0$ decays
exponentially. 
Analysis near $f=1\Leftrightarrow\bar{f}=1-f\ll 1$ then
suggests the following for
$\Phi_0=\bar{\nu}+\bar{\mu}\bar{f}+o(\bar{f})$ (below $d\ge 2\sqrt{\mu}$
is assumed; $f(-\infty)=1$ required): \\
\indent\hspace{3mm}$\bullet$\ $\bar{\nu}>0,\bar{\mu}> 0$ --- for all $d$
(the above $d\ge 2\sqrt{\mu}$ assumed) a unique
profile exists and $\exists x_1: \forall x<x_1\; f(x)\equiv 1$.
\\
\indent\hspace{3mm}$\bullet$\ $\bar{\nu}>0,\bar{\mu}\le 0$ --- a unique
profile whenever $\bar{d}\equiv d(1+\frac 1\Lambda)\ge 2\sqrt{-\bar{\mu}}$
(\ie\ for $|\bar{\mu}|>\mu(1+1/\Lambda)^2$ the spectrum is additionally shrunk);
again identically 1 on a half-line. For $\bar{d}< 2\sqrt{-\bar{\mu}}$
there may exist solutions oscillating around $f=1$, glued to $f=1$ on some $(-\infty;x_1]$.
These violate $0\le f\le 1$, yet may be used in FC in principle. Besides, $f(x_1+)<1$
and there may exist situations when $f$ does not exceed 1 anywhere.
\\
\indent\hspace{3mm}$\bullet$\ $\bar{\nu}<0$ --- no physical solutions. For $\bar{\mu}<0$
and $\bar{d}< 2\sqrt{-\bar{\mu}}$ there may be solutions glued to $f=1$ on some $(-\infty;x_1]$,
but necessarily $f(x_1+)>1$.
\\
\indent\hspace{3mm}$\bullet$\ $\bar{\nu}=0,\bar{\mu}> 0$ --- a unique
solution exponentially approaching 1 as $x\to-\infty\;\forall d$. No
physical solutions if $\bar{\mu}< 0$. 
\\
\indent\hspace{3mm}$\bullet$\ $\bar{\nu}=\bar{\mu}=0$. If
$\Phi_0(f)=\bar{q}\bar{f}^{\bar{r}}+o(\bar{f}^{\bar{r}})$, $\bar{r}>1$,\ $\forall d$ a unique solution
exists. 
The solution of \p{master1}
may be written as
$p=\frac{\bar{q}}\bd \fb^{\bar{r}}\bigl(1+\frac \fb{\Lambda+1}-\frac{\bar{r}\bar{q}}{\bd^2}\fb^{\bar{r}-1}+
O(\fb^2+\fb^{\bar{r}})\bigr)$,
leading to $\fb\sim\bigl((1-\bar{r})\bar{q}\chi/\bd\bigr)^{\frac 1{1-\bar{r}}}$ ($\bar{q}>0$ assumed).
If $\Phi_0(f)\equiv 0$ in some neighborhood of $f=1$ bounded solutions exist
but $f\to 1$ at $x\to -\infty$ \emph{from above}. Again, usable in principle even though $f(\chi)$
is not monotonic.\\
\indent\hspace{3mm}$\bullet$\ $\Phi_0\sim\bar{q}\bar{f}^{\bar{r}}$, but $\bar{r}\in(0;1)$.
$\fb$ vanishes identically on the left of some $\chi_1$, in its right neighborhood
$\fb\sim\bigl(\sqrt{\bar{q}/2(1+\bar{r})}(1-\bar{r})(\chi-\chi_1)\bigr)^{\frac 2{1-\bar{r}}}$.\\
\indent iii) $\Phi_0(f)=\mu f +o(f), \mu<0$ ($\Rightarrow \Phi_0(f)<0$ near $f=0$). For any $d$
a unique profile with $f(+\infty)=0$, exponentially approaching. Behavior near $f=1$ is described
as in ii), depending on $\Phi_0(f\to 1-)$, but because of the unique profile physical near $f=0$ 
the spectrum of $d$ is discrete (or empty).\\
\indent iv) For $\Phi_0(f)=qf^r$ near $f=0$ ($q>0$) there are no solutions of \p{master1}
with $p(+0)=0$ when $r<1$ -- a situation analogous to i). As a further analogy, when $q<0$
there may be a solution (unique up to translations); for this $\exists x_2:\: f=0$ at 
$x>x_2$, $f(x\to x_2-)\approx \left((x_2-x)(1-r)\sqrt{-q/(2(r+1))}\right)^\frac{2}{1-r}$.
Using $r=0$ here yields corresponding behavior in i).
When $r>1$ on the other hand,
there are multiple solutions going to 0 (the ODE\rq s peculiarity; general solution decays as 
$f\approx\left( (r-1)qx/d\right)^{1/(1-r)}$), hence one expects the same behavior
as in ii) (apart from this power-law tail into fuel), depending on the $\Phi_0$ shape near $f=1$.
When $\Phi_0(f)\equiv 0$ in some
neighborhood of $f=0$ one expects a discrete spectrum of $d$, with corresponding
eigenfunctions behaving near $f=1$ according to the $\Phi_0(f)$ near 1, as in ii).

Summing this up, 
KPP-like behavior is quite universal for $\Phi_0(f)$
going to 0 linearly or faster at $f\to 0$.
$\Phi_0(f)\equiv 0$ near $f=0$ leads to discrete spectrum; 
positive $\Phi_0(f)$ decreasing slower than linearly
(or not going to zero) as $f\to 0$ leads to absence of steady-state solutions.
Eigenfunction $f(x)$ has an infinite tail at $f=1$ if $\Phi_0|_{f\to 1}$
decreases linearly or faster. 

It should be stressed that these conclusions
are based on analysis of asymptotic behavior of solutions near $f=0$ or 1;
as with KPP case considered above the fact that there is a 2-parameter set of
solutions going to zero as $f\to 0$, say, is not enough to make claims as to global
behavior of a solution physical near 1 --- it may still become negative or unbounded
near 0.
For the cases with claimed continuous spectrum of eigenvalues the way the second term
in \p{master1} damps the solution near $f=0$ suggests that for any fixed $\Lambda$ the spectrum
contains all reals above some $d_\Lambda$. Some estimates of this lowest eigenvalue (for the
case without expansion), and references can be found in \cite{xin} (I am grateful to Lenya Ryzhik for
pointing out this review to me). Numerical studies of several models (with $(r,\bar{r})=(2,1)$;
$(5,1)$;$(0.5,1)$;$(1,2)$;$(1,0.5)$) agree with general claims of previous paragraphs. A new element
appears (in contrast to KPP) for more general power law decay of the burning rate near $f=1$
and $f=0$: the lower boundary of eigenvalues $d_\Lambda$ may depend on $\Lambda$. Say, for
$\Phi_0=f^2(1-f)$ it was observed that $d=1$ was an eigenvalue for $1/\Lambda=\{0.05;4;20\}$;
$d=0.464$ was an eigenvalue for $1/\Lambda=\{4;20\}$ but not 0.05; and $d=0.215$ was an eigenvalue
for $1/\Lambda=20$ but not 4 or 0.05. Asymptotic behavior of solutions near $f=0$ or 1 is the 
same for any positive $d$, so no estimates for the $d_\Lambda$ follow from local consideration
(near ends of integration interval for the $f$. Unlike the case for KPP model where $d_\Lambda=2$
is pointed out by local analysis.)

It does not seem feasible to use KPP-like profiles in FC
because of the continuous spectrum of the velocities, thus long times of
relaxing to steady state, and large widths with two exponential tails
(thus no way to localize the \fl\ compactly but still have the steep gradient
region well resolved on the grid).

\section{Step-function: velocities and widths}\label{1ddw}
\subsection{Analytic solution}
Analytic solution of \p{artvislamdim}
with constant diffusivity and step-function  $\Phi_0(f)$ (original implementation, \cite{X95})
as found in \cite{zh07} is summarized below. 

Translational invariance is fixed by considering a particular
solution with $f(\chi=0)=f_0$. In the region $\chi>0$ $\Phi_0(f)=0$ (no reaction) and solution is a diffusive
exponential tail (analog of the preheating zone in standard Arrhenius flames) corrected by matter expansion 
term in \p{artvislamdim}, only significant where $f$ is not too small:
\begin{equation}\label{sol2}
f=2\Lambda\left[\Bigl(1+\frac{2\Lambda}{f_0}\Bigr)e^{d\chi}-1\right]^{-
1}\approx\Bigr(\frac 1{2\Lambda}+\frac 1{f_0}\Bigr)e^{-d\chi}.
\end{equation}

Last approximation is for $\chi\gg\frac{1}{d}\ln
\bigl(1+{2\Lambda}/{f_0}\bigr)$. When this region where $f$ exponential decay is perturbed by expansion term
is narrower than characteristic width of exponential decay ($1/d$) flame width can be reasonably characterized
by the slope of the profile $f(\chi)$ at $f=f_0$; there the slope is maximal, thus this way one gets an estimate
of the width from below:
\begin{equation}\label{w1}
w_a=\frac {f_0}{\left|df/d\chi\right|_{f=f_0}}=\left[d\left(1+\frac{f_0}{2\Lambda}\right)\right]^{-1}.
\end{equation}
For comparison, to quantify how long the tail is compared to the rest of the flame, we also employ another
width estimate, 
\begin{equation}\label{w2}
w_b=\frac{\chi(f=f_-)-\chi(f=1)}{1-f_-}
\end{equation}
with $f_-=0.1$. Analytical expression was found for $w_b$; if $f_0\ge f_-$ 
\begin{equation}\label{w2exp}
w_b=\frac{1}{d}\left[\ln \frac{1+2\Lambda/f_-}{1+2\Lambda/f_0}
+d^2\Bigl(1+\frac{1}{2\Lambda}\Bigr)\right].
\end{equation}
This was computed using solution \p{sol2} at $\chi>0$ and value
\begin{equation}\label{xi1}
\chi_1=-d\left(1+\frac{1}{2\Lambda}\right)
\end{equation}
for the rightmost point where $f=1$, found directly from \p{artvislamdim} integrated once over $\chi$.

On $\chi\in(\chi_1;0)$ solution of \p{artvislamdim} was found as 
\begin{equation}\label{genphi}
f/\Lambda=-1-\left(\frac{6\sigma}{d^2\Lambda}\right)^{1/3}\left(\frac{1}{3\sigma}+
\frac{d}{d\sigma}\ln\left(\mathrm{I}_{1/3}(\sigma)+B\mathrm{K}_{1/3}(\sigma)\right)
\right),
\end{equation}
where 
\begin{equation}\label{sig}
\sigma= \left(\frac 2\Lambda \right)^{1/2}\left(d\chi-
\frac{d^2\Lambda}2\right)^{3/2}
\end{equation}
and $\mathrm{I},\k$ are modified Bessel functions; $B$ is an integration constant (real; another
constant was fixed by requiring $df/d\chi$ to be continuous at $\chi=0$). 

This solution has to satisfy boundary conditions following
from (\ref{bndl}--\ref{bndr}),

\begin{equation}\label{bdy}
f=\left\lbrace\begin{array}{ll}
f_0 & \mbox{at } \sigma_0=d^2\Lambda /6\\
1 & \mbox{at } \sigma_1=\left(1+\frac 1\Lambda\right)^3\sigma_0
\end{array}\right. ,
\end{equation}
thus leading to  a transcendental equation defining sought for eigenvalue for $d$ (entering through 
$\sigma_0$ and $\sigma_1$ arguments):
{\setlength\arraycolsep{2pt}
\begin{eqnarray}\nonumber
& &\left(\mathrm{I}_0\Bigl(1+\frac{f_0}\Lambda+\frac
1{3\sigma_0}\Bigr)+\mathrm{I}_0^\prime\right)\left(\k_1\Bigl(1+\frac
1{3\sigma_1}\Bigr)+\k_1^\prime\right) \\ \label{eigd}
& &\qquad -
\left(\k_0\Bigl(1+\frac{f_0}\Lambda+\frac
1{3\sigma_0}\Bigr)+\k_0^\prime\right)\left(\mathrm{I}_1\Bigl(1+\frac
1{3\sigma_1}\Bigr)+\mathrm{I}_1^\prime\right)=0.
\end{eqnarray}}%
Notation is $\mathrm{I}_0=\mathrm{I}_{1/3}(\sigma_0)$, $\k_1=\k_{1/3}(\sigma_1)$, etc.
\ With definitions above this
equation has a unique solution for $d$ for every $f_0,\:\Lambda$.

\begin{figure}[htbp] \begin{centering}
\includegraphics[angle=270,width=5.1in]{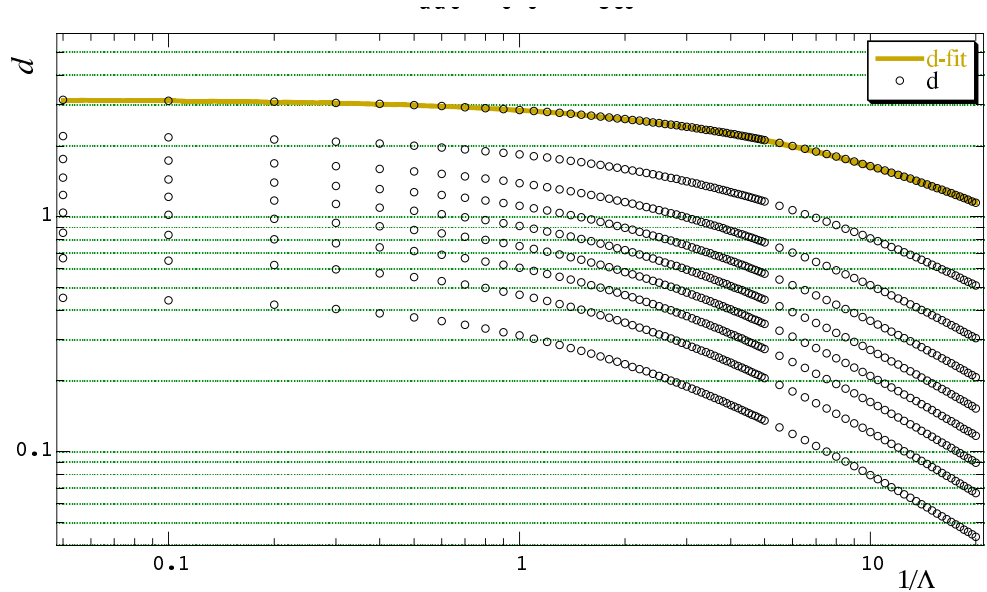}
\caption[Step-function burning rate. Flame velocities]
{Flame velocities. The nine sequences
$d(\Lambda)$
correspond (top to bottom) to $f_0=0.1,$\ldots 0.9 with step 0.1. The worst
$(f_0$=0.1) fit (\ref{mf0}) is shown.
\label{stepd}}
\end{centering}\end{figure}

\begin{figure}[htbp] \begin{centering}
\includegraphics[angle=270,width=5.1in]{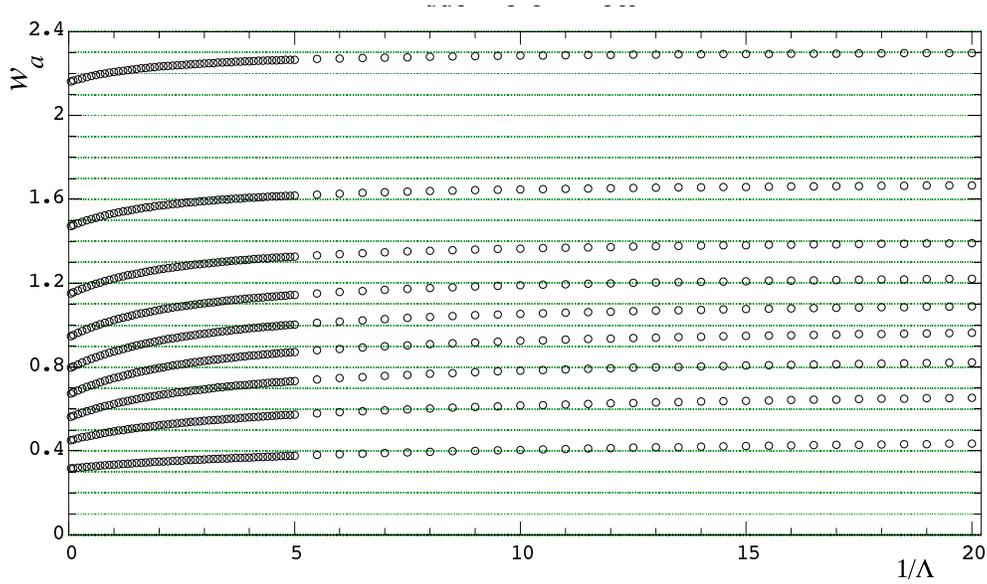}
\caption[Step-function burning rate. Flame width $w_a$]
{Flame width $w_a$ at the same $\Lambda$ and $f_0$.
Larger widths correspond to larger $f_0$.
\label{stepwa}}
\end{centering}\end{figure}

\subsection{Results for velocities and widths}\label{stepdw}
\begin{figure}[htbp] \begin{centering}
\includegraphics[angle=270,width=5.1in]{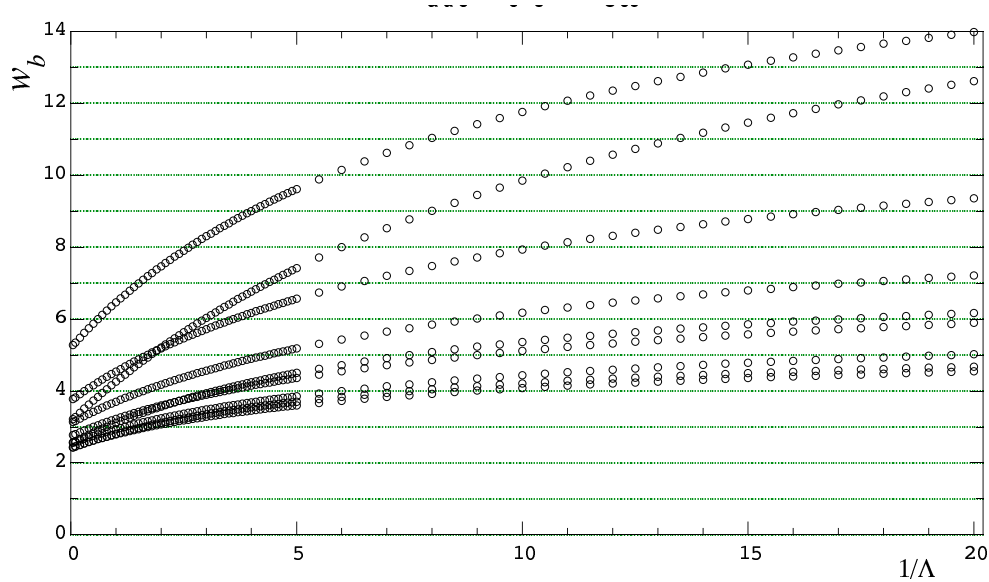}
\caption[Step-function burning rate. Flame width $w_b$]
{Flame width $w_b(f_0,\Lambda)=-\chi|_{f=0.1}^1 \Bigm/0.9$. The order of $f_0$ 
(from larger to smaller $w_b$ at $1/\Lambda=20$) is 0.9, 0.1, 0.8,
0.7, 0.2, 0.6, 0.5, 0.3, $0.4\,$.
\label{stepwb}}
\end{centering}\end{figure}

Numerical solutions of \p{eigd} are presented in Fig.~\ref{stepd}. %
Figs.~\ref{stepwa} and \ref{stepwb} show $w_a$ and $w_b$ for these two models.
Flame profiles as in Fig.~\ref{stepx} were obtained by direct numerical integration of
\p{master1} with $d$ from Fig.~\ref{stepd}. Relative difference between
$d$\rq s found numerically and analytically (\ie\ by solving \p{eigd})
is less than $5\times 10^{-3}$\%. It was that large for $f_0\ge 0.7$;
for $f_0\in[0.3; 0.6]$ the difference was of order $10^{-8}$, the accuracy
required of the $d$ in numerical procedure (accuracy for solving \p{eigd} was set to
$4\times 10^{-16}$ and apparently these, \lq\lq analytic\rq\rq\ errors played
no role), and monotonically
increased to $20\times 10^{-8}$ with $f_0$ further decreasing to $0.1$.
Errors in widths followed similar
trends (and \lq\lq numerical\rq\rq\ widths were consistently larger than
\lq\lq analytical\rq\rq\ ones, whereas velocities were smaller), though
there was additional contribution from crude estimation (up
to the whole grid) from integral curves $x(f)$ obtained from $p(f)$ via
further trapezium integration; the discrepancy in $w_{a,b}$ was less than 
$8\times 10^{-3}$\%.

For the SN Ia deflagration problem matter
expansion is not large, thus it is worth trying to treat $1/\Lambda$ as a small
parameter. A first-order correction to the solution $h_0\equiv d(1/\l=0)^{-2}$
of \p{Lambdainf} is (\cite{zh07})
\begin{equation}\label{dasym}
d^{-2}=h=h_0\left(1+\frac{h_1}{2\Lambda}+O(\Lambda^{-2})\right),\; 
h_1=5h_0-e^{-1/h_0}\frac{2+h_0-h_0f_0}{f_0-e^{-1/h_0}}.
\end{equation}
When $f_0$ is small as well this may be estimated with the aid of
expansion from the end of Sec.~\ref{step:qual} for $h_0$:
\begin{equation}\label{dasymf0}
h_1=5f_0-e^{-1/f_0}\left(\frac 2{f_0}+1-6f_0\right)
-e^{-2/f_0}\left(\frac 4{f_0^2}+\frac 2{f_0\vphantom{f_0^2}}-6-6f_0
\right)+O(e^{-3/f_0}/f_0^3),
\end{equation}
this is usable up to $f_0\approx 0.3$.
Error of using \p{dasym} does not exceed 1\% for $1/\Lambda\le 1$ (but $d=h^{-1/2}$ must be used.
Expanding $d$ up to $O(1/\Lambda)$ leads to far worse agreement.) At $1/\Lambda=3$ the error
grows to 6.7\% for $f_0=0.3$ (1.8\% for $f_0=0.2$); at $\Lambda>0.69$ (SN Ia problem, 
$\rof\ge 3\times 10^7\:\gcm$) it is
within 2.1\% for $f_0=0.3$, 0.16\% for $f_0=0.2$.
At these $f_0$ both expressions for $h_1$ give the same agreement.

\section{Alternative flames with finite widths}\label{stepK}
In this section we present results for flame model \p{artvisfront0} with non-constant diffusivity,
$K_0(f)=f^r$, $r\in(0;1)$, and step-function rate \ref{Phistep}.
It was studied in \cite{zh07} for a convenient in FC feature of flame profiles being localized, 
with no tails (as a further improvement
over flame model due to \cite{X95} studied in the previous section, which produces flames with 
\lq\lq preheating\rq\rq\ exponential tails, and KPP with long tails into both fuel and ash). 

Like in Sec.~\ref{steadyK1} we are looking for a solution of eigenproblem \p{master1} with $p(0)=p(1)=0$ 
(as before $p=-df/d\chi\equiv -\sqrt{\tilde{K}/R}\,df/dx$).
The above $K_0(f)$ leads to $f(\chi)$ $C^1$-smoothly monotonically interpolating
between $f=1\: \forall \chi<\chi_1=-d(1+1/2\Lambda)$ (fixing $f(0)=f_0$;
$f=1-(\chi-\chi_1)^2/2+d(\chi-\chi_1)^3/2+\ldots$ at $\chi\to\chi_1+$) and
$f=0\:\forall \chi>\chi_2$ ($f=(rd(\chi_2-\chi))^{1/r}(1+O(\chi_2-\chi))$ at
$\chi\to\chi_2-$).

$\chi_2$ and the total width $w_c=\chi_2-\chi_1$ may be expressed in elementary
functions of $f_0$, $\Lambda$ and $d$ for rational $r$ (or as incomplete
$\Gamma-$function in general). Values $r=1/2$ and $r=3/4$ seem most adequate.
Corresponding widths are
{\setlength\arraycolsep{2pt}
\begin{eqnarray}\nonumber
w_{c,\frac 12}&=&\frac 1d\int_0^{f_0}f^{r-1}\left(1+\frac f{2\Lambda}\right)^{-1}
\,df\biggr|_{r=1/2}-\chi_1\\ \label{w312}
 &=&\frac 2d\sqrt{2\Lambda}\arctan\sqrt{\frac{f_0}{2\Lambda}}+d\left(1+\frac 1{2\Lambda}\right),\\ \nonumber
w_{c,\frac 34}&=&2^{5/4}\Lambda^{3/4}d^{-1}\left[\textrm{Arctan}\,\frac{(2f_0/\Lambda)^{1/4}}
{1-(f_0/2\Lambda)^{1/2}} 
 +\ln\frac{(f_0/2\Lambda)^{1/2}-(2f_0/\Lambda)^{1/4}+1}
{(f_0/2\Lambda+1)^{1/2}}\right]\\ \label{w334}
 & &+\,d\left(1+\frac 1{2\Lambda}\right).
\end{eqnarray}}%
In the last expression the branch of Arctan to be used is Arctan$\,\in[0;\pi)$.

For any $r$ in the limit $\Lambda\to\infty$ (small expansion)
{\setlength\arraycolsep{2pt} \begin{eqnarray*}
w_{c,r}&=&\frac{f_0^r}{dr}\left(1-\frac{f_0}{2\Lambda}\,\frac r{r+1}+\Bigl(\frac{f_0}{2\Lambda}\Bigr)^2\,\frac
r{r+2}+O\Bigl(\frac{f_0}{2\Lambda}\Bigr)^3\right)+d\Bigl(1+\frac 1{2\Lambda}\Bigr)\\
 &=&\Bigl(\frac{f_0^r}{rd_0}+
d_0\Bigr)+\frac 1{2\Lambda}\left(d_0\Bigl(1-\frac{h_1}2\Bigr)-\frac{f_0^r}{rd_0}\Bigl(\frac{f_0r}{r+1}-
\frac{h_1}2\Bigr)\right)+O(\Lambda^{-2})
\end{eqnarray*}}%
(as in the $r=0$ case $d(f_0,r|\Lambda)\equiv h^{-1/2}=d_0(1+\frac{h_1}{2\Lambda}+\ldots)^{-1/2}$;
$d_0=h_0^{-1/2}$, $h_1$, etc. are now some functions of $f_0$ and $r$);
as $\Lambda\to 0$ $w_c$ diverges as
{\setlength\arraycolsep{2pt}
\begin{eqnarray*}
w_{c,r}&=&\frac{(2\Lambda)^r}{d}\,\frac{\pi}{\sin\pi r}-\frac{f_0^{r-1}}{1-r}\,
\frac{2\Lambda}d+O(\Lambda^r)-\chi_1 \\
 &=&\frac{2^r\pi}{G_0\sin\pi r}\Lambda^{r-1}(1-G_1\Lambda)
+\frac{G_0}2-\frac 2{G_0}\,\frac{f_0^{r-1}}{1-r}+O(\Lambda)
\end{eqnarray*}}%
We used $d=G_0\Lambda(1+G_1\Lambda+o(\Lambda))$ as in the step-function model with a standard diffusion term:
$d(\Lambda)$ dependence is qualitatively very similar. Fits of $d(\Lambda)$ are presented in the next section.

\begin{figure}[htbp] \begin{centering}
\includegraphics[angle=270,width=5.9in]{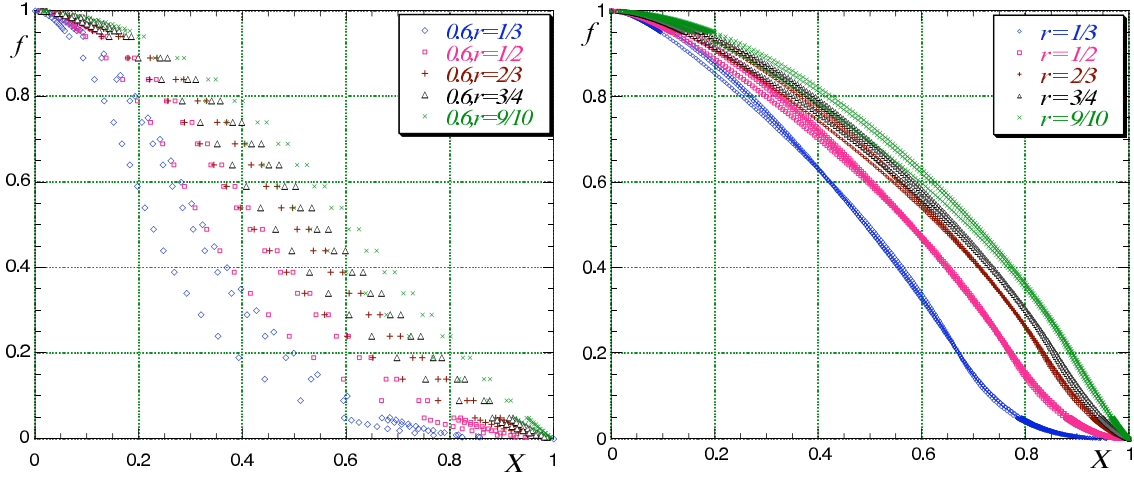}
\caption[Normalized flame profiles for different diffusivity exponents $r$]
{Normalized to unit width flame profiles with $f_0=0.6$ (left)
and $f_0=0.2$ (right). The three curves for each $r$ correspond to $1/\Lambda\in\lbrace
0.16;0.6;3\rbrace$, lower near $f=0$ profiles for larger $1/\Lambda$.
\label{kx}}
\end{centering}\end{figure}

For a range of $\{f_0,r\}$ flame profiles deliver what they were expected to originally.
Their width $w_c$ may consistently serve as a measure of $f$ gradients, and upon coupling \p{artvisfront0} to
the hydrodynamic equations one would really get reasonably uniform heat release within the flame
width. On the contrary, for models with traditional diffusion term $w_b$ was somewhat
arbitrary quantity, most of it (for larger $f_0$ and KPP to a greater degree) might correspond to
\lq\lq tails\rq\rq\ of a profile, and the heat release in FC would remain localized, contrary to
the intention to more-or-less uniformly distribute it over a few cells near the modeled flame front.
Some representative flame profiles are shown in Fig.~\ref{kx}; they are normalized to unit width (that is,
reexpressed in terms of a new $X=(\chi-\chi_1)/(\chi_2-\chi_1)$; resulting
supp$(df/dX)=[0;1]$). Three specific combinations $(f_0;r)$ are of particular interest for use in FC:\\
\indent\hspace{3mm}$\bullet$\ $r=1/2$. This has advantage that $f(x)$ behavior near
$f=0$ is the same as near $f=1$, thus we are unlikely to introduce new problems
(compared to the original $r=0$ scheme). $f_0=0.2$ is then convenient as the $f(x)$ shape
is least sensitive to $\l$ in its range of immediate interest, $1/\Lambda\in[0;2]$.\\
\indent\hspace{3mm}$\bullet$\ $r=3/4$, $f_0=0.6$. For the $\Lambda$\rq s of interest corresponding
flame profiles seem most symmetric overall with respect to $f\mapsto 1-f$; this is perhaps a
better realization of the above idea: as the whole $w_c$ width is to be modeled on a few
integration cells this approximate global symmetry seems more adequate to consider than
the symmetry of minute regions near $f=0$ and $f=1$.\\
\indent\hspace{3mm}$\bullet$\ $r=3/4$, $f_0=0.2$. Profile gradients are still uniform enough
on $[0.1;0.9]$, they drop to zero in a regular manner within reasonable $\Delta\chi\leq 0.25w_c$.
At $f_0=0.2$ the profiles seem least sensitive to $\l$, ensuring consistent performance in all
regions of the star. 

As another possibility it might seem more physical to write the reaction-diffusion equation as
\begin{equation}\label{artvisKrho}
\frac{\partial(\rho f)}{\partial t}+\bnabla(\rho f\mathbf{u}))= 
\bnabla\left[\rho\tilde{K}K_0(f)\bnabla f\right]+\rho R\Phi_0(f),
\end{equation}
with density appearing under divergence in the third term, $\nabla(\rho K\nabla f)$. 
This case was studied as well. Normalized
to unit total width flame profiles are more sensitive to $\l$ than the ones
corresponding to \p{I1}; such a model is also less tractable
analytically. The model with diffusion term of this form and with $K(f)=const$
was also studied as a possible alternative to the original one \p{artvisfront0}.
Asymptotic behavior of $d$ and the widths at small and large $\l$ are the same as
presented in the previous section, hence the decision to stick to the original
(apparently consistently performing) model. Physical transport
coefficients change after flame passing, making it
questionable if putting the $\rho$ under $\nabla$ really makes the model in this paragraph
\lq\lq more physical\rq\rq , in view
of the artificial $K_0(f)$ dependence chosen vs effective $K(f)$ for turbulent burning (the latter
are higher in the ash, qualitatively similar to our $K_0(f)$.) 

\section{Suggestions for implementation}\label{1dsuggest}
\subsection{Fits of the flame speed dependence on expansion}
Analytically found asymptotes for $d(\Lambda)$ (Sec.~\ref{stepdw}) at small $1/\Lambda$
do not provide enough accuracy to be used for flame tracking in 
outer layers of WD (although the errors are within 1\% for $f_0=0.2$ and $1/\Lambda<2$.)
Next order in $1/\Lambda$ seems sufficient
at $f_0\le 0.2$ yet computations become too involved in $r\ne 0$ case
(Sec.~\ref{stepK}). More importantly, flame capturing as presented is a
general method, thus it is desirable to get results with larger range
of validity in a ready to use form. In this section we present fits
neatly interpolating between small and large $\Lambda$ regions and then
summarize the procedure for getting $R$ and $\tilde{K}$ for \p{I1}
in the SN Ia simulations.

\begin{equation}\label{mf0}
d(\Lambda)=\frac{m_1}{1+m_2/\Lambda}+\frac{m_3}{(1+m_4/\Lambda)^2}
\end{equation}
with $m_{1\ldots 4}$ obtained at each $f_0$ to minimize the mean square
deviation (with weights proportional to actual $d(\Lambda)$)
was the simplest fit found.
The values shown in the graph below guarantee 0.2\%
accuracy for each $f_0\in\{0.01;0.025(0.025)0.975\}$ and
$\Lambda\in[10^{-3};10^5]$ studied (for any $\Lambda$ in fact, as comparison
of asymptotes shows). For $f_0>0.3$ agreement is significantly better.
\begin{figure}[htbp] \begin{centering}
\includegraphics[width=5.1in]{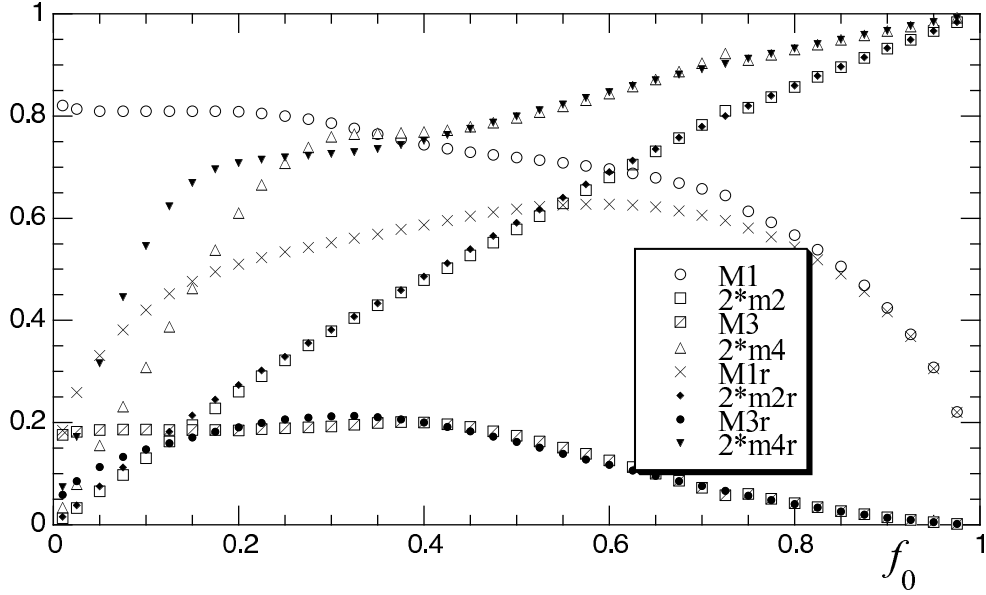}
\caption[Fit $d(\l)$ parameters for $r=0$ and $r=3/4$, dependence on $f_0$]
{Fit \p{mf0} parameters for $r=0$ (the first 4 curves according to the legend) and $r=3/4$ (next 4).
$M_{1,3}=m_{1,3}\sqrt{f_0}$.
\label{mfit}}
\end{centering}\end{figure}
I have not succeeded to find a simple expression for $m_{1\ldots 4}(f_0)$
providing good agreement for all expansions; this is due to delicate
interplay between the two terms near $1/\Lambda=0$. Complicated form of
$m_{1\ldots 4}(f_0)$ is related to different significance of the two asymptotic
regions in the fit: for $f_0=0.01$ the $d(\Lambda)$ becomes reasonably linear
($d/\Lambda\approx \mathrm{const}$) only at
$1/\Lambda>300$, whereas for $f_0>0.5$ this happens at $1/\Lambda\ge 2$.
To overcome this a number of other possible fits were tested. In short,
3-parameter fits were significantly less accurate (though 3\% accuracy was
achieved, uniformly on $\Lambda\in(0;\infty)$), fits with more than 4 parameters did not yield significant
improvements; none of accurate fits considered admitted simple expressions for its
coefficients in terms of $f_0$.

This is not a major issue as $f_0$ is a parameter one can fix at some
convenient value throughout the simulation. $f_0=0.2$ seems most suitable for the finite width flames
with $r=1/2$  or $3/4$, as discussed in Sec.~\ref{stepK}. 
For constant diffusivity and step-function burning rate 
of \cite{X95}, with new coefficients prescription described here, we suggest using $f_0=0.2$, based on the
following consideration.
In \cite{X95} $f_0$ was fixed at 0.3, which essentially yielded thinnest
flames throughout the $\Lambda$ range in the problem (A. Khokhlov,
private communication. Note that this agrees with our findings, cf. Fig.~\ref{stepwb}.)
The width of the \fl\ of the original model is $W\simeq w_b(f_0,\Lambda)\beta\Delta x$; 
$\beta=1.5$ was used, $\Delta x$ is a grid size; it changes with expansion $\l$
proportionally to $w_b(f_0,\Lambda)$, the magnitude of this change may be seen in Fig.~2.6.
\ Now, the prescription I advocate normalizes
width as well as the flame speed\footnote{%
We can now also easily quantify the error in flame speed achieved in \cite{X95}: 
the values of $K$ and $R$ used there result in
actual flame speed larger than the prescribed one by a factor of $d(f_0,\Lambda)\sqrt{f_0}$.
It is reasonably close to 1 at small $1/\Lambda$ and $f_0=0.3$: the flame speed that original model 
produces is 7\% smaller in the WD center. However it is $\sim 1.45$ times smaller when $\rho=3\times 10^7$,
at 0.5C+0.5O composition.
}, thus the logical criterion for $f_0$ to
suggest would be the fastest tail (near $f=0$) decay in the profile normalized
to unit width $w_b$, for the case of traditional FC realization.
This translates to smallest $1/w_bd$. This, however, can be made
arbitrarily small by taking sufficiently small $f_0$; we suggest to
stick to $f_0=0.2$, as this yields $1/w_bd$ least sensitive to $\Lambda$ in $(0.4;\infty)$;
flame profiles normalized to unit $w_b$ 
also exhibit fairly low sensitivity to $\Lambda$. At smaller $f_0$ at small $\Lambda\,(\leq 0.5)$
profile shapes on $f\in [0.1;1]$ become rather nonlinear, making the choice of $w_b$
as a measure for \lq\lq width\rq\rq\ more questionable.

Flame speed when matter expansion may be neglected, a solution of \p{Lambdainf}, 
value of which appears in expansion \p{dasym} for $d$ at small $\l$, 
can be approximated as \[
d_0=\left[\left(1-f_0^{1.4448}\exp(0.58058(1-f_0^{-1})\,)\right)\Bigm/ f_0
\right]^{0.5}, \]
with errors of $\le 0.64\%$ for any $f_0$. For $r=1/2$ and $r=3/4$ 
\[d_0=\left[2\left(1-f_0\right) f_0^{r-1}\right]^{1/2}\] 
provides 1.5\% accurate fits.

\begin{table}[!hbpt]
{\center \begin{tabular}{rrrrrrc}\hline
$f_0$ & $r\;$ & $m_1\;\;$ & $m_2\;\;$ & $m_3\;\;$ & $m_4\;\;$ & $\epsilon_d,\,10^{-4}\%$ \\ \hline
0.1& & 2.9248& 0.081016& 0.23690& 0.32153& 90\ \\ 
0.2& & 1.9588& 0.14574& 0.26929& 0.42824& 66\ \\
0.3& & 1.4620& 0.19389& 0.32605& 0.39759& 11\ \\
0.4& & 1.1889& 0.24193& 0.30495& 0.39044& 6.1\\
0.5& & 1.0312& 0.29358& 0.23117& 0.40545& 11\ \\
0.6& & 0.90646& 0.34328&0.15479& 0.42613& 7.2\\
0.9& & 0.44817& 0.46634&0.015034& 0.48381& 5.3$\vphantom{q_{\int_q}}$\\ 
0.2&3/4& 1.1683& 0.14139& 0.40205& 0.37378& 56\ \\
0.6&3/4& 0.81846& 0.34834& 0.14334& 0.42779& 9.0$\vphantom{q_{\int_q}}$\\ 
0.2&1/2& 1.3969& 0.14439& 0.35064& 0.39886& 52\ \\ \hline

\end{tabular}
\caption[Fit $d(\l)$ parameters $m_{1,2,3,4}$]
{Parameters of fit \p{mf0} for $d(\l)$ for flame models with burning rate \p{Phistep} and constant $K$
($r=0$ effectively, as in the original FC model: first 7 entries) and $K=f^r$ (last 3 lines). 
The last column shows the maximum relative discrepancy between the $d(\Lambda)$ and its fit 
(with $m_{1\ldots 4}$ truncated exactly as in the Table) at $\Lambda\in [1/3;20]$,
$\epsilon_d=(|\Delta d|/d)_\mathrm{max}$.\label{tab1}}}
\end{table}

The fit parameters $m_{1\ldots 4}$ for $f_0=0.1\ldots 0.9$, as well as for the
$\lbrace f_0,r\rbrace$ values suggested in Sec.~\ref{stepK} optimized for
$\Lambda\in [1/3;50]$ are summarized in Tab.~\ref{tab1}. 

\subsection{Prescribing normalizations of diffusivity and reaction rate for propagating the \fl }\label{prescr}
The strategy for using the results of this chapter in FC is as follows.
One picks the favorite flame model; this means a pair $\lbrace f_0,r\rbrace$ for the models considered in
this chapter (based on step-function burning rate, Eq.~\ref{Phistep}, and diffusivity $K=f^r$). 
One finds corresponding dimensionless steady flame 
speed $d$ and width $w$, in each
computational cell. For models studied in this chapter, this is accomplished by taking $m_{1,\ldots 4}$ from
Table~\ref{tab1} corresponding to the $\lbrace f_0,r\rbrace$ used, 
\p{mf0} then defines $d$ based on local $\Lambda$ \p{lam}, and
\p{w2exp} (or \p{w312} for $r=1/2$, or \p{w334} for $r=3/4$) determines dimensionless width
$w$ ($w_b$ or $w_c$). Next, one defines \emph{needed} $D_f(\Delta)$ (based on local
density, gravity, etc; \eg\ \p{St}, $\Delta$ meaning the grid spacing).
\begin{equation}\label{KR}
\tilde{K}=\frac{D_f}d\cdotp\frac{W}w,\quad  R=\frac{D_f}d\Bigm/\frac{W}w,
\end{equation}
(where $W$ is a desired flame width, say $4\Delta$) will then yield scale-factors for the parameters
\p{dimless} appearing in the equation governing $f$ evolution, \p{I1}. This equation with so
defined coefficients when coupled with standard equations of hydrodynamics will lead
to a \fl\ with the desired speed $D_f$ and width $W$ within the approximations adopted in this chapter
(steady 1D burning, pressure and $H\rho$ approximately constant across the flame, discretization effects 
neglected).

\section{Nonstationary numerical tests}\label{nsteady1D}

Here we briefly present numerical verification of the key results
of the previous sections, which were based on steady-state consideration. The goal 
is to see how well the prescribed velocity and width were achieved when prescriptions
of Sec.~\ref{1dsuggest} were used. Flame profile sensitivity to the expansion parameter is also 
studied here in realistic non-steady simulations. 

\subsection{Flame speed and width}
For the simulations presented below {\tt{ALLA}} code developed by A.~Khokhlov
[\cite{X98, X00}] was used. The simulations were performed in 1D without gravity,
with actual WD equation of state, and heat release corresponding to complete
0.5C+0.5O burning to the NSE composition at a given density. The 1D integration
domain (called \lq\lq tube\rq\rq\ below) was closed at one end (reflecting
boundary conditions); 4 cells of hot ash
were placed at this end for ignition, in hydrostatic equilibrium with cold fuel in
the rest of the tube. At the other end outflow conditions were imposed. 

Four models were studied: the three proposed at the end of Sec.~\ref{stepK}, and a model with constant 
diffusivity and $f_0=0.2$. For each model the simulation was run for 4 densities,
$\rho=\{3\times 10^7; 10^8; 3\times 10^8; 2\times 10^9\}\:\gcm$. These yielded expansions
$1/\Lambda=\{1.442; 0.7322; 0.3975; 0.1669\}$ respectively (released heat was 
$q=\{7.399; 5.736; 4.469; 3.382\}\times 10^{17}\:\mathrm{ergs\:g^{-1}}$). The coefficients
in Eq.~\p{artvisfront0} were determined from Eq.~\p{KR} with $D_f$ set constant $80\:\kms$ for all runs,
$W=4$ cells. $d$ was defined using \p{mf0} with relevant coefficients from Table~\ref{tab1}, 
exact expressions for $w_b$ (for Model $\lbrace f_0,r\rbrace=\lbrace 0.2,0\rbrace$) or total width $w_c$ 
(for the three finite width models emphasized in Sec.~\ref{stepK}) were used as $w$ in Eq.~\p{KR}. 

Table~\ref{tab2} summarizes the results.
$D$ is the measured flame speed, determined as $D=\frac{1}{\rho_\mathrm{fuel}}\frac{dm_\mathrm{ash}}{dt}$,
$W_{1,2,3}$ characterize the flame width. These widths were determined in the
following way (following A.~Khokhlov): at each timestep from 2001 to 10000 a
number of cells with $f$ between $f_d$ and $f_u$ was found, and then averaged over 
these 8000 steps. $W_1$ was obtained this way while choosing $(f_d;f_u)=(0.1;0.9)$,
$W_2$ --- with $(f_d;f_u)=(0.01;0.99)$, and $W_3$ with $(f_d;f_u)=(0.001;0.999)$. Steady state
profile had been established by timestep 500; at timestep 10000 the flame was still far enough
from the open tube end. For all the results in this Table the tube
length was 620 km, corresponding to 256 integration cells. The last timestep (10,000) corresponded to
1.8--2.6 s depending on density. 

\begin{table}[!hbpt]

{\center 
\begin{tabular}{|@{$\;$}c@{$\;$}|r@{$\;\quad$}r@{$\;\quad$}r@{$\;\quad$}r|r@{$\;\quad$}r@{$\;\quad$}r@{$\;\quad$}r|}\hline
$\rho$ & $D\vphantom{D^1}$ & $W_1$ & $W_2$ & $W_3$ &
         $D$ & $W_1$ & $W_2$ & $W_3$ \\
\cline{2-9}
 & \multicolumn{4}{@{$\!$}c@{$\!$}}{$(r;f_0)=(0.75;0.2)$}&
     \multicolumn{4}{|@{$\,$}c@{$\,$}|}{$(r;f_0)=(0.75;0.6)$} \\ 

\hline
$3\times 10^7\vphantom{10^{7^7}}$ & 76.4 & 2.69 & 3.97 & 4.96 & 
                 73.2 & 2.56 & 3.92 & 5.20 \\
$1\times 10^8$ & 77.8 & 2.70 & 4.02 & 4.92 &
                 74.2 & 2.59 & 3.99 & 5.24 \\
$3\times 10^8$ & 78.5 & 2.67 & 4.04 & 4.88 &
                 75.0 & 2.61 & 4.01 & 5.22 \\
$2\times 10^9$ & 78.9 & 2.64 & 4.04 & 4.70 &
                 75.6 & 2.61 & 4.01 & 5.12 \\
\hline
 \multicolumn{1}{|c}{}& \multicolumn{4}{@{$\!$}c@{$\!$}}{$(r;f_0)=(0.5;0.2)$}&
     \multicolumn{4}{|@{$\,$}c@{$\,$}|}{$(r;f_0)=(0;0.2)$} \\ \hline
$3\times 10^7\vphantom{10^{7^7}}$ & 76.4 & 2.55 & 3.77 & 4.75 & 
                 76.7 & 3.18 & 5.00 & 6.42 \\
$1\times 10^8$ & 78.2 & 2.58 & 3.85 & 4.76 &
                 78.2 & 3.22 & 5.23 & 6.81 \\
$3\times 10^8$ & 78.9 & 2.57 & 3.88 & 4.75 &
                 78.9 & 3.23 & 5.38 & 7.07 \\
$2\times 10^9$ & 79.4 & 2.54 & 3.91 & 4.61 &
                 79.4 & 3.22 & 5.49 & 7.30 \\
\hline

\end{tabular}
\caption[Flame velocities $D$ and widths $W_{1,2,3}$ at
different densities $\rho$ for 4 flame models.]
{Flame velocities, $D$ (in $\kms$), and widths $W_{1,2,3}$ (in cells) at
different densities $\rho$ (in $\gcm$) for 4 flame models\label{tab2}}}
\end{table}

To see if statistics was sufficient the same simulation
was run with 16 times longer tube (divided into 4096 cells) for 160000 timesteps 
for two of the models.
Flame speed and the three widths agreed within 0.1\% (these were averaged over last 158000 steps
in these long runs) with those in Table \ref{tab2}. Therefore larger than 4 cells
$W_3$ for the 3 finite-width models (as well as $D$ distinct from 80 $\kms$) must be
attributed to discretization: as $f$ changes from 0 to 1 mostly within 4 cells
one generally would expect errors in estimating gradients via finite differences
to affect the flame profile and propagation speed. To quantify this discretization effect
the model with $r=0.75,\: f_0=0.2,\:\rho=3\times 10^7\:\gcm$ was run in 1D domains
512, 1024 and 2048 cells long, with $D=80\:\kms$ and $W$ equal to 8, 16 and
32 cells respectively (these $D$ and $W$, as always, define $\tilde{K}$ and $R$ via \p{KR}, and
are the values for flame speed and width one hopes to get with the simulated flame).
The results are summarized in
Table~\ref{tab3}. These (together with the first quartet in Table~\ref{tab2}) illustrate the trend.
In part, for the flame width of order 16 cells or wider the $W_3$ becomes smaller than 
the prescribed \emph{total} flame width, as it should; the difference between the 
prescribed flame speed $D$ and the actual value also tends to zero when the number of zones
in the flame increases. The same holds for different densities: say, measured flame speed 
for the same $(r=0.75,\: f_0=0.2)$ model at $\rho=2\times 10^9\:\gcm$ is 
$\{78.8;79.4;79.9\}\:\kms$ for $W=\{ 8; 16; 32\}\Delta$ respectively.

\begin{table}[!hbpt]

{\center 
\begin{tabular}{rrrrr}\hline
$W\:$ & $D\vphantom{D^1}$ & $W_1$ & $W_2$ & $W_3$ \\
\hline
8  & 77.3 &  5.25 &  7.48 &  8.40 \\
16 & 79.1 & 10.57 & 14.67 & 15.87 \\
32 & 80.0 & 21.32 & 29.27 & 31.53 \\
\hline
\end{tabular}
\caption[Flame velocities and widths at different prescribed flame widths $W$]
{Flame velocities and widths at
different prescribed flame widths $W=x|^{0+}_{f=1-}$ (in cell sizes, $\Delta$). 
$(r;f_0)=(0.75;0.2)$, $\rho=3\times 10^7\:\gcm$.\label{tab3}}}
\end{table}

This study shows that really the difference between the prescribed flame speed and the
actual one achieved in simulations is due to a small number of cells within the flame.
The discrepancy can be corrected for by tuning the $d$ and $w$ values in \p{KR},
that is adjusting our analytic prescription for the coefficients in \p{I1} or \p{artvisfront0}
with additional ($\Lambda$-dependent) factors. 
This will be done in the next Chapter for flame models studied there. 

\subsection{Flame profiles}
\begin{figure}[htbp]\begin{centering}
\hfill
\begin{minipage}[t]{.485\textwidth}
\begin{center}
\includegraphics[width=2.88in]{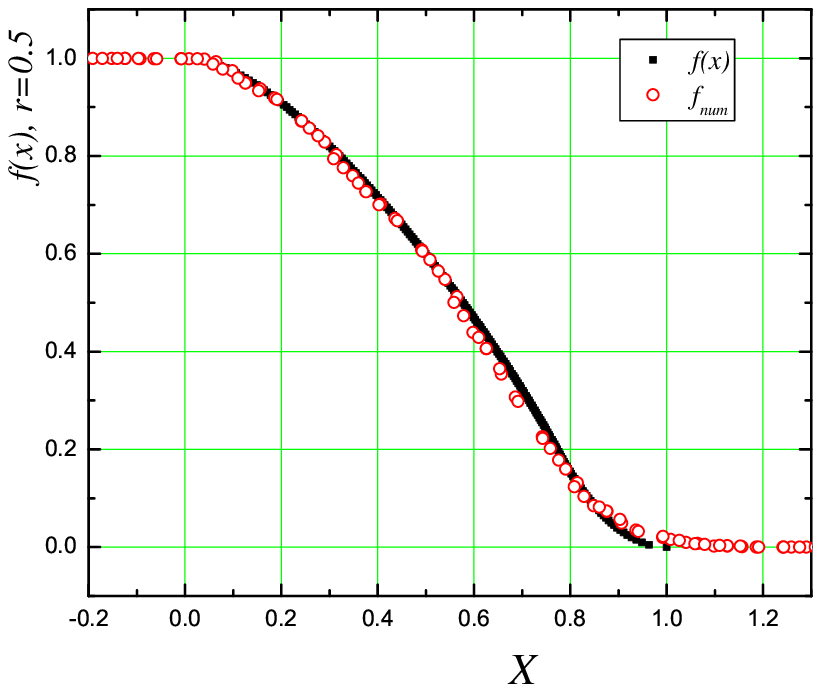}
\end{center}
\end{minipage}
\hfill
\begin{minipage}[t]{.485\textwidth}
\begin{center}
\includegraphics[width=2.88in]{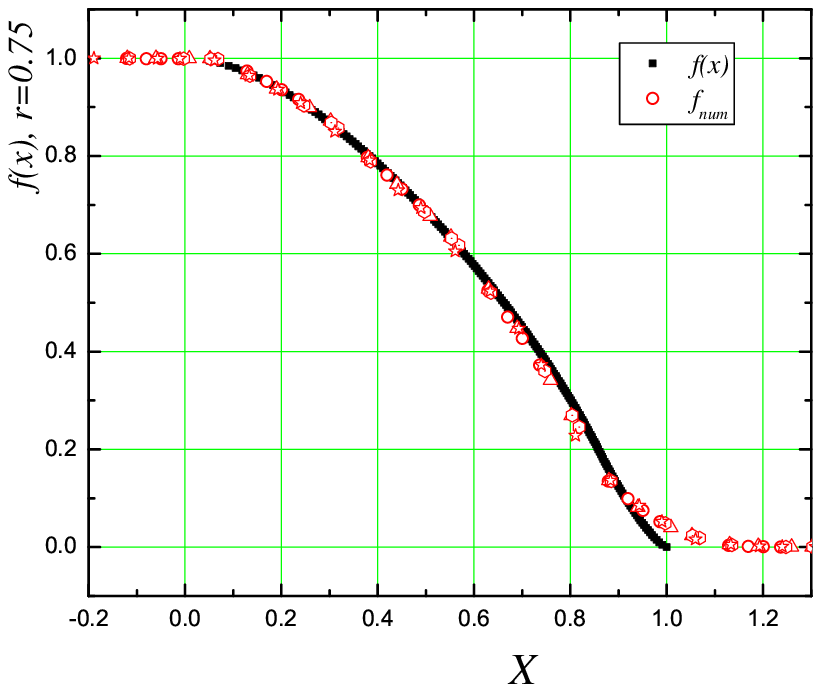}
\end{center}
\end{minipage}
\hfill

\caption[Steady-state and numerical flame profiles normalized to unit steady-state width]
{Theoretical (steady-state; black curve) and numerical flame profiles (red dots)
normalized to unit theoretical width. Numerical profiles 
are represented by values $f(x_i)$ near the flame position for 4 timesteps at 4 densities. 
For the $(r;f_0)=(0.75;0.2)$ model the dots are differentiated according to fuel density: circles correspond 
to profiles at $\rho=2\times 10^9\:\gcm$, hexagons to $3\times 10^8$, stars to $1\times 10^8$, 
and triangles to $3\times 10^7$.
\label{stepKnum}}
\end{centering}\end{figure}

Two of the models in Sec.~\ref{stepK} were proposed for use for the low sensitivity of their flame profiles to
the expansion parameter. Fig.~\ref{stepKnum} shows that this nice property is not spoiled by the discretization
effects, and further clarifies the nature of longer profile tails in discretized setting.
For each model, $f_0=0.2$, and $r=0.75$ or 0.5, the values of $f$ were recorded near the flame position for
4 different timesteps; for each timestep the set of values $f(x_i)$ was first renormalized in $x$
direction by dividing all the $x_i$ by $4\Delta$, thus normalizing the numerical profiles
to unit total width (more precisely, they would have had unit width if there had not been
discretization corrections); then these renormalized profiles were translated in $x$ direction so that they
least deviated from the steady-state profiles (Fig.~\ref{kx}). This procedure was performed with
the 2 models displayed in Table~\ref{tab2}, for the same 4 densities. As Fig.~\ref{stepKnum} shows
the numerical profiles with 4 cells wide flames (1) closely follow corresponding steady-state (continuous)
profile at all times (apart from the longer tail), and (2) as a consequence, are insensitive to density. 
These numerical profiles, further, show about the same density independence for the two models, 
$(r,f_0)=(0.5,0.2)$ and $(0.75,0.2)$, which suggests to stick to the former one because of its more symmetric
profile.

We conclude that all results of this chapter agree for different methods used for their derivation:
analytical versus numerical solution of steady-state (continuous) problem versus direct numerical simulation
using full hydro-solver {\tt{ALLA}}. All discrepancies observed are clearly related to their respective
causes: errors of numerical integration of ODE\rq s in steady-state approach (effects of which shown to
vanish with refining resolution; these errors are within $10^{-2}\%$ for flame velocities and widths 
when bulk integration step is $\Delta f=10^{-5}$, further refined up to $\Delta f_\mathrm{ref}=10^{-8}$
near peculiar points of defining ODE \p{master1} --- the setup we used for most calculations); 
discretization effects in direct numerical estimates in non-steady
setup. Corrections due to discretization were seen to be negligible when the \fl\ is resolved on 
16 or more cells, but they were up to 10\% in obseerved flame speed for some flame models when
prescribed total flame with $W$ was set to 4 cells, and more
than that in flame widths. The effect of discreteness of non-steady simulation on flame width was
seen to be mostly due to longer tails in flame profiles (due to numerical diffusion). These cross-checks
between different approaches provide trustworthy estimates for accuracy of corresponding methods, what
resolution for numerical integration of eigenvalue problem is sufficient, what are discretization
corrections for models with different flame profiles. The same checks are performed in the following
chapters for 2 new flame models, for which the methods described in this chapter are used for similar
calibration, needed for use in FC and for head-to-head comparisons of numerical and physical non-steady 
effects between the models (for which parameters like $D$ and $W$ are fixed the same for all comparisons 
performed). We study numerical noises the flame models introduce in 1D simulations in Chap.~\ref{noise1d}, 
isotropy of flame propagation and flame surface instabilities in 2D and 3D in Chap.~\ref{noisend}.

\chapter{Flames in non-steady simulations: numerical noises in 1D}\label{noise1d}

This chapter and the next one are devoted to studying features of model flames in non-steady simulations. 
Non-steady properties we observe (and describe below)
are visibly separated into physical (thus expected) effects, and effects related to numerics.
The latter are numerical noises, like sound waves generated and flame speed fluctuating
in time, with periods directly related to flame propagation with respect to the grid; anisotropic 
flame propagation in 2D/3D setting, with grid dictated special directions. The numerical artifacts
are clearly undesirable, they disturb and systematically deviate numerical model behavior from physically
sound one, having nothing in common with reality. Finding numerical realization  showing
least possible amount of these numerical artifacts is necessary, and is one of important 
goals of our study.

Other effects we observe also deviate artificial flame behavior in simulations from what
would seem ideal for flame capturing applications; effects of this class, however, are expectable
physical phenomena for any diffusion-reaction flames. 
These effects include dependence of flame
propagation speed upon local flame curvature (Markstein effect, \cite{markstein}); growth of
wrinkles on initially smooth flame surface, hydrodynamic instability of the type (LD for short) studied in 
\cite{lan,dar}. These effects depend to a degree on the model diffusivity and reaction rate functions;
no model, however, is free of these effects. Real physical flame is also an example of reaction-diffusion
system, albeit more complicated; its propagation shows the same physical effects.
%
Ideal flame capturing scheme therefore would be not the one showing no effects of this physical
class, but the one demonstrating these effects with magnitude equal to that of (smeared over grid cell scale)
physical flame region. Results below, Chap.~\ref{noisend}, demonstrate that features of LD instability depend 
strongly on the parameters of the model; we cannot hope that some random flame model physical features will match 
quantitatively those of real physical flame region. Detailed study of real flame is required to find its Markstein 
length, critical length and growth rate with respect to hydrodynamic instability; only then one would be 
prepared to analyze deviations from corresponding features of artificial flame model
(studied in this work), and try to correct for these deviations.

Some instabilities, like Rayleigh-Taylor, depend on density contrast across the flame zone, and 
do not depend significantly on specifics of density distribution in the transient (flame) region, as long
as corresponding instability scale exceeds this transient region width. This is not the case for
a thick \fl\ used in FC, yet it is the task of subgrid model to prescribe renormalized effective
\lq\lq turbulent flame speed\rq\rq\ to correct for this intricate subgrid geometry being smeared out
by the thick artificial \fl. This is not related to the scope of the thesis, this subgrid
prescription should be the same for any flame model; it may only slightly depend on specifics of density
distribution within the \fl, which distinguishes one model from another (assuming the same widths of the 
flames they produce). Response of the flame to front curvature (Markstein effect) and 
LD instability, on the other hand, do depend significantly on specific flame model. They also depend strongly
on the flame width, thus are expected to be different for the model thickened \fl and for the real thin
nuclear (or chemical) flame. What is average manifestation of these real hydrodynamic effects on large, resolved 
scales, whether it is may be comparable to corresponding effects observed for thickened flame with the width of order 
of characteristic width of the convoluted with instabilities physical flame brush --- is an interesting
topic to study, however this is not touched upon in the thesis. 

With qualitative understanding above in mind, we study several flame models in real non-steady simulations
without gravity (to have constant fuel density throughout the simulation, and to avoid complicating the 
observations with RT instability, which is essentially the same for different flame models). This chapter
deals with numerical noises which are 1D in nature. No physical 1D instabilities are known for flame models 
of the type we consider (with only one significant transfer coefficient); we did not observe any in simulations
performed.

We describe and calibrate (using methods of 
Chap.~\ref{steady} two new flame models in Sec.~\ref{noises:models}.
We present results for sound waves produced by these models, as well as by the model studied in 
Chap.~\ref{steady} (based on step-function burning rate \p{Phistep}, and power-law diffusivity 
$K_0(f)=f^r$, $r=3/4$) in 1D simulations in Section~\ref{noises}, after describing numerical methods used.
A feature is identified in the burning rate (namely discontinuity), which causes noise in 
simulations; this noises are too intensive for the model used in \cite{X95} as well as in other models
studied in Chap.~\ref{steady}, based on step-function burning rate. The 2 new models introduced in 
Sec.~\ref{noises:models} produce acceptable level of noise; this was a criterion for selecting them 
for further study in the next chapters, where  2D and 3D behavior is analyzed.

\section{Models studied: definition and steady state parameters}\label{noises:models}
\subsection{Definition of the models}\label{noises:defmod}
As discussed in Chap.~\ref{steady} certain features of artificial flame system are desirable for flame capturing applications.
In part, we want the system to have a unique eigenvalue for $D_f$, and want flame profiles to have finite width 
(without exponential or power-law tails), with reasonably constant $df/dx$ within the \fl . This limits possible $K(f)$ and 
$\Phi(f)$ in \p{I1} suitable for flame capturing. 

As a brief overview, two realizations of \p{I1} were used in literature before \cite{zh07}. Diffusivity is constant in both,
$K(f)=\tilde{K}=\mathrm{const}$, the realizations differ by reaction rate forms. The model originally proposed in 
\cite{X95} has step-function $\Phi(f)$, \p{Phistep}. This yields \fl\ having an  exponential tail into fuel 
(this tail, where $f<f_0$, represents \lq\lq preheating zone\rq\rq);
$D_f$ eigenvalue is unique and exists for all $\tilde{K},R,f_0$. Another flame model tried in astrophysical literature
(\cite{Dursi03}), based on KPP burning rate [\cite{KPP}] $\Phi(f)=Rf(1-f)$, has continuous $D_f$ spectrum 
(for any $D_f>2\sqrt{\tilde{K}R}$ there exist a solution of \p{artvislam} satisfying physical boundary conditions), 
and two infinite tails in eigenfunctions $f(x)$: $f$ approaches 0 and 1 exponentially as $x$ goes to $+\infty$ or $-\infty$ 
respectively. 

In Sec.~\ref{stepK} we proposed a new flame model based on the original step-function burning rate, but with diffusivity 
dependent on $f$:
\begin{equation}\label{PhistepK}
\Phi(f)=R\Phi_0(f)=\left\lbrace
\begin{array}{ll}
  R &\: \mathrm{for\ } f_0<f<1\\
  0 &\: \mathrm{for\ } f<f_0
\end{array} \right.\; ; K(f)=\tilde{K}f^r.
\end{equation}
This diffusion-reaction model has a unique steady-state speed $D_f$, like the original one (\cite{X95}; effectively a limit 
case of \p{PhistepK} with $r=0$), and for $r>0$ corresponding flame profile is finite: the region, \fl, where $f$ is 
neither 0 nor 1 has finite width. We proposed to use this model with 
\begin{equation}\label{rf0}
f_0=0.2,\; r=0.75,
\end{equation}
as these parameters led to flame profiles having the same shape for any $\l\in [0.25;\infty)$. This encompasses all
physically interesting range of fuel densities for deflagration in a WD, from central density of 
$\sim 2.2\times 10^9\:\gcm$ ($\l\approx$ 6.2) down to $\sim 3\times 10^6\:\gcm$ (considering half-carbon,
half-oxygen composition). We will present results for this $(f_0,r)$ pair only, as this type of model turns out to be quite 
noisy for any $(f_0,r)$ (including original model in \cite{X95}, corresponding to pair $(f_0,r)=(0.3,0)$), 
significantly worse than the other two models we will study (and thus of limited interest for use in flame tracking). We refer to this model as Model A in the following.

The second model we analyze in this paper is based on a modification of KPP burning rate with constant diffusivity, 
\begin{equation}\label{skpp}
  \begin{array}{lcl}
    \Phi(f)&\! =&\! R(f-\e_f)(1-f+\e_a)\:(0<\e_f,\,\e_a<1),  \\
     K(f)&\! =&\!\tilde{K}=const,
  \end{array}
\end{equation}
dubbed \lq\lq shifted KPP\rq\rq\ (sKPP for short) at FLASH center meetings where it was invented, due to the nature of corrections to 
KPP burning rate, significant only when $f$ is close to either 0 or 1; these corrections effectively cut exponential tails 
of original, \lq\lq not shifted\rq\rq\ KPP-flame profile (corresponding to $\e_f=\e_a=0$) rendering flame localized in space,
like model \p{PhistepK}. Besides, 
this model has unique flame propagation speed, another advantage over original KPP model\footnote{%
These properties, as well as asymptotic behavior of the flame profile of this model (namely, $f(x)$ is parabolic near 
both flame boundaries, $f=0$ and 1), may be found in qualitative analysis in Sec.~\ref{qual:gen}.}
. We will describe the results for 
this model for $\e_f=\e_a$ case: these 2 parameters have similar effect on model characteristics due to the symmetry of its 
burning rate $\Phi(f)$, and there was found no significant advantage when using different values in the pair, $\e_f\neq\e_a$. 
$\e_f=\e_a=10^{-3}$ produces optimal results in terms of 2D/3D behavior and 1D noises (see below).

The third model studied, Model B, the one we recommend for use in flame capturing based on its behavior in multidimensional simulations,
is specified by
\begin{eqnarray}\label{syntf}
  \Phi(f) &=& Rf^{s_f}(1-f)^{s_a}(f-c(\l))\\ \label{syntk}
     K(f) &=& \tilde{K}f^{r_f}(1-f)^{r_a},
\end{eqnarray}
with
\begin{equation}\label{expon}
  s_f=1,\: s_a=0.8,\: r_f=1.2,\: r_a=0.8,
\end{equation}
and $c(\l)\in[0.005,0.3]$ determined locally as a function of expansion parameter for fuel/ash at given local pressure
and fuel composition. This model has a unique steady-state flame propagation speed (when $c(\l)>0$); flame profiles are 
localized, and do not have long tails (in contrast with model \p{skpp}); they look similar to profiles of model 
(\ref{PhistepK}--\ref{rf0}). Term $c(\l)$ was selected to minimize flame propagation anisotropy
and hydrodynamic flame surface instabilities in 2D at different fuel densities (Chap.~\ref{noisend}). 
Reasonable results were achieved with $c(\l)$ defined as a spline: 
\begin{eqnarray}\label{clam}
 c &=& 0.005\; \mathrm{for\ } 1/\l<0.515 \\
 c &=& 0.3\;   \mathrm{for\ } 0.81<1/\l<1.5 \\
 c &=& 0.2\;   \mathrm{for\ } 1/\l>1.9,
\end{eqnarray}
and continuously linearly changing between these $\l$ intervals (see Appendix for explicit formulae). 
Exponentials \p{expon} were chosen based
on steady-state considerations and non-steady properties. $s_f, s_a\gtrsim 0.7$ for $\Phi(f)$ to go to 0 fast enough 
at $f\to 0$ and 1 (to avoid significant 1D noises, see Sec.~\ref{noises}). $s_a<1$ for the model to have a unique 
eigenvalue for flame speed. $r_f>1$ for the flame not to have infinite tail into fuel. Somewhat larger $r_f$, up to 2,
produce acceptable results as well, yet no improvement can be obtained in 2D behavior. $r_a=0.8$ was chosen for reasonably
symmetric flame profile, with nearly constant $df/dx$ within. Flame profiles for this model (Model B), and Model sKPP 
\p{skpp} are shown in Fig.~\ref{fkb} below.

\subsection{Steady-state and numerical calibration of the models: method}
\label{nstdy:calibr}
To use flame model in simulations for flame tracking one has to know how to prescribe model parameters in
such a way so that the model flame propagated with required velocity, and had prescribed width. For this calibration 
we use the same approach as in Chap.~\ref{steady}: by factoring burning rate and 
diffusivity into constant dimensionful scale-factors and ($f$-dependent in general) dimensionless form-factors
(Eq.~\ref{dimless}), 
\begin{equation}\nonumber
  \Phi(f)=R\Phi_0(f),\: K(f)=\tilde{K} K_0(f)
\end{equation}
we get an eigenproblem (Eqs.~\p{artvislamdim}, \p{bndl}, \p{bndr}) for finding dimensionless flame speed 
$d=D_f/\sqrt{\tilde{K}R}$ and flame profile $f(\chi)$ in terms of dimensionless coordinate $\chi=x\sqrt{r/\tilde{K}}$
along flame propagation. This problem is solved by numerical integration of the ODE and using Newton-Raphson method
for obtaining the eigenvalue of $d$, for which boundary conditions are satisfied. The procedure and resolutions used are 
the same as discussed in Chap.~\ref{steadynum}.

To quantify the profile shape, whether the slope of $f$ within the flame profile is approximately constant, or if the
profile has long tails, with regions of large and small gradients of $f$ having comparable spatial scales, we calculate 
widths defined in 3 different ways:
{\setlength\arraycolsep{2pt}
\begin{equation}\label{w123}
  \begin{array}{ll}  
    w_1=&\left. \chi\right|_{f=0.9}^{0.1}\equiv\chi(f=0.1)-\chi(f=0.9)\\
    w_2=&\left. \chi\right|_{f=0.99\vphantom{\int_{q}}}^{0.01\vphantom{\int^{T^b}}}\\
    w_3=&\left. \chi\right|_{f=0.999}^{0.001}.
  \end{array}
\end{equation} }
Having found $d$ and $w$ (in any convenient sense, \eg\ any of $w_{1,2,3}$) one has a simple way to normalize the flame model
to yield prescribed flame speed $D$ and width $W$ (same definition of \lq\lq width\rq\rq\ must be used for physical width $W$
for simulations, and dimensionless width $w$; say, both between $f=0.1$ and $f=0.9$). Namely (see Sec.~\ref{prescr}), 
one must use \p{I1} for
flame tracking, with burning rate and diffusivity \p{dimless}, where scale factors are determined through Eq.~\ref{KR}:
\begin{equation}\label{KRchap3}
  \tilde{K}=\frac{D_f}d\cdotp\frac{W}w,\quad  R=\frac{D_f}d\Bigm/\frac{W}w.
\end{equation}
To summarize, for any flame model (\ie\ selected for use dimensionless $\Phi_0(f)$ and $K_0(f)$) one finds corresponding 
$d$ and $w$, and then, using \p{KR}, finds normalization factors for burning rate and diffusivity, $R$  and $\tilde{K}$, 
which lead to prescribed \fl\ \p{I1} speed and width.

Same parameters $d$ and $w$, defining proper normalization of $K(f)$ and $\Phi(f)$, may be found directly numerically, by  
simulating flame propagation in 1D, finding $\tilde{K}$ and $R$ correctly yielding some chosen $D$ and $W$, and then 
reversing \p{KRchap3} to find $d$ and $w$; these latter can be used later on for getting $\tilde{K}$ and $R$ yielding any 
physically motivated flame speed. Such numerically found $d$ and $w$ differ slightly from exact $d$ and $w$ obtained
by solving eigenvalue problem: the former, numerically found $d$, is essentially the eigenvalue of discretized system 
\p{artvislamdim}, where spatial derivatives of $f$ are represented by finite differences on a grid, with total flame width
of just $\sim 4$ grid spacings. Flame profile of discretized problem also deviates from continuous one, exact eigenfunction
of steady-state boundary value problem. Most notably, one gets longer tails near $f=0$ and $f=1$, thus $w_{1,2,3}$
larger than corresponding widths of steady-state non-discretized flame (at least for models with well-localized flame 
profiles). Effects of discretization on flame profiles and 
propagation speed for model \p{rf0} were presented in Sec.~\ref{nsteady1D}; in part, flame speed agrees with steady state one
up to $\sim 1\%$ when total flame width (from $f=0$ to 1) was 16 zones or larger; better agreement is observed for 
smaller expansion. 

When flame width is kept constant throughout the 
simulation (in units of grid spacing) it is preferable to use numerically found parameters $d$ and $w$ for discretized 
problem, specifically ones found numerically by modeling 1D flame with the same flame width as the one required for actual 
multidimensional simulation. This will 
correct for systematic discretization effect (its principal component, which manifested itself in 1D simulations
resulted in $d$ and $w$ deviating from those for continuous, steady-state, model). 

For direct numerical (non-steady) estimation of flame speed and width, as well as for almost all other numerical studies 
presented in this and the next chapters, dimensionally split piecewise-linear code {\tt{ALLA}} (\cite{X98, X00}) was used. 
For all runs performed the fuel was taken as half ${}^{12}\mathrm{C}$, half (by mass) ${}^{16}\mathrm{O}$ mixture; real 
degenerate equation of state was used, gravity was set to zero. We assumed that the fuel transformed to nuclear statistical 
equilibrium composition (depending on local pressure) and all the heat released in the process was released within the \fl . 
Physically this is not the case at lower densities in a WD
(outer layers, $\rho\sim 2\times 10^7\:\gcm$ and below), yet it is the expansion parameter $\l$, 
dependence of flame characteristics on which we are interested in; all the results found are presented as functions of $\l$, 
and not, say, of fuel density (this way, for any fixed $\l$, they must be universal for any equation of state, providing 
assumptions of pressure and $H\rho$ being constant across the flame are satisfied). For SN~Ia problem at lower densities 
it is thus straightforward to find corrections for our results: one just needs to use real heat release within the flame 
(slightly smaller than what we used, based on burning to NSE composition) to find correct value of expansion parameter $\l$, 
and use our results (such as $d(\l)$, $w(\l)$, 
asphericity of flame surface in 3D, or Markstein number) for such found $\l$.

\subsection{Results for flame profiles and velocities}
Results of steady-state calculations of $d$ and $w$ for Models A (\ref{PhistepK}--\ref{rf0}), sKPP (\ref{skpp}) and B
(\ref{syntf}--\ref{clam}) are presented in Table~\ref{tab31}, together with the same quantities 
found numerically.  
As discussed above, the $d$ and $w$ found analytically, or by finding the eigenvalues and corresponding flame profile width
through numerical integration of \p{artvisfront0} (we call this \lq\lq steady-state technique\rq\rq, as opposed to direct 
numerical simulations, where physical quantities are discretized on computational grid), will yield correct prescribed flame 
speed $D$ and $W$ (when using \p{I1} with parameters determined by \p{KR} for \fl\ evolution) only when $W$ is sufficiently 
large ($\sim 16$ cells or more, as tested for model A in Sec.~\ref{nsteady1D}). Systematic deviation from prescribed $D$ and 
$W$ grows as $W$ decreases. To correct for this deviation we numerically obtained $d_\num$ and $w_\num$ yielding correct 
prescribed $W_1$ and $D$. These depend on desired $W_1$ (tending to steady-state values as $W_1\to\infty$), and should yield 
the required width and speed in 2D and 3D setting
(as long as the front may be considered sufficiently flat, and providing its corresponding width is close to $W_1$ used to 
obtain $d_\num$ and $w_\num$ in 1D), with better accuracy than parameters $d_\st$ and $w_{\st}$ found through steady-state 
technique.

For model A $W_1$ was chosen so that to produce (almost total) width $W_3=6$, to use the results of Sec.~\ref{stepK},
where simple analytical expressions were obtained for flame speed and total width ($f=0$ to 1) for this model. For sKPP value 
$W_{1,\num}=4$ was chosen based on 2D performance: for smaller  $W_{1}$ surface instabilities are too large at densities 
below $\sim 10^8\: \mathrm{g\:cm^{-3}}$. 
$W_{1}=3.2$ for model B provided reasonable performance for all densities studied, from $5\times 10^9$ down to 
$8\times 10^5\:\gcm$. Flame speed required in simulations was $D=80\:\mathrm{km\:s^{-1}}$, which is a reasonable value for 
laminar deflagration velocity near the center of a WD (\cite{TimWoo92}). In outer layers, at smaller densities, laminar 
deflagration speed decreases, yet one typically uses a value of \lq\lq turbulent flame speed\rq\rq\ 
$D\sim \stur\simeq 0.5\sqrt{\ma g \Delta}$ as prescribed $D$ there, which by far exceeds the laminar deflagration 
speed, due to large gravitational acceleration a fair distance from the star center ($\ma$ denotes Atwood number, 
Eq.~\ref{At}). 

\begin{table}[!hbpt]
{\center \begin{tabular}{@{$\;$}lr@{$\;\,$}|@{$\;\,$}rrr|@{$\;$}r r@{$\;\;$}r@{$\;\;$}r@{$\;\;$}r@{$\;$}}\hline 
Model    & $1/\l$   &$d_\st$&$w_{1,\st}$&$w_{3,\st}$ &
$W_{1,\num}$&$d_{\num}$&$w_{1,\num}$&$w_{2,\num}$&$w_{3,\num}$ 
  \\ \hline
    A    & 0.1670   & 1.498 & 1.225    & 1.837&
   3.56  & 1.424    & 1.209      & 1.819  & 2.035  \\
    A    & 0.3980   & 1.411 & 1.289    & 1.919&
   3.48  & 1.333    & 1.268      & 1.897  & 2.198  
  \\
    A    & 0.7337   & 1.306 & 1.376    & 2.030&
   3.44  & 1.222    & 1.351      & 2.000  & 2.358  
  \\
    A    & 1.442    & 1.140 & 1.540    & 2.240&
   3.41  & 1.041    & 1.491      & 2.184  & 2.619  
  \\
    A    & 8.571    & 0.5564& 2.401    & 3.445&
   3.16  & 0.4304   & 2.014      & 2.888  & 3.830  
  \\ \hline

sKPP     & 0.1670   & 1.681 & 8.125    &21.31&
4.0      & 1.674    & 8.115      & 15.99  & 21.58
   \\
sKPP     & 0.3980   & 1.664 & 8.893    &23.39&
4.0      & 1.596    & 8.585      & 16.86  & 22.82  
   \\ 

sKPP     & 0.7337   & 1.642 & 9.993    &26.44&
4.0      & 1.515    & 9.319      & 18.27  & 24.82
   \\ 

sKPP     &  1.442   & 1.605 & 12.27    &32.83&
4.0      & 1.396    & 10.91      & 21.49  & 29.34
   \\ 

sKPP     &  8.571   & 1.420 & 33.04    & 92.19&
4.0      & 0.9592   & 22.76      & 46.77  & 65.00
   \\ \hline

    B     & 0.1670  & 0.3178 & 2.159   & 3.424&
   3.2    & 0.3121  & 2.147      & 3.441  & 4.145  
  \\
    B     & 0.3980  & 0.2990 &  2.245  & 3.595&
   3.2    & 0.2921  & 2.236      & 3.583  & 4.310  
  \\
    B     & 0.7337  & 0.1850 &  2.627  & 4.096&
   3.2    & 0.1751  & 2.553      & 4.058  & 4.842  
  \\
    B     & 1.442   & 0.1221 & 2.865   & 4.455&
   3.2    & 0.1129  & 2.744      & 4.317  & 5.161  
  \\
    B     & 8.571   &0.0664 & 3.495   & 5.831&
   3.2    & 0.05816 & 3.119      & 4.920  & 6.016  
  \\

\hline

\end{tabular}
\caption[Steady-state flame speeds ($d_\st$) and widths ($w_{1,\st}$, $w_{3,\st}$), compared to the ones found via 
direct numerical simulations]
{Steady-state flame speeds ($d_\st$) and widths ($w_{1,\st}$, $w_{3,\st}$), compared to the ones found via 
direct numerical simulations. 
$W_1$ represents physical flame width
in number of cells between $f=0.9$ and $f=0.1$ used in numerical estimates. $\rho_\mathrm{fuel}$ corresponding to
$1/\l$ shown for a WD with 0.5C+0.5O composition are $2\times 10^9$, $3\times 10^8$, $10^8$, $3\times 10^7$ and
$8\times 10^5\:\gcm$ respectively.
$D_\num=80\:\kms$ for all the models. Shift parameters of sKPP model are $\e_a=\e_f=10^{-3}$. 
\label{tab31}}}
\end{table}

\begin{figure}[htbp] \begin{centering}
\includegraphics[width=4.0in]{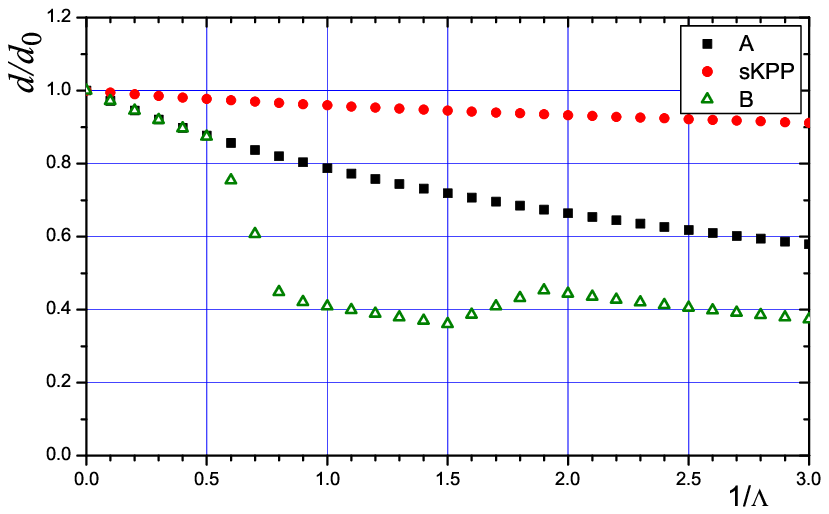}
\caption[Flame speed scaling as a function of expansion parameter $\l$ for models A, B and sKPP, found through steady-state 
technique]{Flame speed scaling as a function of expansion parameter $\l$, found through steady-state technique
for 3 models, A, B and sKPP with $\e_a=\e_f=10^{-3}$. The speeds presented were rescaled by their respective values
at zero expansion, $d_A(1/\l=0)=1.5703$, $d_{sKPP}(1/\l=0)=1.6954$, $d_B(1/\l=0)=0.33335$. Nonmonotonic dependence
of $d_B$ on $\l$ is due to burning rate dependence on expansion through nonmonotonic $c(\l)$.
\label{dn_akb}}
\end{centering}\end{figure}

\begin{figure}[htbp] \begin{centering}
\includegraphics[width=4.0in]{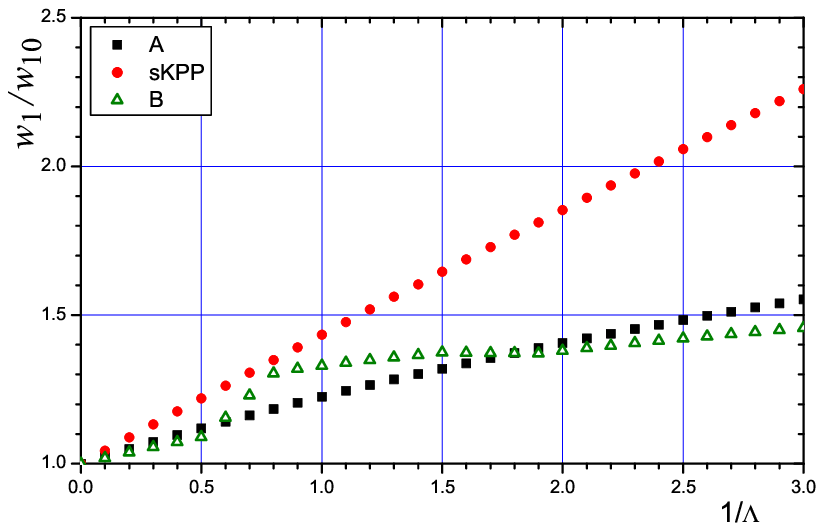}
\caption[Steady flame width $w_1$ scaling as a function of expansion parameter $\l$, for models A, B and sKPP]  
{Steady-state flame width $w_1=\chi(f=0.1)-\chi(f=0.9)$ scaling as a function of expansion parameter $\l$, 
for 3 models, A, B and sKPP with $\e_a=\e_f=10^{-3}$. These widths are rescaled by their values
at zero expansion, $w_{1A}(1/\l=0)=1.1762$, $w_{1sKPP}(1/\l=0)=7.5687$, $w_{1B}(1/\l=0)=2.0924$.
\label{w1n_akb}}
\end{centering}\end{figure}

\begin{figure}[htbp] \begin{centering}
\includegraphics[width=4.0in]{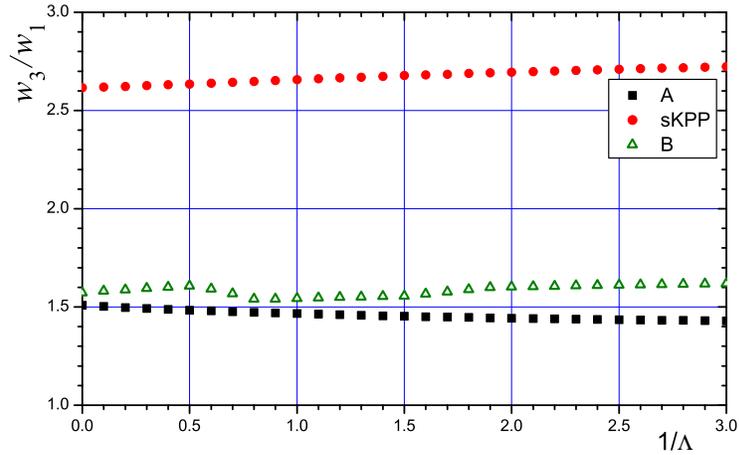}
\caption[Ratio of steady widths $w_3/w_1(\l)$ for models A, B and sKPP]
{Ratio of two dimensionless flame widths defined differently, $w_3/w_1$ 
($w_3=\left. \chi\right|_{f=0.999}^{0.001}$, 
$w_1=\left. \chi\right|_{f=0.9}^{0.1}$,
$\chi$ is a dimensionless flame coordinate) as a function of expansion parameter $\l$, 
found through steady-state technique
for 3 models, A, B and sKPP with $\e_a=\e_f=10^{-3}$.
\label{w31_akb}}
\end{centering}\end{figure}

Dependence of steady state dimensionless speeds $d$ and width $w_{1}$ on expansion parameter $\l$ is further illustrated 
in Figs.~\ref{dn_akb}--\ref{w1n_akb}. Shift parameter $\e_a=\e_f=10^{-3}$ was used for sKPP model. 
The curves were rescaled to coincide at $1/\l=0$ (no expansion) to better represent the form of dependence on $\l$. 
Absolute independent of $\l$ normalizations are not of much 
interest for model use in simulations, as prescription \p{KR} takes care of adjusting $K$ and $R$ for this.
The form of the dependence on $1/\l$, on the other hand, is more difficult to account for, and should be fitted
on a model-by-model basis. Fig.~\ref{w31_akb} shows $w_3/w_1$ ratio as a function of expansion parameter. This ratio is an 
indicator of flame profile shapes. This ratio is larger for sKPP model (long tails in flame profile), 
which is a drawback, as the need to resolve the region where gradient of $f$ is large does not allow one to use flame model 
parameters $\tilde{K}$ and $R$ yielding too narrow $W_1$, and the total flame width then, illustrated by $W_3$ value
is necessarily significantly larger for sKPP model than for models A and B. 

\begin{figure}[htbp] \begin{centering}
\includegraphics[width=4.3in]{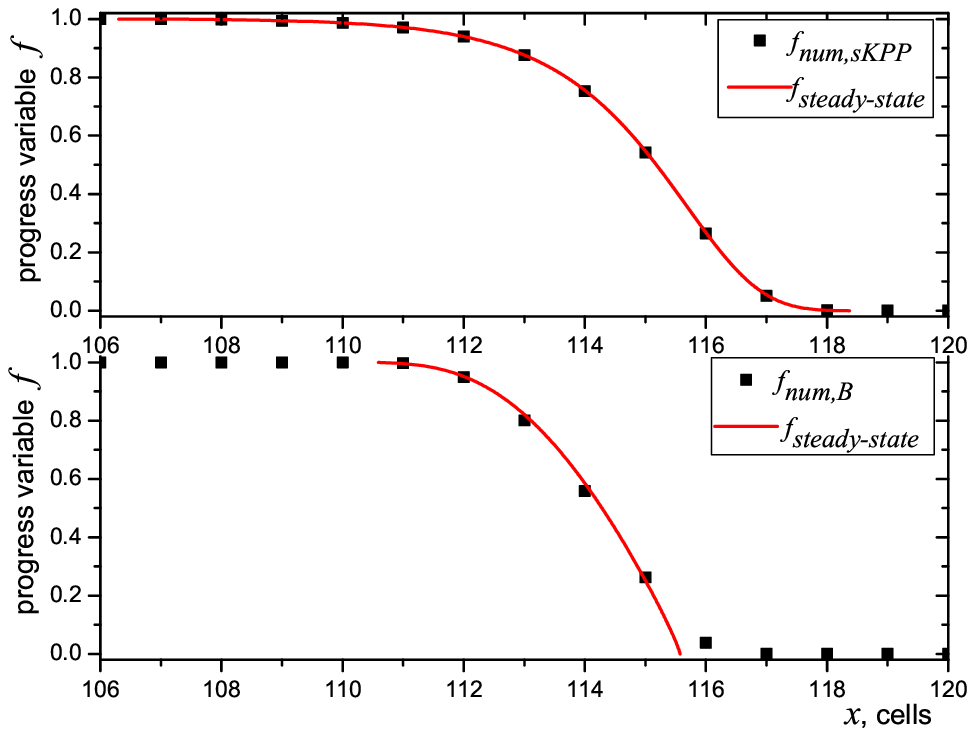}
\caption[Steady and numerical flame profiles for models B and sKPP at $1/\l=1.442$]
{Flame profiles for models B and sKPP (black squares), $\rho_\mathrm{fuel}=3\times 10^7\:\gcm$, $1/\l=1.442$. 
Numerical profiles were recorded in 1D simulations at time step 5000. $W_{1sKPP}=4$ and $W_{1B}=3.2$ cells, the values 
used in multidimensional simulations. Steady state profiles (continuous line, shown only on the 
interval where $f_\st\neq 0$ or 1.) were obtained by rescaling abscissas of steady profiles by 
$W_{1,\num}/w_{1,\st}$ 
(so that the rescaled profile width matched numerical one), and then translating in $x$ direction for best fit. 
Note long tails in sKPP profile.
\label{fkb}}
\end{centering}\end{figure}

For implementation in simulations using FC, Sec.~\ref{prescr}, we fitted the dependence $d(\l)$ and $w_1(\l)$ for Model B.
As parameter $c(\l)$ in artificial reaction rate for Model B is defined as a spline, Eq.~\p{clam}, the fit for speed and width
is also given by different formulae on different intervals of $\l$; it can be found in Appendix, Eq.~\p{Bcalib}.

These fits yield flame speed in 1D setting differing by at most 0.8\% from required 80 km/s in a range $1/\l\in[0.11;8.6]$.
Notice that as expansion increases dimensionless speed is expected to vanish $d(\l)\propto\l$, as it was the case for Model A; 
thus the fit should be modified (last part, at large expansions) if it is to be used at $1/\l\gtrsim 8.6$.

\section{Results for 1D numerical noises}\label{noises}
As the flame propagates over the grid one would expect certain amount of acoustic noise generated. Because of discreteness 
of the grid there is no exact continuous translational invariance of the problem; say, total burning rate (in simulation), 
summed over grid zones in the flame region (where burning occurs) depends not only on continuous flame profile modeled, 
but on the precise positioning of that profile with respect to the grid as well. This leads to sound waves generation 
with period corresponding to time needed for the flame to move across one grid spacing. Depending on boundary conditions 
these waves may bounce back and forth, perhaps seeding and facilitating development of real physical instabilities of 
the system, thus complicating the simulation and making the results questionable. In this section we look at local 
physical quantities distribution in 1D domain (simulation tube), as well as time dependence of certain integral quantities, 
like burning rate. 

All models are studied
under the same conditions: same computation setup, quiet initial conditions (imitating steady state distribution of 
variables in simulation), flame speed is required to equal $80\:\mathrm{km\:s^{-1}}$ (which is achieved by choosing flame
model parameters $\tilde{K}$ and $R$ as described in Sec.~\ref{noises:defmod}. Numerically found values for $d$ and $w$ 
for chosen numerical flame width were used). Flame widths are different for different models, but are 
required to be equal for each individual model for each run (unless specified otherwise) regardless of expansion
parameter; these individual widths were chosen (for model B and sKPP) based on performance in 2D simulations 
(see next chapter), and are summarized, as $W_{1,\num}$, in Table~\ref{tab31}. 

\begin{figure}[htbp] \begin{centering}
\includegraphics[width=4.6in]{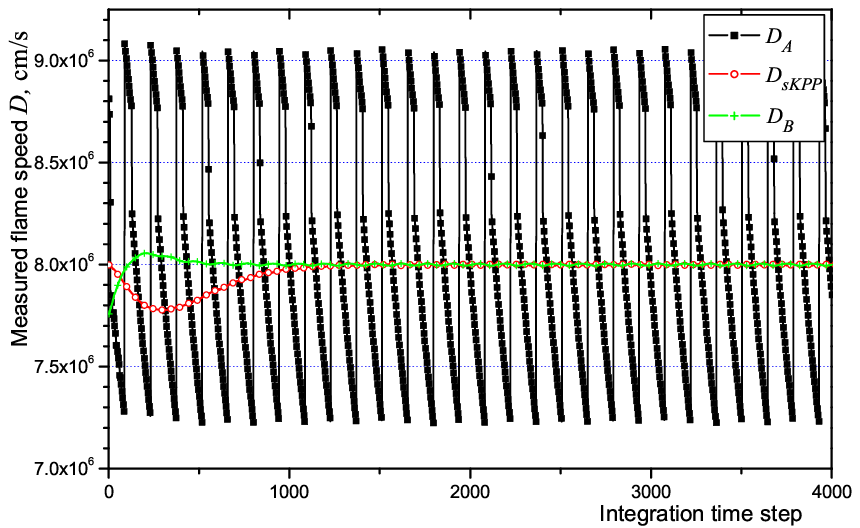}
\caption[Observed instant flame speed as a function of time]
{Flame speed $D(t)=\frac{1}{\rof}\frac{dm_\mathrm{ash\vphantom{,}}}{dt}$ as a function of integration time step.
Fuel density is $2\times 10^9\:\gcm$.
\label{st}}
\end{centering}\end{figure}

Flame speed as a function of integration time is presented in Fig.~\ref{st}. It is apparent from the plot that while 
average in time flame speed is indeed equal to the prescribed value for all models, flame speed for model A fluctuates
by more than 10\% around the average, which is not a desired feature. Also note in this figure that transient effects 
due to not completely perfect initial conditions are completely relaxed for all models within 1200 integration timesteps,
and that transient deviation in flame speed is within 3\% for all models.

\begin{figure}[htbp] \begin{centering}
\hfill
\begin{minipage}[t]{.485\textwidth}
\begin{center}
\includegraphics[width=2.88in]{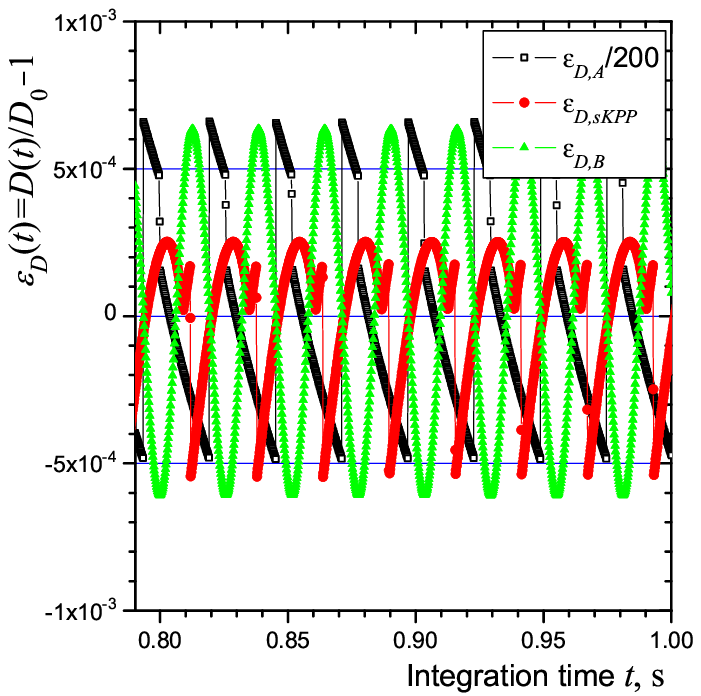}
\end{center}
\end{minipage}
\hfill
\begin{minipage}[t]{.485\textwidth}
\begin{center}
\includegraphics[width=2.88in]{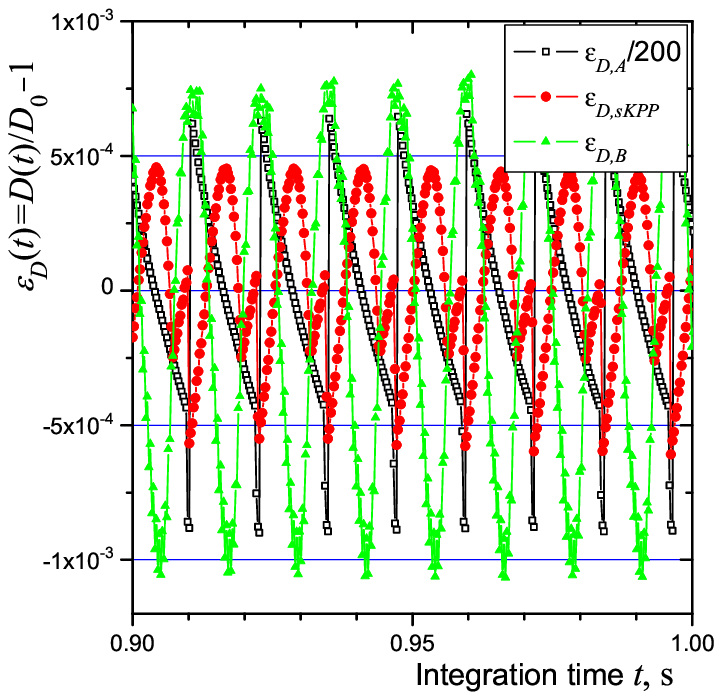}
\end{center}
\end{minipage}
\hfill
\caption[Relative instant flame speed deviation $D(t)/D_0-1$ from its prescribed value $D_0=80\:\kms$ for fuel density 
$2\times 10^9$ and $3\times 10^7\:\gcm$. Models A, B and sKPP]
{Relative flame speed $D(t)$ deviation from its prescribed value $D_0=80\:\mathrm{km\:s^{-1}}$ as a function of 
time. Fuel density is $2\times 10^9\:\gcm$ ($1/\l=\rho_\mathrm{fuel}/\rho_\mathrm{ash}-1=0.1670$) for the left figure, 
and $3\times 10^7\:\gcm$ ($1/\l=1.442$) for the right one. For model A this deviation was divided by a factor of 200 so
that the rescaled deviation had approximately the same scale as deviations for the other two models.
\label{est}}
\end{centering}\end{figure}

The pattern of fluctuations is presented in greater detail in Fig.~\ref{est} for expansion parameters typical
in SN~Ia problem at the beginning and closer to the end of burning in flamelet regime. 
Relative deviations from prescribed value
was rescaled by a factor of 200 for model A there for it to have the same scale in the figures as the other models.
The fluctuations are plotted against physical time, and the range of time shown for the two densities
is inversely proportional to flame front speed with respect to grid, $D(1+1/\l)$. Thus the flame propagates through
the same distance within the time shown for the two plots, approximately 8 grid spacings. And 8 periods of flame speed
fluctuations are clearly seen in each figure, pattern of fluctuations in time is very similar for each period, but 
patterns for different models differ drastically.

\begin{figure}[htbp] \begin{centering}
\includegraphics[width=3.8in]{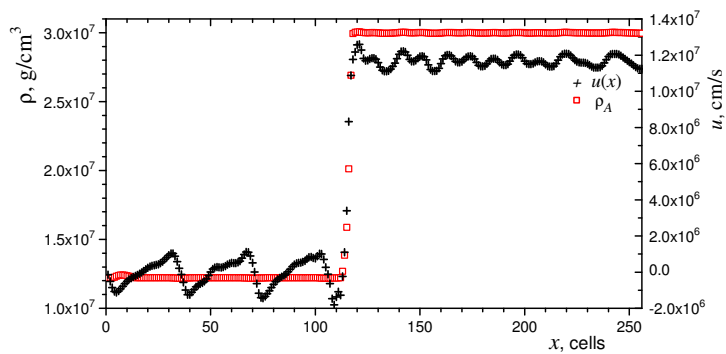}
\caption[Distribution of density and matter velocity in 1D simulation tube for model A, $\rof=3\times 10^7\:\gcm$]
{Distribution of density and matter velocity in the tube at timestep 5000 for model A.
Fuel density is $3\times 10^7\:\gcm$. Note large fluctuations in matter velocity.
\label{a3e7}}
\end{centering}\end{figure}

\begin{figure}[htbp] \begin{centering}
\includegraphics[width=3.8in]{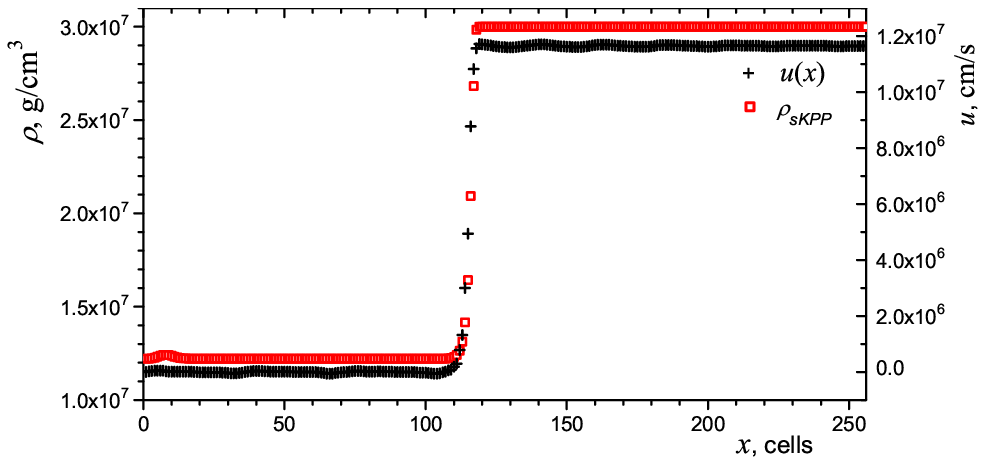}
\caption[Density and matter velocity in the tube at timestep for model sKPP, $\rof=3\times 10^7\:\gcm$]
{Density and matter velocity in the tube at timestep 5000 for model sKPP.
Fuel density is $3\times 10^7\:\gcm$. 
\label{k3e7}}
\end{centering}\end{figure}

\begin{figure}[htbp] \begin{centering}
\includegraphics[width=3.8in]{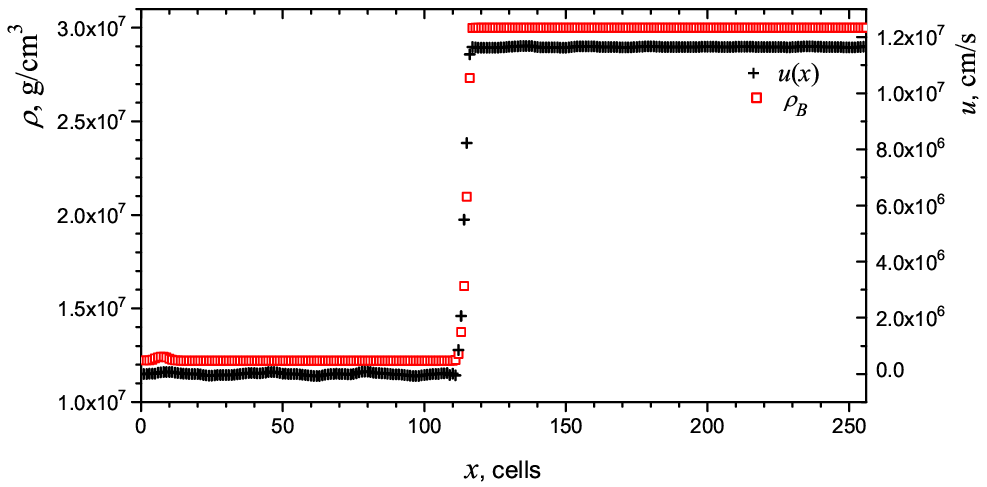}
\caption[Density and matter velocity in the tube at timestep for model B, $\rof=3\times 10^7\:\gcm$]
{Density and matter velocity in the tube at timestep 5000 for model B, fuel density 
$3\times 10^7\:\gcm$. 
\label{b3e7}}
\end{centering}\end{figure}

\begin{figure}[htbp] \begin{centering}
\includegraphics[width=3.8in]{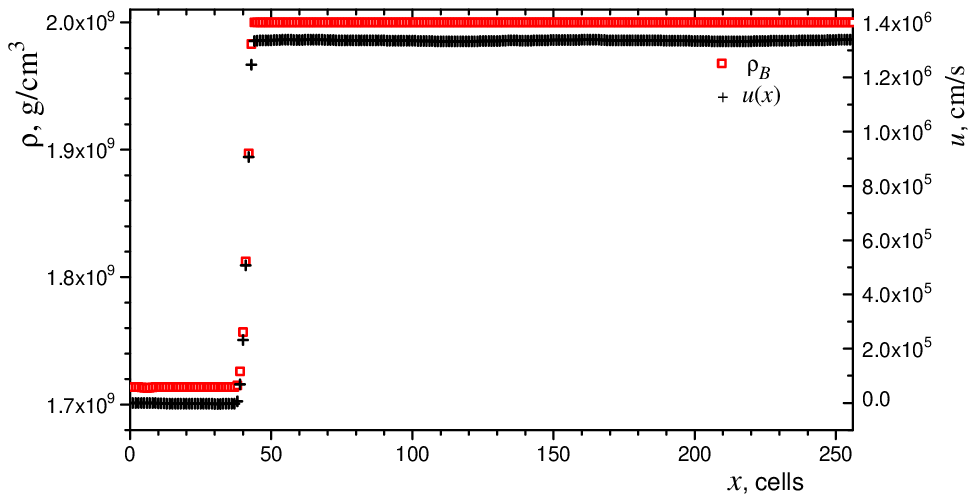}
\caption{Density and matter velocity in the tube at timestep 5000 for model B, fuel density 
$2\times 10^9\:\gcm$. 
\label{b2e9}}
\end{centering}\end{figure}

Next figures show distribution of physical variables in the tube at a fixed moment of time (timestep 5000). Matter 
densities and velocities are presented in Figs.~\ref{a3e7}--\ref{b2e9}. Sound waves are visible for all models
superimposed on density and velocity jump across the flame. Their wavelength is the same for all models at
a given density, and is different in the fuel and ash (as generation period is the same, but sound speed differs).
Say, 6 wavelengths are visible in the figures at $\rho_\mathrm{fuel}=3\times 10^7\:\gcm$ in the fuel, 
between the flame and the right end of the tube. The pattern is complicated due to reflections off the boundaries.
Reflections off the open (right) boundary were observed to be weak, which facilitates relaxation processes.

\begin{figure}[htbp] \begin{centering}
\hfill
\begin{minipage}[t]{.482\textwidth}
\begin{center}
\includegraphics[width=2.85in]{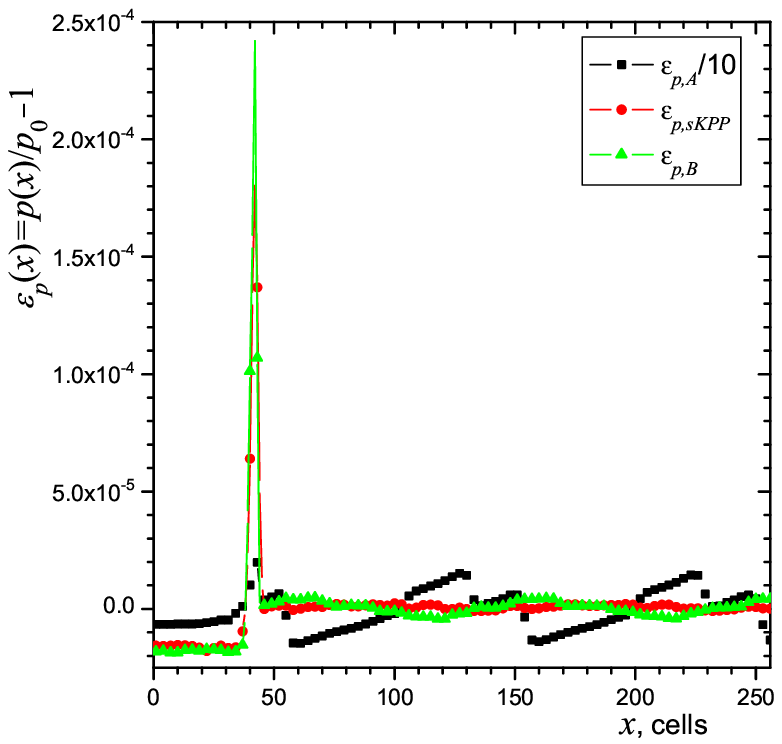}
\end{center}
\end{minipage}
\hfill
\begin{minipage}[t]{.452\textwidth}
\begin{center}
\includegraphics[width=2.65in]{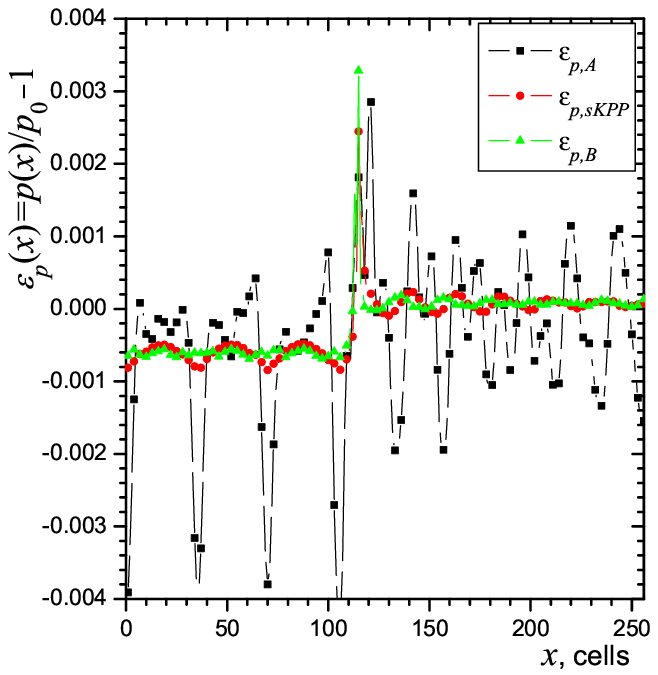}
\end{center}
\end{minipage}
\hfill
\caption[Distribution of relative pressure deviation $p(x)/\langle p_\mathrm{fuel}\rangle$ for models A, B and sKPP at 
$\rof=2\times 10^9$ and $3\times 10^7\:\gcm$]
{Distribution of relative pressure deviation (from its mean value in the fuel $p_0$) in the tube at timestep 5000.
Fuel density is $2\times 10^9\:\gcm$ in the left figure, and $3\times 10^7\:\gcm$ in the right one.
\label{sp}}
\end{centering}\end{figure}

Distribution of pressure is shown in Fig.~\ref{sp}. We show relative deviation from average pressure in the fuel:
the latter is calculated by averaging over all cells where reaction progress variable $f<10^{-4}$. Pattern of
distribution of the pressure in the sound waves resembles that in density and velocity distributions. It is  
apparent that average pressure in the ash (left of the flame) is lower than that in the fuel (by $\sim\rof u^2/c^2$, 
$c$ denoting sound speed).
Also note a strong peak in the flame zone. Simulations with larger flame width, $~20$ cells and above, show several
short pressure waves within the flame, with amplitude rapidly decreasing (as one leaves the flame; 
within a few zones) to the amplitude of pressure waves outside the flame zone; for models with step-function burning
rate it is obvious from that pattern that pressure waves are generated in the cell where burning rate changes discontinuously.
Heat starts releasing abruptly in the cell when $f$ value reaches $f_0$, generating strong pressure pulse. While
exact numerical code used, boundary conditions, have effect on non-physical, numerical noise behavior in simulations, 
certain features in flame model have universal effect on noisiness of simulations where these models are used. These
observations on noise generation show that models utilizing discontinuous burning rate functions are 
to produce significant noise. Table~\ref{tab32} quantifies our observations. On this grounds model A, with $~10\%$
fluctuations in flame speed as a function of time, and in matter velocity distribution in \lq\lq steady\rq\rq\ 1D flame 
propagation seems unacceptable for use in flame capturing.

Model B has continuous burning rate, which goes to zero as $f\to 0$ or 1. Burning rate \p{skpp} is also continuous, however
there is a jump by shift parameter $\e_f$ or $\e_a$ when $f$ reaches 0 or 1, which must be a source of noise. Yet 
with values $\e_f=e_a=10^{-3}$ this jump is $\sim 1000$ times smaller than in model A (or original model, \cite{X95});
as a result 1D noises are comparable for sKPP and model B. We concentrate on studying these 2 models further on.

\begin{table}[!hbpt]
{\center \begin{tabular}{@{\ }l@{\ \ }l@{\ \ }rrr||llrrr@{\ }}\hline 
Model    & $1/\l$   & $\e_D$     & $\e_{u,\mathrm{ash}}$    & $\e_p$       &
Model    & $1/\l$   & $\e_D$     & $\e_{u,\mathrm{ash}}$    & $\e_p$       \\ \hline

B        & 0.167   &     4.39   &   1.72   &  0.0237  &
sKPP     & 0.167   &     2.24   &   0.577  &  0.00794    \\

B        & 0.398   &     2.28   &    6.69   &    0.154   &
sKPP     & 0.398   &     2.22   &    1.55   &   0.0358      \\

B        & 0.734   &     1.91   &    12.03  &     0.421  &
sKPP     & 0.734   &     2.25   &     3.18  &     0.111     \\

B        & 1.44    &    6.38    &    21.1   &     0.79  &
sKPP     & 1.44    &    2.9     &    25.6   &     0.95     \\


B        & 2.33    &   5.83     &   18.4    &     2.07  &
sKPP     & 2.33    &   3.85     &   13.9    &     1.52     \\

B        & 5.05    &   14.6     &   10.7    &    3.26   &
sKPP     & 5.05    &   10.6     &   10.4    &    3.18      \\

B        & 8.57    &  21.5      &   70.4    &    45.1   &
sKPP     & 8.57    &  17.1      &   120     &    85.2      \\

A        & 0.167   &     726    &    69.6   &   0.958   &
A,$W_3=24$ &0.167  &     167    &    15.8   &   0.226      \\

A        & 1.44    &     733    &     225   &   14.2    &
A,$W_3=24$ &1.44   &     228    &    64.3   &   4.06       \\

\hline

\end{tabular}
\caption[Numerical noises in 1D, in flame speed, matter velocity and pressure. models A, B and sKPP]
{Numerical noise in 1D simulations. Each $\e$ written needs to be multiplied by $10^{-4}$ to yield actual relative dispersion.
Average flame speed was $80\:\kms$, flame width as in
Table~\ref{tab31} ($W_1=3.2$~cells for model B, 4 cells for sKPP, $W_3=6$~cells for A), except 2 entries for model A,
for which $W_3=24$~cells. Note that making flame wider for model A still does not render noises acceptable.
Simulation box was 256 cells long for expansion parameter $1/\l<1.5$ and 2048 cells long for larger expansion.\newline
For each run dispersion of $D(t)$ was found over $N=8000$ steps from 2001 to 10000, its relative value 
$\e_D=\sqrt{\langle D^2(t)\rangle_t-\langle D(t)\rangle_t^2}/\langle D(t)\rangle_t$ is shown (here $\langle\ldots\rangle_t$
stands for averaging over the $N$ timesteps). 
$\e_p$ represents relative dispersion in pressure in the fuel averaged over time: 
at each timestep relative dispersion of pressure in the fuel was calculated, 
$\e_p(t)=\sqrt{\langle p(x)^2\rangle_f-\langle p(x)\rangle_f^2}/\langle p(x)\rangle_f$, $\langle\ldots\rangle_f$ means
taking  average in space over fuel cells, defined as cells where progress variable $f<0.0001$; these $\e_p(t)$
were than averaged over $N=8000$ last timesteps to yield $\e_p$ shown in the table.
$\e_{u,\mathrm{ash}}$ represents relative dispersion in matter velocity in the ash in the same manner: at every 
timestep  
$\e_{u,\mathrm{ash}}(t)=\sqrt{\langle u(x)^2\rangle_a-\langle u(x)\rangle_a^2}/\langle u(x)\rangle_a$ was defined
($\langle\ldots\rangle_a$ meaning averaging over cells with ash, where $f>0.9999$);
these were averaged over the 8000 timesteps to yield $\e_{u,\mathrm{ash}}$ shown.
\label{tab32} }
}
\end{table}


\chapter{Flames in 2D and 3D simulations}\label{noisend}
In addition to numerical noises that
manifest themselves in 1D as well, there are two physical phenomena that should affect flame propagation in more than one
dimensions. First, normal flame speed of a curved flame generally depends on the front curvature. This is usually called 
Markstein effect, though this term also refers to specific form of that dependence, namely flame speed depending linearly
on flame curvature: $D_R=D_{R=\infty}\left(1-\frac{Ma}{R}\right)$.
$R$ here denotes radius of curvature of the flame; $D_R$ stands for normal propagation speed of a flame with radius of curvature $R$;
proportionality parameter $Ma$ is called Markstein length. Such linear dependence was observed in experiments and simulations
(\cite{markstein}). See next chapter for more detail, and for studying this effect numerically on model flames. 

Second, when fuel and 
ash densities are different, Landau-Darrieus (LD) instability leads to growth of small perturbations on smooth flame surface,
resulting in wrinkling the flame, and formation of corners later on. Perturbations of any wavelength are unstable in idealized
case of infinitely thin flame with no Markstein effect (\cite{dar,lan}), that is when flame speed does not depend 
on curvature. When $Ma>0$, for infinitely thin flame, only perturbations with wavelength exceeding certain critical value 
$\lcr$ grow. 
This is what is observed for all our models. Besides Markstein length found positive, the flame width is significant 
($\sim 4$ cells, comparable to initial radius of burnt bubble in simulations). Theory of LD-type instability of 
finite-thickness flames is much more involved 
(see, \eg\ \cite{matmat}). It can be expected that the critical wavelength exceeds a few flame widths; its exact value, 
as well as instability growth rate should depend on exact distribution of density and heat production within the flame.
Numerical effects may additionally distort flame surface. The aim of this section is to study how the shape of initially 
spherical bubble changes with time.

\section{Simulation setup}
As for 1D simulations of Chap.~\ref{noise1d} we use uniform grid with the same grid spacing across all our simulations. 
The simulations were run in square $N\times N$ cells box in 2D, cubic $N\times N\times N$ box in 3D; $N$ ranging from 64 
(box side 132~km) to 2048 (4224~km) for 2D simulations, and from
64 to 256 for 3D ones. Normalizations $\tilde{K}$
and $R$ of flame model diffusivity and reaction rate are chosen, as in 1D simulations described above, via \p{KR}, to
yield \fl\ speed $80\;\mathrm{km\:s^{-1}}$ and width $W_1$ as shown in Tab.~\ref{tab31}, unless specified otherwise. 

For all simulations one of two types of boundary 
conditions is used. In first type of simulations deflagration is initiated by a sphere of hot ash 
placed fully in the interior 
of the square/cubic grid (except for a few runs described below 
initial flame center coincides with box center)%
; outflow boundary conditions are imposed on all 
boundaries. 
Initial distribution of reaction variable $f$  and physical quantities is set smooth, corresponding to
quasi-steady state spherical flame burning outward. Initial conditions are characterized by initial
spherical \fl\ center position $(x_0,y_0)$, and radius $R_0$; the latter signifies here radius of the surface 
$f=0.9$. 
Explicitly, abscissa of 1D steady-state flame profile $f(\chi)$ 
(steady progress variable as a 
function of dimensionless coordinate, eigenfunction of \p{artvislamdim}) is rescaled so that flame width ($f=0.1$ to $f=0.9$) 
were $W_1$ (the width of the flame we want to achieve) and then translated; in precise terms
 $f(\chi)\mapsto c(r):$ $c(r)=f\left( (r-r_0)\frac{w_1}{W_1}\right)$, 
translation $r_0$ chosen such that $c(R_0)=0.9$. Initial reaction progress variable distribution is then set spherically 
symmetrically around the center of what we loosely call \lq\lq a sphere of burned material at $t=0$\rq\rq, according to this rescaled 
profile $c(r)$, $r$ being a distance from the flame center. So, if the center of some cell is $R_0$ from the flame center, 
initial value of reaction progress variable is 0.9 in that cell; if it is $R_0+W_1$ off the center, progress variable is 0.1 there, etc. 
$f$ is zero for all cells more than total flame width away from the initial \lq\lq sphere of ash\rq\rq, and is 1 at the center of that 
sphere (unless the flame is too wide, and its 
tail reaches the center). Other thermodynamical variables, and matter velocity are distributed accordingly, mimicking their 
distribution for quasi-steady state spherical flame burning outward from the center (though neglecting the difference in 
flame profile $f$ of spherical and planar flame). 

Simulations of second type are run in quadrant (2D) or octant (3D): burning is initiated from the corner of the cube, by 
filling a sector of a sphere centered at that corner with ash. As for the full cube simulations, this wording means smooth 
initial distribution of thermo- and hydrodynamical variables mimicking quasi-steady state 2D/3D burning from the corner, 
with reaction variable $f(r=R_0)=0.9$ and $f(r=R_0+W_1)=0.1$ at $t=0$; matter velocity directed radially off the ash sector 
vertex (in computation cube corner), matter resting at the corner. 
Reflecting boundary conditions are imposed (throughout the run) on 2 sides/3 faces crossing in that corner, and outflow 
boundary conditions on the rest of simulation box sides/faces. Such quadrant/octant simulations are more economical and must 
provide the same results as full cube simulations\footnote{%
full cube containing 4 times more cells in 2D, 8 times more in 3D; due to symmetry all 8 full-cube
octants behave the same, and presumably the same way as one isolated octant with reflecting boundary conditions on 3 faces 
passing through its vertex.}, 
thanks to the problem symmetry about the cube center when central ignition is used. These are
customarily used in literature for this reason. Quite severe numerical artifacts are often observed in such octant 
simulations, with system behavior along octant reflecting faces significantly different from bulk behavior.
We use both types of setup precisely to check how different results one gets in full-cube and octant simulations, due to, 
say, LD instability seeding (or modification) by the boundaries, or maybe just imperfectly realized boundary conditions --- in
our simpler setting, with simple chemistry and without gravity.

\section{Flame behavior in 2D, theory and observations}
\begin{figure}[htbp] \begin{centering}
\includegraphics[width=4.0in]{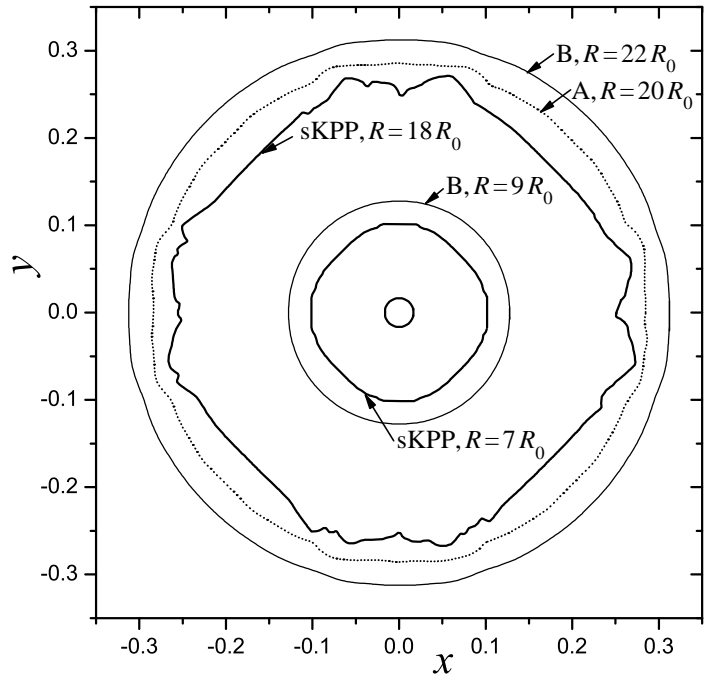}
\caption[Evolution of initially circular flame in 2D for models A, B and sKPP at $\rof=3\times 10^7\:\gcm$]
{Flame for the 3 models, represented by surfaces where reaction variable $f=0.5$. Coordinates are in units of box size, \eg\ 
$x=\pm 0.5$ corresponds to right/left boundary. Fuel density is $3\times 10^7\:\gcm$.
The outermost surface is that of Model B, at timestep 14000, when flame radius is $R=22R_0$; initial radius 
$R_0=30\; \mathrm{km}$ for all runs shown in this Figure. This flame is also shown at earlier time, when its radius is $9R_0$. 
Innermost circle represents flame surfaces at $t=0$. For Model sKPP surfaces are 
shown (thick solid lines) for timesteps when $R=7R_0$ and when $R=18R_0$. Note that global asphericity due to anisotropic flame 
speed is the most prominent feature at early times, and at later times small-scale LD-type instabilities severely distort flame 
surface. Dotted line shows flame position for Model A when its radius $R=20R_0$.
\label{2d3mods}}
\end{centering}\end{figure}

Fig.~\ref{2d3mods} shows how initially spherical flame evolves with time, for the 3 models studied with expansion parameter
$1/\l=1.442$ ($\rho_\mathrm{fuel}=3\times 10^7\:\gcm$). One can observe two features of flame evolution: first, some global anisotropy 
develops, it looks like flame propagates faster along axes than along the diagonal of the grid, for Model A and sKPP; second, small scale 
wrinkling of the flame surface is observed, resembling LD instability development. It is seen that Model sKPP, despite having largest 
width $W_1$ (and by far largest $W_3$) demonstrates the largest distortion of spherical surface as deflagration progresses. The fact that 
Model A behaves better shows that these 2D distortions are unrelated to 1D noises, by far strongest for Model A. This rules out
possible explanation of 2D behavior by relating it to flame speed fluctuations in time, different in different directions. Pressure
waves due to numerical noise are by far most intensive for Model A in 2D/3D as well, yet these waves are also not the most important factor
in disturbing flame surface, as comparison of flame surfaces for the models demonstrates.

These observed phenomena are detrimental for use in flame capturing. Of course, real flame region is subject to its own instabilities,
it interacts with real physical sound waves and turbulence; all of these distort its surface, initially spherically flame will quickly
become non-spherical in real world, as our model flames do. However, the fact that the amount of distortion strongly depends on the model,
even for comparable \fl\ widths (making any flame thinner aggravates asphericity development, so when sKPP model width is made to match that
of Model B its behavior becomes further worse) shows that there is no hope that some random model instability development would miraculously 
match that of real physical flame region. Anisotropy in flame speed is clearly a numerical artifact and is not observed in nature. There is 
no control over small scale wrinkles developing on \fl\ surface (apart from making the \fl\ wider; see results below). Therefore the best model
is clearly the one preserving spherical flame surface as closely as possible, and this is how model B was designed. 

Evidence below suggests
that small-scale perturbations have the nature of LD-instability. It is then understandable that actual distribution of matter density
within the \fl\, in part, influences smallest unstable scale $\lcr$, as well as instability growing rate, which naturally
explains why different models behave differently. From the figures below it follows that sKPP has the smallest $\lcr$, and 
instabilities grow fastest. Models A and B yield flames having density distribution within the flame closer to linear,
this seems to make them more stable to LD-like instability. As observed in \cite{X95}
approximately linear density distribution is also the case for real flame zone (averaged over scales corresponding to typical numerical grid 
sizes) in RT-driven deflagration simulations, in SN~Ia simulations in part. Hence there is more hope that Model B (and not sKPP) would 
behave more similar to real flame region in terms of flame perturbations evolution.

\begin{figure}[htbp] \begin{centering}
\includegraphics[width=4.0in]{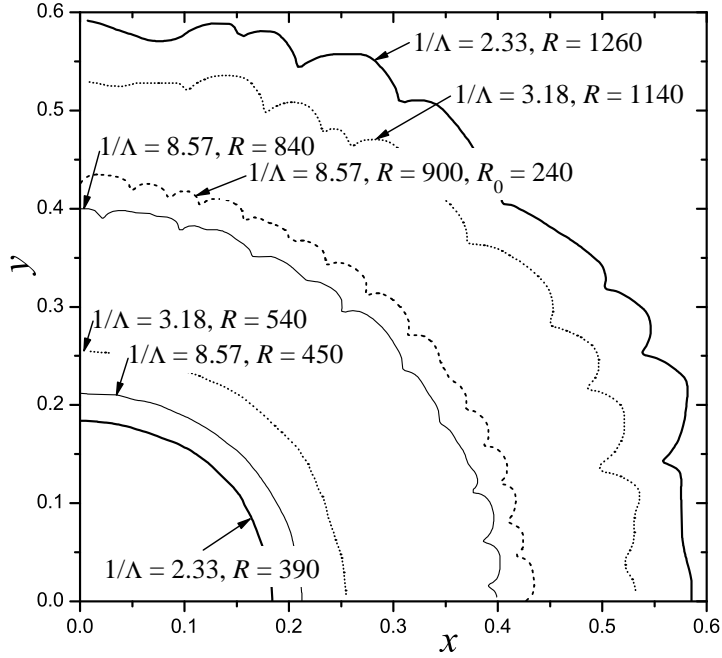}
\caption[Flame surface for model B at large expansion, $\rof\le 10^7\:\gcm$ at late times. Developed LD-type instability]
{Flame surface for Model B at late times, for high expansion, corresponding to fuel densities $8\times 10^5$, $5\times 10^6$
and $10^7\;\gcm$. For each density two flame positions are shown (indicated by flame radius written next to 
corresponding surface, in km): 
when perturbations by LD-type instabilities start to be visible,
and when they are well developed. Notice that characteristic lengths of perturbations increase as expansion across the 
flame decreases; also that the flame retains reasonably spherical shape (globally) until late enough times: 
$R=540\;\mathrm{km}$ is the same as the radius of sKPP flame in Fig.~\protect{\ref{2d3mods}}. Initial radii of the 
flames were $R_0=120\;\mathrm{km}$ for $1/\l=2.33$, $R_0=90\;\mathrm{km}$ for 
$1/\l=3.18$, $R_0=30\;\mathrm{km}$ for $1/\l=8.57$. For $1/\l=8.57$ 
results of one more simulation are shown (one flame surface when its radius reached $900\;\mathrm{km}$, short dash). 
This simulation was run in full square (with central ignition; only one quadrant is shown, $x=0.5$ corresponds to 
simulation box boundary for this last surface, whereas $x=1$ is right box boundary for rest of the runs), initial 
radius $R_0=240\;\mathrm{km}$. 
It is apparent from comparison that characteristic lengthscales of surface 
perturbations are determined solely by expansion, and do not show any significant correlation with initial radius.
No large scale structure anisotropy of flame surface is visible for the central ignition run, even though its surface 
shown is just $15\%$ of the box size off the box boundary. Finally, at times shown there is not much difference 
between large scale anisotropy for central and corner ignition; it is more noticeable at later times, and 
perturbations generation on reflecting boundaries can be seen, but overall the difference between quadrant and 
full-square runs is not large (for Model B) until so late times when open boundaries start playing a role (this was 
checked for full-square simulations in up to $2048\times 2048$ boxes for some densities).
\label{2dBq}}
\end{centering}\end{figure}

Fig.~\ref{2dBq} shows flame surface for model B in simulations in quadrant $1024\times 1024$ cells at late times, when radius is large enough
to exceed minimal unstable perturbation length, and enough time has passed for the instability to develop. In this figure small-scale 
surface features evolution, seen in Fig.~\ref{2d3mods} for Model sKPP, looks like typical LD instability development. Notice that,
as expected, characteristic length of perturbations increases with fuel density, as expansion across the flame decreases. 
In Fig.~\ref{2d3mods}, at 
$1/\l=1.442$ just first signs of instability development can be seen, with length of order of a quarter of the flame circumference,
$\sim 400$ cells. Thus, even though at any density flame with large enough radius will become LD unstable, for densities 
$\rho_\mathrm{fuel}\gtrsim 3\times 10^7\:\gcm$ time required for instability development exceeds typical multidimensional simulation
durations. More careful quantitative analysis shows (some numbers are presented below) that there is not much difference between 
quadrant and full-square runs (ones with central ignition) in terms of surface perturbations as long as flame arc length in quadrant does 
not exceed a few $\lcr$, characteristic lengths of LD-type perturbations. 
After that boundaries do contribute to instability growth in their vicinity, yet for the durations our runs had (enough
for the flame to pass $\sim 1000$ cells) there was not seen apparent difference far enough from the axes for full-square and quadrant 
simulations (for Model B). This may be seen for a few runs shown in Fig.~\ref{2dBq}.

\begin{figure}[htbp] \begin{centering}
\includegraphics[width=4.0in]{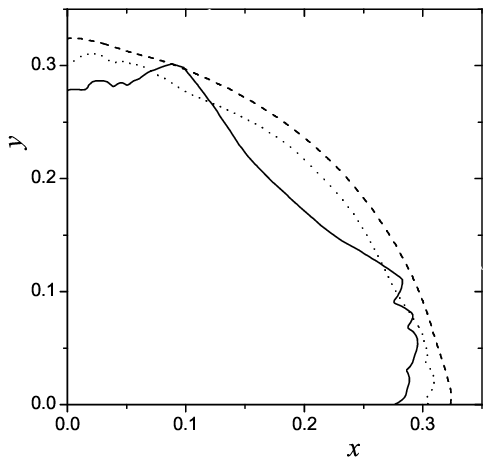}
\caption[Flame surface for Model sKPP with $W_1=4$, 8 and 16 cells; $\rof=3\times 10^7\;\gcm$]
{Flame surface for Model sKPP with $W_1=4$, 8 and 16 cells, $\rof=3\times 10^7\;\gcm$. 
One quadrant of full-square simulation ($1024^2$) 
is shown; $x=0.5$ represents right boundary of the box.
Solid line represents $f=0.5$ curve in a $W_1=4$ flame when its average radius is 600~km. Dotted line shows the 
curve defined the same way for $W_1=8$ flame when its radius reached 630~km. Dashed line shows $W_1=16$ flame with 
radius 660~km. Note that LD-type wrinkles are fully developed for $W_1=4$ flame, but for $W_1=16$ flame just
first signs of instability development are seen at the radius shown; lengthscale of the wrinkles increases with flame 
width. Also note that the flame is still noticeably closer to the center along grid diagonal than along the axes even 
for $W_1=16$.
\label{2dKq}}
\end{centering}\end{figure}

At smaller expansions all models behave better in 2D, no asphericity is seen by naked eye in simulations with fuel density of above 
$3\times 10^8\:\gcm$ ($1/\l<0.4$; some quantitative results are presented in the next subsection). At larger expansions, however, sKPP flames 
behavior deteriorates rapidly, making sKPP model use in FC dangerous at physically interesting densities (down to a few times $10^6\:\gcm$, 
although there is 
less time for instability development at smaller densities, as corresponding regions are reached later in simulations with ignition close 
to star center). When sKPP model shift parameters $\e_a,\:\e_f$ are increased the flame becomes more stable to LD-type instability 
(being another piece of evidence suggesting that long sKPP profile tails may be related to predisposition to LD instability for sKPP model: 
these tails are shorter for larger parameters $\e_a,\:\e_f$); global flame anisotropy is not improved significantly with increasing flame
 parameters. 1D noises become objectionable at larger shifts, reaching, \eg\ $1.8\%$ average dispersion in flame speed at $\e_a=\e_f=0.1$, 
$1/\l=1.442$ (0.54\% at $\e_a=\e_f=0.03$, 0.38\%
at $\e_a=\e_f=0.01$). Values $\e_a=\e_f=10^{-3}$ 
yield 1D noises close to those for Model B (at corresponding flame widths we use); sKPP is studied with these $\e_{a,f}$ values below, unless 
indicated otherwise. There might be a hope that making sKPP flames slightly wider could drastically improve their stability. As 
Fig.~\ref{2dKq} shows, LD-like instability is really suppressed for wider flames, expectedly; however, global flame propagation anisotropy 
decreases rather slowly with increasing flame width. Hence we are forced to conclude that for use at larger expansions, $1/\l\sim 0.6$ and
larger, Model B is the only viable candidate of all the models we considered.

\section{Numerical results for flame shape evolution, 2D}
Here we trace how flame shape changes with time for models B and sKPP, for a range of fuel densities. Besides comparing
model to model in similar conditions we also want to compare flame asphericity development for different initial flame
radius, and for different distances to the boundaries. 
By varying initial radius we want to better understand the role of initial conditions, whether instabilities readily
grow from the very beginning of the computation (and then this growth will be further accelerated in astrophysical
simulations with gravity), or whether there is some initial interval of time when flame radius is small and LD-type
instabilities decay (then, in part, unavoidable imperfections of initial conditions will not play that significant
role in flame behavior at later times, which is desirable); what the dependence of global flame propagation
anisotropy on flame radius is. To check the role of boundaries on flame surface evolution we compare behavior of the
models at different resolutions; while keeping grid spacing the same this effectively moves boundaries further away
from flame surface, whereas flame width and initial radius remain fixed, both in terms of number of cells and in
physical units. 

In Table~\ref{tab41} flame geometry is characterized by 2 numbers (called flame asphericity parameters below). First, 
by interpolation we find a set of points $(r_i,\phi_i)$ (in polar coordinates) on the surface $f=0.5$.
Among these we find the two with extremal distances from flame center $r_\mathrm{max}$ and $r_\mathrm{min}$, average 
radius over all points, $R=\langle r_i\rangle$, and the best fit of $r_i$ distribution with 
$\tilde{r}_i=r_0+r_1\cos 4\phi_i$. The two numbers we use to describe flame surface
are $\Delta r/R=(r_\mathrm{max}-r_\mathrm{min})/R$ and $r_1/r_0$ (All values written in Tab.~\ref{tab41} are 
multiplied by a common factor of $10^3$.)
These numbers give rough idea about flame surface: if it is circular both numbers
are zero; if $r_1/r_0$ is close to one half of $\Delta r/R$  one would conclude that large-scale anisotropy of flame 
propagation speed is the main feature, dominating over amplitudes of small-scale LD-type wrinkling; if $r_1/r_0$ is 
significantly less than one half of $\Delta r/R$ the wrinkles yield more contribution into $\Delta r/R$ than 
systematic large-scale anisotropy of flame surface.

As observed in the previous section, flame shape in figures, for
Model sKPP $r_1/r_0>0$, flame surface is in average closer to its center along diagonal of the grid than along grid 
axes. For Model B this systematic anisotropy is controlled by parameter $c(\l)$ in the burning rate. If, at a given 
expansion, term $c(\l)$ is used that is larger than its optimal value (which we approximate as \p{clam}) the flame 
propagation speed along grid diagonal exceeds that along grid axes, and vice versa; this anisotropy is related to 
dependence of density distribution within the \fl\ on $c(\l)$ (governing reaction rate distribution within the \fl). 
The values we suggest in \p{clam} were chosen this way, to optimize flame speed anisotropy.

It was checked in simulations with different resolutions that boundaries start playing a role in disturbing flame
surface when the flame passes approximately half way from the center to box boundary face. Numbers $\Delta r/R$ and 
$r_1/r_0$ at that point start deviating from those observed in the runs with larger box (in terms of number of cells; 
the flames have the same width, and are compared at the same radius for smaller and larger boxes), 
by a factor of order 1.4--2, indicating that actual perturbation started developing somewhat earlier, and have grown 
to an extent such that deviations in flame asphericity parameters become significant. 

\begin{table}[!hbpt]
{\center \begin{tabular}{@{$\,$}l@{$\,$}r|@{$\,$}rr|@{$\,$}rr|@{$\,$}rr|@{$\,$}rr@{$\,$}}\hline 
Model    & $1/\l\;$   &\R{\frac{\Delta r}R}{1} &\R{\frac{r_1}{r_0}}{1\;}&\R{\frac{\Delta r}R}{2} &\R{\frac{r_1}{r_0}}{2\;} &
                       \R{\frac{\Delta r}R}{3} &\R{\frac{r_1}{r_0}}{3\;}&\R{\frac{\Delta r}R}{5} &\R{\frac{r_1}{r_0}}{5\;\vphantom{5_;}}\\ \hline

sKPPq$,\,N=512$& 0.167&1.36                        & 0.787   &   3.42                        &  1.38   &
                       5.45                        &  2.76   &   8.40                        &  4.19   \\

sKPPq    & 0.398    &  2.16                        &  -1.23  &   3.07                        &  -1.65   &
                       3.34                        &  -1.12  &   1.58                        &  0.083   \\

Bq       & 0.398    &  2.93                        &  -1.74  &   8.61                        &  -4.80  &
                       11.8                        &  -5.96  &   11.4                        &  -5.59  \\

sKPP     & 0.734    &  8.57                        &   4.00  &   19.8                        &   9.87  &
                       29.4                        &   11.4  &   49.0                        &   18.7  \\


B$,\,N=512$& 0.734  &  3.12                        & 0.331   &    5.42                       &  1.60   &
                       8.06                        & 1.34    &      -                        &     -   \\
 
Bq$,\,N=512$& 0.734 &  2.37                        & 0.246   &    5.28                       &  1.49   &
                       9.85                        & -2.33   &    11.8                       & -1.88   \\ \hline

sKPP     &  1.44    &  13.3                        &  4.79   &    54.6                       &  24.5   &
                           99.2                    &  36.4   &    197                        &  53.6   \\


sKPP$,\,W_1=8$&1.44 &  7.09                        &  2.03   &   37.5                        &  17.1   &
                       41.9                        &  19.4   &   60.9                        &  23.8   \\

sKPP$,\,W_1=16$&1.44&  3.58                        &  -0.988 &   26.1                        &  10.2   &
                       36.5                        &  13.5   &   47.2                        &  17.0   \\

sKPP$,\,\e=0.01$&1.44& 6.41            &  1.62   &   30.0                        &  14.1   &
                       39.6                        &  13.2   &   79.3                        &  22.2   \\

sKPP$,\,\e=0.03$&1.44& 3.63            &  -0.811 &   16.0                        &  6.40   &
                       13.5                        &  4.37   &   34.9                        &  3.63   \\
%
sKPP$,\,x_0,y_0=0.1$& 1.44& 14.6                   &  5.21   &   50.6                        &  23.9   &
     96.4                                          &  30.8   &    204                        & 55.0    \\

sKPP$,\,y_0=0.2$&1.44&   11.6                      &  4.23   &   57.1                        &  22.5   &
                        116                        &  32.1   &    314                        &   33.5   \\ \hline

B        &  1.44    &  8.17                        & -4.58   &      2.77                     &  0.59   &
                       3.75                        & -1.17   &      8.80                     &  0.046   \\ 

B, $R=90$ &  1.44   &  6.86                        & -3.64   &      9.47                     & -4.51   &
                       12.1                        & -5.68   &      11.0                     & -3.11  \\

B, $R=180$&  1.44   &  4.03                        & -1.78   &      7.89                     & -3.61   &
                       9.69                        & -4.41   &      11.9                     &  4.31  \\

B, $R=15$ &  1.44   &  7.61                        &  3.69   &      17.6                     &  9.43   &
                       8.78                        &  3.79   &      9.57                     &  3.24  \\ 

Bq, $N=512$& 1.44   &  7.93                        & -4.68   &      2.66                     &  1.09   &
                       2.93                        & -1.08   &      7.89                     &  0.14   \\ 

%
\hline
B$,\,R=90,\,N=2^{11}$ & 2.33&  12.3                & -6.23   &       18.2                    & -9.20   &
                                 23.5              & -9.35   &       50.5                    & -10.5   \\ 

B$,\,R=90$          & 2.33&  12.4                  & -6.23   &       17.9                    & -8.93   &
                                 26.5              & -9.25   &       48.6                    & -1.86   \\

B          & 2.33   &   20.0                       & -11.7   &       15.8                    & -7.09   &
                                 18.5              & -3.92   &       34.9                    & 0.09   \\ 

B$,\,R=90$ & 3.18   &   14.1                       & -6.76   &       20.3                    & -10.6   &
                                 24.7              & -11.9   &       43.0                    & -12.3   \\ 

B          &   8.57 &   4.31                       & -2.22   &       6.68                    & -2.95   &
                                 12.7              & -5.05   &       34.6                    & -5.69   \\ 


\hline

\end{tabular}
\caption[Asphericity of flame surface in 2D simulations at different times, for models B and sKPP, at different densities and initial 
conditions]
{Asphericity of flame surface at different times, for models B and sKPP. Each $\frac{\Delta r}R$ shown needs to be multiplied with a 
factor $10^{-3}$ to yield actual $\frac{\Delta r}R$; same for $r_1/r_0$.
Each pair of asphericity parameters is shown for 4 different moments in time, when flame surface radius reached 
$R_1=R_0+30\:(\mathrm{km})$ (the $1^\mathrm{st}$ pair of columns), $R_2=R_0+90$ (the $2^\mathrm{nd}$ pair), $R_3=R_0+270$ 
(the $3^\mathrm{rd}$ pair), and $R_5=R_0+600$ (the last one).
Unless specified otherwise each model is assumed to yield $D=80\:\mathrm{km\:s^{-1}}$, and width $W_1$=3.2 cells for 
Model B, and 4 cells for sKPP. Initial flame radius $R_0=30\:\mathrm{km}$ (unless different $R_0$ is specified; 
$R_0$ and $R$ are in kilometers in this Table.) Default setup is central ignition, $1024^2$ grid cells; 
simulations in quadrant with corner ignition are indicated by letter \lq\lq q\rq\rq\ at model name; 
if the side of a simulation box is different from $N=1024$ grid cells, actual $N$ is specified next to model name. 
For default $R_0=30\:\mathrm{km}$ the moments of time shown
  roughly represent moments when boundary conditions start playing a role for central ignition simulations
in boxes with sides 128, 256, 512 and 1024 cells respectively (this corresponds to the time when flame surface is 
about its radius away from the boundaries; this is true regardless of whether the flame is LD-unstable or not (at radii shown),
as may be seen on the results in the Table for Model sKPP at $1/\l\ge 0.734$ or Model B at $1/\l\ge 2.33$. 
\label{tab41}}}
\end{table}

\begin{figure}[htbp] \begin{centering}
\includegraphics[width=4.0in]{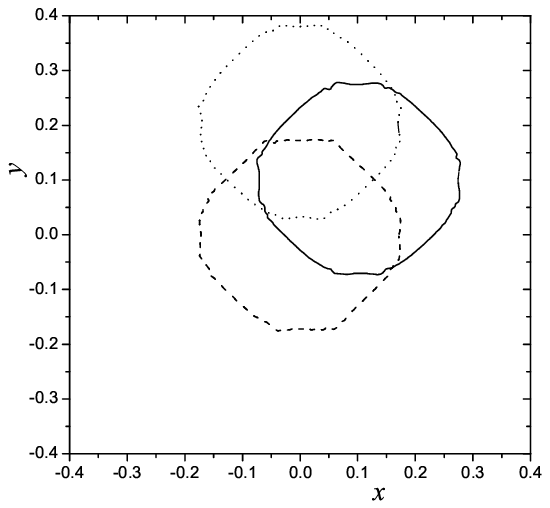}
\caption[Flame surface for Model sKPP with different locations of a seed flame at $t=0$; $\rho=3\times 10^7\;\gcm$]
{Flame surface for Model sKPP with different locations of a seed flame at $t=0$. $\rof=3\times 10^7\;\gcm$, 
$D=80\:\mathrm{km\,s^{-1}}$, $W_1=4$~cells, $R_0=30$~km, each flame is shown when its radius reached 360~km 
(at timestep about 7500).
Dashed line shows the flame with central ignition, $x_0=y_0=0$ (box boundaries are at $x,y=\pm 0.5$); solid line
corresponds to the flame with $x_0=y_0=0.1$, and dotted one to the flame with $x_0=0, y_0=0.2$. Notice that the first 
signs of large-scale distortions are seen on the flame closest to the boundary, due to matter
speed asymmetry, yet small scale perturbations on all three flames show essentially the same amplitude.
\label{offcenter}}
\end{centering}\end{figure}

Asphericities summarized in Tab.~\ref{tab3} suggest that the role of boundary conditions is under control, and 
seemingly the same for Model B and sKPP, and whether the flame is stable or not. Numerical noises in the bulk, 
far from the flame surface are comparable for 
models B and sKPP, all this evidence suggests that it is not some intricate interaction with boundaries which makes 
sKPP flame speed anisotropic, but rather local properties of sKPP flame. Small scale ripples generation may be more 
susceptible to background noises in sKPP model, this partly explaining fast growing amplitudes of those. Regardless of 
whether there is any significant contribution to wrinkle growth due to waves reflected from the boundaries, Model sKPP
is unsuitable for simulations with large expansions of matter across the flame. For clarity, nonetheless, we performed
several runs with off-center ignition to verify whether instability growth is mostly governed by local physics near the
flame, or whether better boundary conditions may, to extent, improve sKPP flames behavior. In those full-square runs
initial conditions corresponded to quasi-steady spherical 2D burning, with flame center not coinciding with the center
of the cube. Results for flame centered at points $(0;0.2)$ and $(0.1;0.1)$ (in units of box size: $x=\pm 0.5$
are two of the box boundaries in these units) for expansion $1/\l=1.44$ are shown in Tab.~\ref{tab41} together with
results for central ignition. It is seen that asphericity parameters do not differ drastically for these 3 setups. 
Same holds true for smaller expansions, when asphericities are less, as well as for Model B.
Fig.~\ref{offcenter} shows flame surfaces for the 3 setups: small scale wrinkles look very similar, regardless of 
relative distances to the boundaries.

\begin{figure}[htbp] \begin{centering}
\includegraphics[width=3.5in]{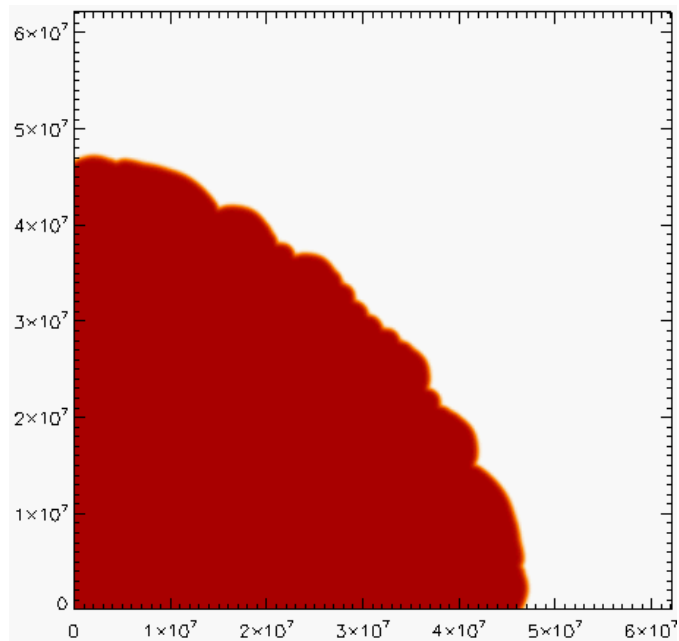}
\caption[Flame surface for Model sKPP, $\rof=3\times 10^7\;\gcm$, computed in quadrant $256\times 256$ cells with {\tt FLASH} code]
{Flame surface for Model sKPP, $\rof=3\times 10^7\;\gcm$, $D=80\:\mathrm{km\,s^{-1}}$, $W_1=4$~cells, computed in quadrant 
$256\times 256$ cells with {\tt FLASH} code.
\label{KPPFLASH}}
\end{centering}\end{figure}

In the next section we present some results for growth or decay of a sinusoidal perturbation on a planar 2D flame.
Those quantitatively confirm that the small-scale instability we observe on circular flames are really of LD type.
This suggests that the results should not qualitatively depend on a particular code used for simulations, as they are 
a manifestation of a real physical phenomenon, which should be observed with any code. 
Large scale flame speed anisotropy, as we saw, depends to some extent on flame width, it is likely to depend on a 
particular way derivatives are estimated by the code (larger stencils used would have effect of additionally smearing 
the flame, thus stabilizing it to some extent), yet the qualitative result that diagonal propagation
is slower than that along the axes (for sKPP) is likely to hold true for any code. It would be interesting to 
systematically check our results with a different numerical scheme. While we have not performed systematic code-to-code
comparison, typical results obtained with code {\tt ALLA} were confirmed with {\tt FLASH} code (\cite{fryxell}), 
for Model sKPP developing global asphericity and LD-type instabilities with short unstable length, and Model B 
showing much better behavior in this respects at lower fuel density. Derivatives are estimated with $4^\mathrm{th}$ 
order accuracy in {\tt FLASH} code (vs $2^\mathrm{nd}$ order in {\tt ALLA}). Fig.~\ref{KPPFLASH} shows a typical flame
surface simulated in quadrant $256\times 256$ with {\tt FLASH} code, with the same flame model parameters as discussed
above, with $1/\l=1.44$. Same features of the model are observed.

\section{Perturbed planar 2D flames, LD instability}
\label{Mark1D}
Here we present results of a study of the instability development on sinusoidally-perturbed planar flames in 2D. 
We simulate such
flames in rectangular domains with reflecting boundary conditions on 3 of the sides, and outflow boundary condition
on the fourth side, towards which the flame propagates. We observe that for domains with a width below certain critical value 
$L_\mathrm{cr}$ (which depends on expansion parameter and the flame model used) the perturbation decays as the flame 
propagates. For wider domains certain non-planar flame surface develops, which steadily propagates into the fuel. 

\subsection{Theoretical background}
\label{Mark:theory}
The behavior we observe, flames in wide domains developing non-planar front shape,
qualitatively agrees with known from experiments behavior of terrestrial flames; however, no analytical studies
have been done in the regime relevant in FC, with fairly wide flames and non-negligible expansion of matter across the flame.

First theoretical studies of this type of instability, linear analysis of its development (when the perturbation is small), 
were done by \cite{lan} and \cite{dar} in approximation 
of infinitely thin flame, with flame velocity independent of its geometry and background hydrodynamical flows; the result is
that perturbations with any wave number $k$ grow exponentially, with the rate $\omega$ proportional to $k$:
\begin{equation}\label{landar}
\omega_\mathrm{LD}(k)=kD\frac{-\sigma+\sqrt{\sigma^3+ \sigma^2- \sigma}}{\sigma+1}.
\end{equation}
$D$ here denotes flame speed, $\sigma=\rof/\rho_\mathrm{ash}=1/\l+1$. 
As discussed above this approximation is invalid at small perturbation lengths, comparable with the flame width. Real flames are
generally stable with respect to such short-wavelength perturbations. 

A simple (and physically sound) modification of
the original LD consideration leading to stable perturbations with large enough $k$ is to take into account flame speed dependence on
the flame curvature. Assuming linear dependence on curvature $1/R$, $D(R)=D_{R=\infty}\left(1-\frac{Ma}{R}\right)$, still in the
limit of infinitely thin flame, modified dispersion relation is given by (\cite{markstein})
\begin{equation}\label{landarma}
\omega=\frac{kD}{\sigma+1}\left[ -\sigma(1+Ma\,k)+\sqrt{\sigma^3+\sigma^2-\sigma+(Ma\,k-2\sigma)Ma\,k\sigma^2}\right].
\end{equation}
For positive $Ma$, which is the case for all flame models we consider for FC applications and for most chemical flames in practice%
\footnote{$Ma$ may be negative in, \eg , systems with significantly different diffusivities of different reactants when 
the most diffusive reactant is deficient (for example
for lean hydrogen-oxygen flames), cf. \cite{MatCuiBec03} for theoretical study. Infinitely thin flames with negative $Ma$ are unstable
for any perturbation wavelength, as are flames in original LD approximation, with $Ma=0$.
}, perturbations with
\begin{equation}\label{lcrma}
2\pi/k<\lcr(Ma) = 4\pi(\l+1)\,Ma
\end{equation}
decay. This is understood qualitatively, as positive $Ma$ stabilizes the flame by decreasing propagation velocity on convex parts
of the flame surface (the ones that are displaced forward into the fuel, with respect to a plane unperturbed flame surface), and 
increasing the propagation velocity into fuel on concave parts, which are behind the unperturbed surface. This effect tends to
suppress growth of the instabilities (due to main component of LD instability, with $D(R)$ dependence neglected), and it
does stabilize the smooth flame surface at sufficiently short perturbations (when curvature is larger, and this curvature effect is
more pronounced). 

Linear analysis was also performed up to the first order in flame width (in \cite{matmat, pelce82, MatCuiBec03}, under different conditions),
thus elucidating the effect of flames being not infinitely thin, and obtaining some estimates for Markstein lengths. These works
only deal with Arrhenius-type reaction rates, thus the results are not directly applicable to our flame models; besides, especially for 
sKPP, the \fl\ width is quite close to observed minimal unstable wavelength, thus the first order terms in asymptotic expansion in small $Wk$
are not apriori likely to provide a trustworthy description of instability development\footnote{%
Markstein numbers, which are the ratios of Markstein lengths and corresponding flame widths, are typically between $-1$ and 6 for chemical
flames, according to the cited theoretical estimates, and to experimentally found values. Markstein numbers for our model flames
are between 0.2 and 0.5 for Model B (larger values corresponding to larger expansion), and about 3 times smaller
for Model sKPP with $\e_a=\e_f=10^{-3}$ (and go to zero as sKPP shift parameters $\e_a,\e_f$ go to zero.) See next chapter for more detail.
}. 
According to our simulations 
(summarized below) the critical width of the tube, at which
stable regime of flame propagation changes from planar front to certain nonlinearly stabilized shape, closely corresponds to an estimate
based on Eq.~\p{lcrma}
for model B; this quantitative agreement proves that the small-scale instability
we observed in 2D in the previous section is of the hydrodynamic, LD-type nature.

\subsection{Numerical results, comparison to theoretical estimates}
\begin{figure}[htbp] \begin{centering}
\includegraphics[width=4.0in]{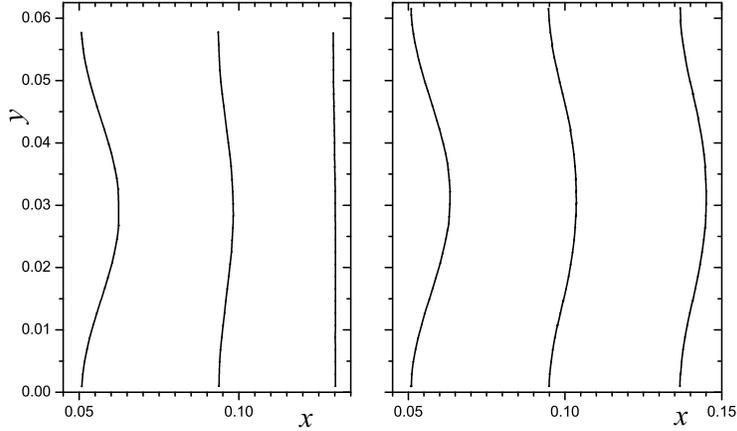}
\caption[Perturbed planar flame evolution at $1/\l=1.44$ for subcritical $L=30\Delta$ and supercritical $L=32\Delta$ domain width]
{Perturbed planar flame evolution at $1/\l=1.44$ for subcritical $L=30\Delta$ and supercritical $L=32\Delta$ domain width. 
$x$ and $y$ are in units of the domain length, which is 512 cells; $y$-scale is the same for both $L$. 
Flames are shown at timesteps 0, 7000 and 14000. The flame at $t=7000$ was translated by $-0.25$ in $x$-direction, the one at $t=14000$
was translated by $-0.5$, to show all three on the same plot.
\label{2dpert}}
\end{centering}\end{figure}

Results for critical tube side $L_\mathrm{cr}$ orthogonal to flame propagation, its smallest value at which initial perturbation develops 
into certain nonplanar shape (instead of relaxing to planar surface, which happens at smaller $L$), are summarized in Tab.~\ref{tab42}. 
We used 512 cells long computation domains for larger expansions, corresponding to fuel densities in a 0.5C+0.5O WD of $10^8\:\gcm$ 
and below, and 2048 cells long domains for smaller expansions; initial perturbation was one sine wave on the domain width
for most of the runs, with amplitude equal to $0.1 L$. 
Initial distribution (in the flame propagation direction) of progress variable $f$ and physical 
quantities corresponded to 1D steady flame profiles. For larger
expansion, $1/\l\ge 1.44$, uncertainty in $L_\mathrm{cr}$ is about 2 cells, the smallest amount by which we could change the tube side;
the smallest supercritical width observed is shown in the table.
At these larger expansions the steady flame shape significantly deviates from planar (see Fig.~\ref{2dpert} for example), and the critical 
tube width is easy to recognize. 

\begin{table}[!hbpt]
{\center \begin{tabular}{@{$\;$}c@{$\;\:$}rrrrrrrrrr@{$\;$}}\hline
$1/\l$                    & 0.167 & 0.249 & 0.398 & 0.734 & 1.44 & 2.33 & 3.18 & 3.95 &5.05 & 8.57 \\
$\lcr(Ma)$                & 60.3  & 43.4  & 30.6  & 30.8  & 30.0 & 19.7 & 19.0 & 18.8 &18.7 & 19.3 \\ 
$L_\mathrm{cr}^*$         & 68    & 48    & 32    & 34    & 32   & 24   & 26   & 20   &  16 & 16   \\ 
$L_\mathrm{cr}^*/5,W_1=16$&       &       &       &       &      & 17.6 & 16.0 &      &14.4 & 13.6 \\
\hline

\end{tabular}
\caption[Minimal tube width at which planar flame is hydrodynamically unstable. Numerical results and analytical estimates in Markstein 
approximation]
{Minimal tube width (in cells) at which planar flame is hydrodynamically unstable, at different expansions $1/\l$. 
$\lcr(Ma)$, the $2^{nd}$ row, is a theoretical prediction 
\p{lcrma} in Markstein approximation (of infinitely thin flame, with linear Markstein law for flame speed dependence on the front 
curvature); the values of Markstein length were taken from Tab.~\ref{tab51}. The third row shows estimates of the critical tube width 
from above from direct numerical simulations (evolution of initially sinusoidal front in 2D tube); these are the smallest values we found 
to yield non-planar (nonlinearly stabilized) steadily propagating flames. The estimate from below is 2 cells smaller for 
$1/\l\ge 1.44$, and 4 cells smaller for the lower expansions shown. Flame speed is $80\:\kms$, flame width $W_1=3.2$ cells. 
The last row shows same upper estimates for $L_\mathrm{cr}^*$ obtained for $W_1=16$ cells wide flames
($V=80\:\kms$), to understand whether somewhat different relation between $L_\mathrm{cr}^*$ and estimate $\lcr(Ma)$ at higher expansion
represents a physical effect (corresponding to increased role of finite flame width), or numerical discretization effect (observed in
Sec.~\ref{nsteady1D} to be more pronounced at larger expansion, for various quantities). This latter $L_\mathrm{cr}^*$ was divided by 5,
so that to offset the effect of Markstein length $Ma$ (and thus corresponding $\lcr(Ma)$) being 5 times larger for this extra wide \fl.
\label{tab42}}}
\end{table}

For smaller expansion, on the other hand, nonlinearly-stabilized non-planar shape at near-critical
tube width only deviates by a few cells from the plane, for tube widths of tens of cells. This deviation scales approximately
as $1/\l$ when this expansion parameter is $\ll 1$, in accord with analytical studies (\cite{Siv77, MS}), this making recognition of
critical tube width less certain. Besides, relaxation times become significantly longer at smaller expansion: linear growth rate in LD
approximation scales as $0.5Dk/\l$; Eq.~(\ref{landarma}) yields growth rate scaling $\propto 1/\l^3$ (assuming $Ma$ going
to some constant value at zero expansion) near critical tube width $L_\mathrm{cr}=\lcr(Ma)$ \p{lcrma}:
$\omega_{Ma}(L_\mathrm{cr}+\delta L)=\frac{D\,\delta L}{16\pi\,Ma^2\,\l^3}$. The newest version of the code allowing stable computation 
for over 300000 timesteps required for densities of $2\times 10^9\:\gcm$ and above became unavailable to us in the middle of this study, 
thus uncertainties for $\lcr$ at smaller expansions are estimated to be of order 4 cells. Only one run was finished at fuel density
$5\times 10^9\:\gcm$ ($1/\l=0.118$, $\lcr(Ma)=82$ cells); the domain width of 72 cells was found subcritical.

It is interesting to observe agreement within about 10\% between numerically found values for the critical tube width and theoretical 
estimate in Markstein approximation at expansions (1.44 and below), corresponding to fuel density of $3\times 10^7\:\gcm$ in SN~Ia 
problem. It is even better at expansions $1/\l<0.734$ if Markstein numbers found in direct numerical simulations (see Tab.~\ref{tab51}) 
are used. At higher expansion agreement is somewhat worse.
\ The fact that at smaller expansions observed $L_\mathrm{cr}$ is somewhat larger that theoretical one may be attributed to numerical 
diffusion tails in $f$ profiles, thus yielding somewhat larger Markstein length than the one found in continuous quasi-steady state
technique. At smaller expansions, however, numerically estimated $L_\mathrm{cr}$ becomes smaller than $\lcr(Ma)$. To check whether
we understand the effect of \fl\ discreteness properly we reran some simulations at small densities with distinct normalization of
the \fl\ governing parameters, $\tilde{K}$ and $R$, to yield \lq\lq flames\rq\rq\ with 5 times larger width, for which the effects of
discreteness must be less pronounced. As Markstein length (and thus $\lcr(Ma)$) scales proportionally to the flame width we divided
observed upper estimate $L_\mathrm{cr}^*$ for such wider flames by 5, for direct comparison with the rest of the numbers in the table.
The effect of changing the flame width is significant (as it was for other quantities related to the \fl\ in Sec.~\ref{nsteady1D}
at large expansions), yet in the predicted direction: critical perturbation length decreases when discretization effects become smaller.
Thus we are tempted to conclude that the discrepancy observed at higher expansions is due to real physical deviation from Markstein
approximation, due to finite flame width (moreover, representing a larger fraction of $L_\mathrm{cr}$ at these larger expansions); 
this deviation seems to monotonically increase with expansion. 


For model sKPP with $\e_a=\e_f=10^{-3}$ and $W_1=4$ cells the critical perturbation wavelength is about 2.5 times smaller at small expansions 
(in Markstein approximation --- as are corresponding Markstein lengths, cf. Tab.~\ref{tab51} for Markstein numbers). 
These are even smaller at larger expansions,
7 times smaller than those for Model B at $1/\l=1.44$. We really observe steadily propagating non-planar sKPP flames (corresponding
to a full wavelength, not reorganizing into a half-wave configuration) in domains 10 cells wide at $\rof=3\times 10^7\:\gcm$, 8 cells wide 
at $\rof\le 5\times 10^6$; $L_\mathrm{cr}\approx 8$ at $\rof=10^7$ --- a seemingly stable non-planar configuration in 8 cells wide domain
starts transforming into a half-wave configuration after running through 120 cells (this flame relaxes to a stable half-wave configuration 
when it runs through 250 more cells). Flame profile, $f$ distribution across the flame, in such stationary flames is clearly distorted; 
the flame is wider where it is convex towards the fuel. At smaller expansions $L_\mathrm{cr}$ is noticeably larger than according to 
Markstein approximation, say it is 30 cells at $\rof=3\times 10^8\:\gcm$, almost the same value as for Model B (with 
standardly used for the latter $W_1=3.2$ though, \ie\ thinner than the sKPP flame).

\subsection{Discussion}
Steady flame shapes we observe in supercritically wide domains visually resemble the ones reported in \cite{RastMat06}, the only study 
we are aware of dealing theoretically/numerically with non-linear stabilization of flames with finite expansion. Results for flame shapes
are compared there to analytical results of \cite{MS} for small expansion, with convincing agreement. Methodology of \cite{RastMat06}
is quite different from ours: the authors try to study non-linear stabilization in Markstein approximation, with infinitely thin flame
(tracked with Level Set technique), Markstein effect introduced by hand in prescribing flame speed as a function of
locally computed curvature. Results for critical wavelength are not reported there with reasonable accuracy, although
for a few expansions it is noted whether existence of stable non-planar shape is in agreement with linear stability analysis, Eq.~\p{lcrma}.
Our observations agree with those in \cite{RastMat06} in that the most stable flame shape has exactly one maximum for 
supercritical tube width; this also agrees with analytical results in small expansion case\footnote{%
These observations do not seem conclusive though. Say, our results might suggest that a configuration with several wavelength corresponding
to a stably propagating flame shape, arranged end-to-end on a tube width, may be stable with respect to sufficiently small perturbations.
Such configurations were obtained by starting with several sinusoidal wavelengths on a tube width as an initial condition; this evolved to
described configuration, rather that to one-maximum one; this configuration in several runs propagated (quasi-)stably through entire 
2048-cell long domain. Moreover, certain small perturbations not respecting the discrete symmetry of multi-wave configuration seemed to
decay, without transforming the multiwave configuration into an absolutely stable one with one maximum on tube width. In other runs, 
however, with superficially similar initial conditions (each wave having supercritical length, in part) perturbations started 
to grow asymmetrically, with the shape transforming to a one-wave or a half-wave configuration. \cite{RastMat06} started with random 
flame shape, with no symmetries, in simulations dedicated to getting one-wave configuration in the process of instability developing
towards its nonlinear stabilization; thus no results are presented to judge about local stability of multi-wave configurations.
This is in contrast to analytical results in small-expansion analysis, where instability of any configuration distinct from a one-maximum 
one is proven mathematically. Those results, however, are obtained for basically a different problem, with all terms beyond first order in 
$1/\l$ dropped (similarly to our estimates for $d(\l)$ based on expansion in $1/\l$ at the end of Sec.~\ref{stepdw}). It is plausible that 
for the full problem there may be islands of stability in perturbation space near multiwave-configurations, vanishing as $1/\l\to 0$,
or perhaps islands existing only when finiteness of the flame width is taken into account. This requires further study.}.
 
We used reflecting boundary conditions on
long boundaries of the computation domain, whereas \cite{RastMat06} use periodic ones. This makes it possible for us to observe 
another stable propagation mode, different from that shown in Fig.~\ref{2dpert}, namely, the one with a half of the steady flame shape 
occupying the whole tube width. Such a mode is expected to be more stable than the one with the whole profile (symmetric with respect 
to the longitudinal symmetry axis of the tube). However the times needed for transition from a (quasi-)stable whole-wavelength shape 
to stable half-wavelength shape, with just numerical noises in our simulations as a trigger, were usually long; only in a few 
runs could we observe such a transition, in tubes with width not sufficient to support a stable non-planar profile with one wavelength, 
but wide enough to accommodate a half of the critical wavelength. Since this happened more readily at larger expansions, for some 
of the corresponding results we report in Tab.~\ref{tab42} we used asymmetric initial conditions, with half of a sine wave on tube width 
(maximum on one side, minimum on the other). 

To sum up, results of this section show that the small-scale instability we observed in 2D simulations is indeed a hydrodynamic
instability of LD-type. It is masked for Model sKPP with numerical artifacts, like global flame propagation anisotropy, and 
demonstrates significantly different characteristic instability lengthscales and growth rates in different directions. For Model B
numerical artifacts are almost absent, and characteristic for LD instability shapes are easily recognized on 2D flames. 

Dominating lengthscales of perturbations of cylindrical flames show somewhat stronger dependence on expansion parameter than observed above
critical wavelengths of perturbations on planar flames (we refer to Model B through the end of this section). The weak dependence of the 
latter is in accord with linear results in Markstein approximation: 
Eq.~\ref{lcrma} suggests rather weak dependence of $\lcr$ at expansions $1/\l\ge 2.33$: dependence 
through factor $1+\l$ (weak at small $\l$) is further offset with (accidentally?) correlated dependence $Ma(\l)$. 
Theoretical (linear) results for LD-type instability for spherical flames (see, \eg , \cite{zelbook}, Ch.~6) in Markstein approximation help
to qualitatively understand characteristic lengthscales of the perturbations, as well as timescales associated with their development.
In contrast with nonlinear stabilization of flames in the tube, where the stable shape (at least in the limit of small
expansion) steadily propagating has 1 maximum, for spherical flames only higher harmonics are unstable; the ones represented by spherical
functions $P_n^m(\cos\theta)e^{im\phi}$ with index $n$ less than certain critical value grow slower than the flame expands, and do not 
lead to the surface distortion 
as time progresses (all harmonics grow according to a power law in 3D and 2D, not exponentially as perturbations on a planar flame). 
This critical index is $n_\mathrm{cr}=12$ for positive Markstein length and $1/\l\approx 3$. For larger and smaller
expansion the critical index is larger. 

For any supercritical harmonic $n$ a perturbation having corresponding distribution over the flame 
surface will grow after the flame has reached certain radius $r(n)$. 
Markstein effect will not inhibit growth of any supercritical harmonic at
late enough times, as the linear size of corresponding perturbation scales proportionally to the flame radius, and will eventually become
large enough for Markstein stabilizing effect to become non-substantial. The harmonic that becomes unstable at the smallest flame
radius will dominate for some time, however we are not aware of any theoretical studies for characteristic perturbations structure at 
late times. According
to flame surfaces in Fig.~\ref{2dBq} for the times studied linear perturbation lengthscales stay approximately constant with time, cells 
expanded with the flame keep subdividing into smaller cells. Characteristic linear perturbation scale is about 50 cells for $1/\l=8.57$, 
and larger for smaller expansions, according to the figure. Ratio of this scale to $\lcr$ for planar flames is about 3 at this
expansion, and somewhat larger at smaller expansions. 

To reiterate, theoretical results we summarized correspond to a spherical flame in 3D in Markstein approximation, thus specific numbers,
like $n_\mathrm{cr}$, first harmonic to start growing, etc. are not expected to describe well results of our simulations of cylindrical
flames in the previous section. Qualitative picture, however, presented in the previous paragraph seems to apply to 2D as well as 3D setup 
(following the derivations in 3D), and agrees with the features of small-scale instability development we observed in numerical simulations.
One particularly interesting number to check is a critical radius of the flame when the first unstable mode appears. According to linear 
analysis in 3D in \cite{zelbook} this radius is $r_\mathrm{cr}=Ma\,\tau^*$, where dimensionless $\tau^*$ depends on expansion, and is 
about 60--70 for $1/\l\in[2.33;8.57]$, about 80 for $1/\l=1.44$, 120 for $1/\l=0.734$, 300 for $1/\l=0.398$, $>600$ for $1/\l=0.167$.
For $1/\l=1.44$ this yields $r_\mathrm{cr}\approx 110$ cells, or 230 km; this seems to reasonably well agree with numerical results in 3D we 
present in the following section. 

\section{3D results for flame shape evolution}
LD-type wrinkling of flame surface is expected to grow faster in 3D and may become an issue for Model B at larger expansions; 
it is likely to remain a problem for Model sKPP. 

Our approach is the same as described for 2D simulations. 
Flame propagation is initiated either from the
center of the cubic computation box with outflow boundary conditions on all faces, or from the corner, with reflecting boundary conditions 
on the 3 faces passing through that corner, and outflow boundary conditions on the other three faces; we refer to the latter setup as 
a simulation in octant. Initial distribution of physical variables and reaction variable $f$ is spherically symmetric around the flame 
center; a sphere (a sector for octant simulations) around the center contains hot ash of the reaction at rest; away from the initially 
burned region physical variables correspond to fuel, moving radially outwards, according to quasisteady solution for a spherical flame 
in 3D; in the intermediate region, the \fl\ itself, physical quantities correspond to a mixture of fuel and ash, $f$ has some 
intermediate value monotonically changing from 1 in the burned sphere to 0 in the fuel ahead of the flame,
according to 1D steady-state profile. The distance from the initial \fl\ center to the point where $f=0.9$ is called initial flame 
radius $R_0$. 

The largest computational box we use for 3D simulations is $256^3$; LD-type instabilities do not have enough time to develop to the extent 
close to that in 2D simulations. To characterize flame geometry we use several numbers akin to those we used in the previous section. 
First, $\Delta r/R$ (defined the same way as for 2D, by dividing the difference $\Delta r$ between extremal radii on $f=0.5$ surface by 
the average distance $R$ from the flame center over all points on the surface) characterizes overall sphericity. To get an idea about 
large scale deviations from the spherical shape, and in which directions these are maximal, we consider cross-sections of $f=0.5$ surface 
by equatorial plane ($z=0$), by planes containing $z$ axis and forming angle $\pi/4$ or $\pi/8$ with $xz$ plane, and finally by 2 planes 
parallel to $xy$ plane, $z=R/2$ and $z=R\sqrt{3}/2$. We mark results for these planes with index 0, 4, 8, 2 and 3 respectively. For each 
plane we consider a slice, subset of points on the surface $f=0.5$ that are at most $R/10$ away from the crossing plane. For points on each 
slice we compute fit parameters $r_0$ and $r_1$ similarly to what we did for 2D flames. For instance, for plane 2, parallel to $xy$ plane, 
we fit cylindrical coordinates $(\rho_i=\sqrt{x_i^2+y_i^2}, \phi_i)$
of points in the corresponding nearby slice with $\tilde{\rho}_i=r_{02}+r_{12}\cos 4\phi_i$. 

The ratios $r_1/r_0$ for the slices described are listed
in Tab.~\ref{tab43} for models B and sKPP for several densities, for several moments of time for each density; these ratios give an idea 
about flame surface geometry. Moments of time shown were chosen as in the previous section. We remind that, according to 2D comparisons, 
open boundaries start influencing results for flame surface, for $256^3$ runs, when its radius reaches approximately 120 km (about 60 cells) 
for central ignition, and approximately 300 km for corner ignition (octant simulations).

\begin{table}[!hbpt]
{\center \begin{tabular}{lrr|rrrrrr}\hline 
Model    & $1/\l\;$ & $R\;$ & $\frac{\Delta r}R\:$    &$\frac{r_{10}}{r_{00}}\;$&$\frac{r_{14}}{r_{04}}\;$&
$\frac{r_{18}}{r_{08}}\;$   &$\frac{r_{12}}{r_{02}}\;$&$\frac{r_{13}}{r_{03}}\;$\\ \hline

Bo,$\,N=128$& 0.167 & 120   &  8.81                   & -3.24                   &  -2.12                  &
 -2.56                      &  -4.14                  &  3.94 \\

Bo,$\,N=128$& 0.167 & 195   &  6.37                   & -2.10                   &  -1.41                  &
 -1.65                      &  -2.84                  &  1.92 \\

Ko,$\,N=128$& 0.167 & 195   &  7.87                   &  3.71                   &   1.73                  &
  2.56                      &  2.34                   &  -8.99 \\

sKPP     & 0.398    & 120   &  6.22                   & -1.24                   &  -1.73                  &
 -1.66                      &  -1.97                  &  4.76 \\

sKPP     & 0.398    & 210   &  7.53                   &  3.23                   &  0.77                   &
 2.08                       &  -0.78                  &  4.33 \\ \hline

sKPP     & 0.734    & 60    &  27.4                   &  4.34                   &  9.88                   &
 5.59                       &  10.9                   &  -7.62 \\

sKPP     & 0.734    & 120   &  55.3                   &  12.5                   &  18.1                   &
 10.4                       &  20.7                   &  4.07 \\

sKPP     & 0.734    & 195   &  75.6                   &  15.9                   &  24.8                   &
 15.1                       &  27.6                   &  5.02 \\

B        & 0.734    &  60   &  17.0                   &  0.27                   &  7.01                   &
 1.39                       &  5.69                   &  1.11 \\

B        & 0.734    & 120   &  32.3                   &  2.82                   & 13.4                    &
 4.48                       &  16.5                   &  7.32 \\

B        & 0.734    & 195   &  43.0                   &  6.10                   &  15.9                   &
 6.24                       &  17.5                   &  15.9 \\

Bo,$\,N=128$& 0.734 & 120   &  31.7                   &  2.94                   & 13.3                    &
 3.90                       &  16.9                   &  7.51 \\ 

Bo,$\,N=128$& 0.734 & 195   &  42.0                   &  6.11                   & 15.8                    &
 5.59                       &  16.9                   &  8.76 \\ \hline

sKPP     & 1.44     & 60    &  34.9                   &  6.17                   &  9.42                   &
 7.41                       &  9.48                   &  5.16 \\

sKPP     & 1.44    & 120    &  118                    &  23.8                   &  25.9                   &
 23.9                       &  46.0                   &  14.9 \\

sKPP     & 1.44    & 195    &  75.6                   &  15.9                   &  24.8                    &
 15.1                       &  27.6                   &  5.02 \\

Bo       & 1.44    &  60    &   14.1                  &  -5.14                 &  -1.78                   &
 -3.49                      &   -                     &  -0.8 \\

B        & 1.44    & 120    &   31.9                  &  3.59                   &  7.57                    &
 4.23                       &     -                   &  0.99 \\

B        & 1.44    & 195    &   46.7                   &  10.7                   &  10.7                   &
 9.63                       &      -                   &  36.9 \\

B,$\,R_0=90$&1.44  & 195    &   13.8                   &  3.26                   &  2.26                   &
 2.08                       &   1.73                   &  -0.36 \\

B        & 1.44    &  60    &   13.1                  &  -4.60                 &  -1.29                   &
 -3.07                      &   1.19                  &  -11.0 \\

Bo       & 1.44    & 120    &  29.2                   &  3.73                   &  7.02                    &
 4.18                       &  11.2                   &  7.96 \\

Bo       & 1.44    & 195    &  30.9                   &  3.42                   &  7.22                   &
 3.93                       &  12.4                   &   7.30\\

Bo       & 1.44    & 300    &  31.9                   &  3.25                   &  7.35                   &
 3.11                       &  14.3                   &  2.54\\ \hline

\end{tabular}
\caption[Asphericity parameters at different times (indicated by flame radius $R$), for models B and sKPP in 3D]
{Asphericity parameters, multiplied by 1000, at different times (indicated by radius $R$ the flame has reached), 
for models B and sKPP 
(abbreviated to K in some places) in 3D. Flame velocity is $D=80\:\mathrm{km\:s^{-1}}$, width $W_1$=3.2 cells 
for Model B, 4 cells for sKPP. Default initial flame radius is $R_0=30\:\mathrm{km}$, box size $256^3$ (side
$528\:\mathrm{km}$), central ignition by default. If other than default values were used, they are indicated 
next to the model name; octant simulations are indicated by letter \lq\lq o\rq\rq . 
\label{tab43}}}
\end{table}

As in 2D simulations, when expansion is small, $1/\l<0.4$, both models behave well, asphericity parameters stay within 1\%
even when flame surface is close to box boundaries. According to theoretical estimate in Markstein approximation, quoted in the 
previous section, first unstable modes start growing at flame radius exceeding 200 cells at such expansions (for Model B),
and growth rate is small at small expansion. 
Already at $1/\l=0.734$, corresponding to fuel density $10^8\:\gcm$
in SNIa problem (with half carbon, half oxygen fuel) asphericity parameters grow almost linearly with time for both models, 
although even extremal $\Delta r/R$ remain within 5\% for model B (as they do for this model for $1/\l=1.44$). 
One can observe that $r_{1i}$ is positive
and grows for all times shown for model B, in contrast to the behavior for 2D simulations. 
It is noteworthy that for Model B results for octant 
and full-cube simulations are quite close to each other, with somewhat worse agreement for cross-section 3 (explicable, as flame surface
forms sharp angle with crossing plane in this section, thus small 3D perturbations would look larger in the section). This implies that
reflecting boundaries do not alter significantly LD instability of adjacent flame surface, thus octant simulations may be reliably used 
for modeling SNe~Ia with central ignition (as far as \fl\ propagation anisotropy and LD instability are concerned).  

\chapter{Markstein effect}\label{markstein}
In numerical simulations of flame propagation in 2D and 3D we observed that the instantaneous flame speed 
was somewhat different from the one measured for 1D flame in a tube with the same diffusivity and burning rate in 
governing equation \p{I1}. It was smaller at the beginning of the simulation (at least for small expansion); Markstein effect is
a possible physical explanation of this phenomenon. Flame front is stretched by geometrical effects when it is not planar; density 
and velocity distribution across such a curved flame differ from those in planar steady flame, as a result so does instantaneous 
normal propagation speed. On general grounds one might expect corrections to flame propagation speed for small curvatures as expansion
\begin{equation}\label{mark}
D_{r_0}=D_{r_0=\infty}\left(1-\frac{Ma}{r_0}+\frac{Ma_2}{r_0^2}+\ldots \right)
\end{equation}
for the speed $D_{r_0}$ of a flame with radius of curvature ${r_0}$ in inverse powers of (large) ${r_0}$. In 3D this expansion could be more 
involved, with two principal radii of curvature of the flame surface $r_{01}$, $r_{02}$ appearing on the right hand side. Such systematic
deviation of the speed of the curved flame from planar one was observed in experiments (see \cite{markstein}); the leading 
term, $Ma/r_0$ is enough to account for curvature corrections in those. In 3D setting $1/r_0=1/r_{01}+1/r_{02}$ must be used
in the term linear in curvature (the only linear combination invariant under rotations). 

Negative sign at $Ma/R$ in \p{mark} was chosen for convenience: with such a convention
$Ma>0$ for our model flames; some authors use notation with all pluses in \p{mark}. Notation $Ma_2$ (as well as $Ma_3$
and so on for coefficients of next terms in expansion \p{mark}) is ours, 
such  corrections of higher order in curvature are not studied in the literature.

In this chapter we study this curvature effect, first in quasi-steady state approach we describe, and then numerically for the flame models 
presented in the previous chapter. We compare the results obtained using these 2 methods, and conclude that for model B physical Markstein
effect dominates numerical effects, as well as corrections to effective flame propagation speed due to hydrodynamic instability development 
at expansions of $1/\l=1.44$ and below.

\section{Quasi-steady state technique}\label{markstein:intro}
Here we present a general method for obtaining Markstein lengths of diffusion-reaction flames. No assumptions of small reaction zone 
thickness are made (unlike existing studies; for the model flames we consider, and are interested to apply the technique for, \lq\lq
preheating\rq\rq\ and \lq\lq reaction\rq\rq\ zones have the same spatial scale). For a spherical flame corresponding quasi-steady velocity 
(slowly changing with time as the flame radius changes) is found as an eigenvalue of a set of differential equations for quasi-steady
distribution of matter velocity and quantities governing reaction rates (species concentrations and the temperature for physical
combustion systems, reaction progress variable $f$ for the model flames we considered before). Results for model flames are presented
in the next sections.

For clarity we present the method for a system with reaction rate dependent on concentration of one species $f$ and on temperature
$T$ (a typical approximation in studying premixed flames in chemical and astrophysical simulations). For convenience we suppose reaction
rate $\Phi=R\Phi_0$ represented as a function of enthalpy density $H$ instead of the temperature. Neglecting viscosity the system
is described by Euler equation for matter velocity $\mathbf{u}$, 
\begin{equation}\label{euler}
 \frac{\partial \mathbf{u}}{\partial t}+(\mathbf{u}\bnabla)\mathbf{u}=-\frac{\bnabla p}\rho + \mathbf{g},
\end{equation}
mass continuity equation for evolution of the density $\rho$ distribution
\begin{equation}\label{rhocont}
 \frac{\partial \rho}{\partial t}+\bnabla(\rho\mathbf{u})=0
\end{equation}
and the following equations describing species diffusion and thermal conduction:
{\setlength\arraycolsep{2pt}
\begin{eqnarray}\label{I1f}
    \frac{\partial f}{\partial t}+\mathbf{u}\bnabla f&=&
         \bnabla(K\bnabla f)+R\Phi_0(f,H) \\ \label{I1H}
    \frac{\partial H}{\partial t}+\mathbf{u}\bnabla H&=&
         \bnabla(\kappa\bnabla H)+qR\Phi_0(f,H). 
\end{eqnarray}}%
Here $q$ represents specific heat release of the reaction. $K=\tilde{K}K_0(f,H)$ is diffusivity of the species described by variable $f$, 
$\kappa=\tilde{\kappa}\kappa_0(f,H)$ is heat diffusivity; form-factors $K_0$, $\kappa_0$ and $\Phi_0$ are dimensionless.
We assume $f$ normalized as for the model flames of the previous chapters, 
namely $f=0$ corresponding to unburned fuel, and $f=1$ to reaction product (say, a state with deficient reactant completely depleted).
No external volume force $\rho\mathbf{g}$ is supposed to be present below. 
We consider isobaric burning regime (quasi-steady deflagration with matter velocities much smaller than the sound speed), as we did in 
Chap.~\ref{steady}.

For model flames with reaction rate dependent on reaction variable $f$ only
the system is simplified, in the same way as it is simplified in the case of unit Lewis number $\Le$ (see below); essentially, 
Eq.~\p{I1H} decouples from the rest of the system.
With several species taken into account in reaction rates the needed modifications are also straightforward: an extra equation
of type \p{I1f} is added to the system for each additional species; 
to the final eigenvalue problem (see below, (\ref{rhocontr}--\ref{I1Hr}) 2 ordinary first-order equations are added for each
such species, no new complications are introduced.

The system of combustion equations has a solution describing a spherical flame propagating outwards. It approximately describes
deflagration in a system consisting of pure fuel at rest at $t=0$ and ignited at one point, after certain quasi-steady distribution
of matter velocity and thermodynamic quantities is established, on timescales comparable to sound crossing time of the system. 
Such centrally symmetric distributions ($f(r)$, etc.) propagate with the flame, changing slowly due to geometry (flame radius) 
changing --- in contrast to a planar flame there is no exact translational symmetry (in $r$) for such a system, thus no steady 
solutions in the strict sense. We refer to system behavior like this as 
quasi-steady spherical flame propagation --- when all quantities depend only on the distance from the flame center and time, and their 
changing with time in the frame coexpanding with the flame, that is expanding with so defined instantaneous flame velocity $D(t)$, 
is much slower than in the rest-frame. For any quantity ($f$, $\rho$, $H$, radial matter velocity $u_r$, etc.) in coexpanding coordinates
$(r,t)\mapsto({\eta},t)$, ${\eta}=r-\int D(t)dt$ partial time derivatives are small at fixed location ${\eta}$ with 
respect to expanding flame surface. For corresponding time derivatives in laboratory coordinates $(r,t)$ 
\begin{equation}\label{qsteady}
 \left. \frac{\partial f}{\partial t}\right|_r=\left. \frac{\partial f}{\partial t}\right|_{{\eta}} 
-D(t)\left. \frac{\partial f}{\partial{\eta}}\right|_t\approx -D(t)\left. \frac{\partial f}{\partial{\eta}}\right|_t,
\end{equation}
that is the changes due to profiles slowly reorganizing with changing geometry are negligible in comparison with changes due to flame 
propagation. We rewrite the system (\ref{euler}--\ref{I1H}) below in comoving quasisteady coordinate ${\eta}$ thus getting rid
of explicit time dependence (it will enter only implicitly through flame radius $r_0$ dependence on time) based on approximate equality
like \p{qsteady} holding for $\rho$, $H$, $u_r$. Such a substitution for time derivatives yields the eigenproblem, which determines 
the quasi-steady flame propagation speed $D$ we seek for.

In dimensionless variables 
{\setlength\arraycolsep{2pt}
\begin{eqnarray*}
 \tilde{\eta}&=&\eta\sqrt{R/\tilde{K}} \\
 \tilde{r}&=&r\sqrt{R/\tilde{K}} \\
 \tilde{u}&=&u_r/\sqrt{R\tilde{K}} \\
         d&=&D/\sqrt{R\tilde{K}} \\
 \tilde{H}&=&H/q \\
   \lambda&=&\tilde{\kappa}/\tilde{K} 
\end{eqnarray*}}%
after quasi-steady substitution of the form \p{qsteady} equations (\ref{rhocont}--\ref{I1H}) are rewritten as
{\setlength\arraycolsep{2pt}
\begin{eqnarray}\label{rhocontr}
    (\tilde{u}-d)\frac{d\rho}{d\tilde{\eta}}&=&-\rho\tilde{r}^{-\alpha}\frac{d}{d\tilde{\eta}}(\tilde{u}\tilde{r}^{\alpha}) \\ \label{I1fr}
    (\tilde{u}-d) \frac{df}{d\tilde{\eta}}&=&\frac{d}{d\tilde{\eta}} \frac{K_0df}{d\tilde{\eta}}+\frac{\alpha K_0}{\tilde{r}}
          \frac{df}{d\tilde{\eta}}+ \Phi_0(f) \\ \label{I1Hr}
    (\tilde{u}-d) \frac{d\tilde{H}}{d\tilde{\eta}}&=&\lambda\left[ \frac{d}{d\tilde{\eta}} \frac{\kappa_0d\tilde{H}}{d\tilde{\eta}}+
          \frac{\alpha \kappa_0}{\tilde{r}} \frac{d\tilde{H}}{d\tilde{\eta}}\right] + \Phi_0(f).
\end{eqnarray}}%
In the above $\alpha+1=$ dimension of the problem: $\alpha=1$ for a cylindrical flame, $\alpha=2$ for a spherical flame ($S^2$) in 3D.
Using corresponding equation of state and assumed isobaricity one expresses $\rho$ as a function of $\tilde{H}$ in Eq.~\p{rhocontr},
thus making the system closed. We used $\rho=\mathrm{const}/H$ as we did in the previous chapters.

Boundary conditions for this problem we use are 
{\setlength\arraycolsep{2pt}
\begin{eqnarray}\label{bndrl}
   \tilde{r}=\tilde{r}_0\equiv r_0\sqrt{R/\tilde{K}}:& & f=1,\; df/ d\tilde{\eta}=0,\; d\tilde{H}/ d\tilde{\eta}=0,\; \tilde{u}=0; 
     \\ \label{bndrr}
   \tilde{r} \to +\infty:& & f\to 0,\; \tilde{H}\to H_0/q \equiv H(P_\mathrm{fuel},T_\mathrm{fuel})/q.
\end{eqnarray}}%
It is through these boundary conditions that the flame radius $r_0$ enters the eigenproblem, thus yielding its eigenvalue $d$ 
dependent on $r_0$. In prescribing boundary condition at $r=r_0$ we assumed that the flame does not have infinite tail into ash.
Formulation of quasi-steady problem for systems with such an infinite tail encounters several problems, we will not touch them here.

As in 1D case the system simplifies when $\kappa=K$ (\ie\ Lewis number equals 1; we see, below, that 
these equal transfer coefficients may still depend arbitrarily on $f$ and $H$ for the simplification to hold; $\Le\approx 1$ in
terrestrial flames, in nearly ideal gases). In this case we just reproduce a known result (\cite{z-fk}) that distributions of $f$ 
and $H$ is similar --- for any flame geometry (not necessarily spherical): Eq.~\p{I1H} in this case coincides with \p{I1f} after
rescaling $H$ by $q$, and boundary conditions (for $f$ and $H$ going to constants behind and in front of the flame) then yield 
algebraic relation $H=H_0+qf$; one is left with a reduced system of equations (\ref{rhocont}--\ref{I1f}). The same simplification
occurs in the case of model flames studied for Flame Capturing applications, when $H=H_0+qf$ is postulated by the technique.

We present results for $D(r_0)$ in the next section, for model flames studied in the Chap.~\ref{steady}.

\section{Quasi-steady results for models used in Flame Capturing}
\label{markstein:tech}
Below we stick to a reduced form of the eigenvalue problem, with $H=H_0+qf$. $\rho H=\mathrm{const}$ at constant pressure is assumed. 
Although the system (\ref{rhocontr}--\ref{I1fr}) is not tranlationally invariant in $\tilde{r}$ (unlike the case for a planar front) due to
$\tilde{r}$ explicitly appearing, $\tilde{r}$ does not change significantly within the flame width at large radii\footnote{%
We have tried estimating $Ma$ in the approximation of large $\tilde{r}_0$, treating terms $\sim 1/\xi$ with positive powers 
in Eq.~\ref{Markn} as perturbations, 
using 1D flame profiles $f(\tilde{r})$ in estimating such terms. We only present exact treatment of the problem here for brevity,
as approximate consideration is not significantly simpler. Very rough fully analytical estimates may be obtained assuming, say, linear
flame profile in perturbation terms; this leads to about 20\% error in $Ma$ for finite-width flames.}.
For numerical solution 
we found it convenient to integrate over $f$ instead of $\tilde{r}$, similarly to what we did in Sec.~\ref{steadynum} in 1D. 
This is especially successful for flame models with finite total profile widths. For such models we integrate the following system:
{\setlength\arraycolsep{2pt}
\begin{eqnarray}\label{Markn}
   \frac{dn}{df}&=& \alpha K_0(f)\xi^{-1/\alpha}-\frac{\Phi_0(f)K_0(f)}{n}+d\left(1+\frac{f}{\l}\right)\left(1+\frac{I}{\xi}\right)\\
   \label{Markxi} \frac{d\xi}{df}&=& -\alpha \xi^{1-1/\alpha}\frac{K_0(f)}{n} \\ \label{MarkI}
   \frac{dI}{df}&=& \alpha \xi^{1-1/\alpha}\frac{K_0(f)}{n}\frac{f}{f+\l}.
\end{eqnarray}}%
Here \[ \xi=\tilde{r}^\alpha,\; I=\frac{\tilde{r}^\alpha(\tilde{u}-\tilde{u}_0)}{d(H_0+qf)},\; n=pK_0(f);\quad p=-\frac{df}{d\tilde{\eta}};\]
$\tilde{u}_0$ is a constant such that $I\to 0$ at $f\to 0$ ($r\to+\infty$).
Eq.~(\ref{MarkI}) is \p{rhocontr} rewritten in terms of $I$, \p{Markxi} is a rewritten definition of $n=-K_0(f)\,df/d\tilde{\eta}$,
and \p{Markn} is Eq.~\p{I1fr} in these new variables. 

Boundary conditions at $f=1$ (left boundary of the flame, $\tilde{r}=\tilde{r}_0$) are 
\begin{equation}\label{Markl} 
  f=1:\quad \xi=\tilde{r}_0^\alpha,\; n=0. 
\end{equation} 
At $f=0$ (for finite flames this is the right boundary of the flame, located at finite radius $\tilde{r}_0+w_c$; for flames with an infinite 
tail into fuel $\tilde{r}=\infty$ at $f=0$) the boundary conditions are 
\begin{equation}\label{Markr} 
f=0:\quad I=0,\; n=0.
\end{equation}
This system was integrated from $f=1$ to $f=0$. Values of $I_0=I(f=1)$ and $d$ were estimated based on 1D results (initial seeding), 
and then solved for exactly using Newton-Raphson algorithm to satisfy the 2 boundary conditions at $f=0$. 
 
System (\ref{Markn}--\ref{MarkI}) is singular at $f=1$ as $n(1)=0$. Because of this integration was actually started at 
$1-f_\epsilon$ for certain small positive $f_\epsilon$ (0.01 of bulk integration step); asymptotic expansions for $\xi,\: n,\: I$ 
were used as initial values at $f=1-f_\epsilon$.
For model B (Eq.~\ref{syntf}) these are 
{\setlength\arraycolsep{2pt}
\begin{eqnarray}\label{asympb}
n\equiv pK_0(f)&=&f_\e^{s_a+r_a}/\phi_0\\ \nonumber
\xi &=& \left( \tilde{r}_0+\frac{\phi_0}{1-r_a}f_\e^{1-r_a}\right)^\alpha \\ \nonumber
I   &=& I_0-\alpha f_\e^{1-r_a} \frac{\tilde{r}_0^{\alpha-1}\phi_0}{(1-r_a)(1+\l)}; \\ \nonumber
\phi_0&\equiv& \frac{d}{1-c(\l)}\left(1+\frac{1}{\l}\right)\left(1+\frac{I_0}{\tilde{r}_0^\alpha}\right).
\end{eqnarray}}

For models A (Eq.~\ref{PhistepK}), original model of \cite{X95} (same expression for $\Phi_0(f)$, $r=0$ in the expression for
diffusivity $K$) and sKPP \p{skpp} initial asymptotic values used are
{\setlength\arraycolsep{2pt}
\begin{eqnarray}\label{asymp}
n&=&\nu_0f_\e^{1/2}\Bigl(1+\mu_0f_\e^{1/2}\Bigr)\\ \nonumber
\xi &=& \left[ \tilde{r}_0+\frac{2}{\nu_0}f_\e^{1/2}\Bigl( 1-\frac{\mu_0}{2}f_\e^{1/2}\Bigr)\right]^\alpha \\ \nonumber
I   &=& I_0- 2f_\e^{1/2}\frac{\alpha\tilde{r}_0^{\alpha-1}}{\nu_0(1+\l)}\left[ 1+\Bigl(\frac{\alpha-1}{\nu_0\tilde{r}_0}-\frac{\mu_0}{2}
 \Bigr) f_\e^{1/2}\right]; \\ \nonumber
\nu_0&=& \sqrt{2\Phi_0(1)}=\left\lbrace
\begin{array}{ll}
  \sqrt{2} &\; \mathrm{for\ A,\ Khokhlov\ (1995)} \\
  \sqrt{2\e_a(1-\e_f)} &\; \mathrm{for\ sKPP}
\end{array} \right.\\ \nonumber
\mu_0&=& -\frac{1}{\nu_0}\left[ d\Bigl(1+\frac{1}{\l}\Bigr)\Bigl(1+\frac{I_0}{\tilde{r}_0^\alpha}\Bigr)+\frac{\alpha}{\tilde{r}_0}\right].
\end{eqnarray}}

\begin{figure}[htbp] \begin{centering}
\includegraphics[width=4.0in]{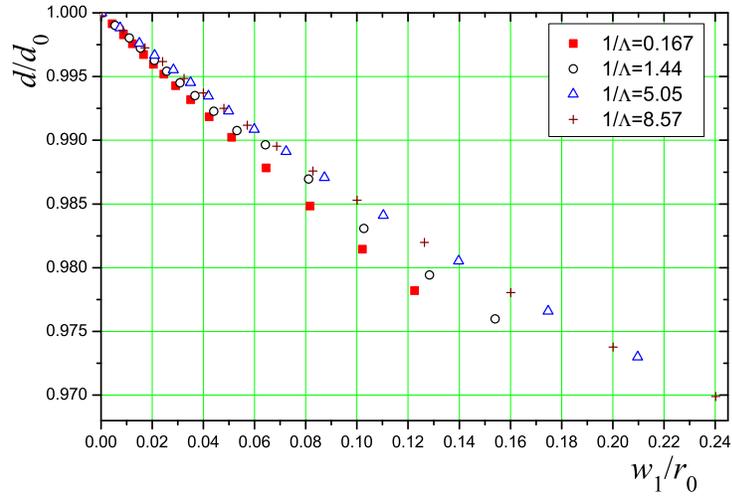}
\caption[Dependence of flame speed $d$ on curvature for Model A, quasi-steady calculation]
{Dependence of flame speed $d$ on curvature, presented in units of inverse flame width $w_1$ for
Model A. $d_0=d(r_0=\infty)$.
\label{adr}}
\end{centering}\end{figure}

Resulting curves for $d(r_0)$ are shown in Fig. 5.1 for Model A, and in Fig. 5.1 for Model B, for 4 expansions each. 
$d(r_0)/d(r_0=\infty)$ is plotted against $w_1/r_0$, the slope of the curves at $w_1/r_0=0$ thus gives Markstein number $M$. 
Deviations from linear Markstein law can be seen at larger curvatures. Values of Markstein numbers are shown in Tab.~\ref{tab51} 
in the next section, compared to these numbers estimated in direct numerical simulations.

\begin{figure}[htbp] \begin{centering}
\includegraphics[width=4.0in]{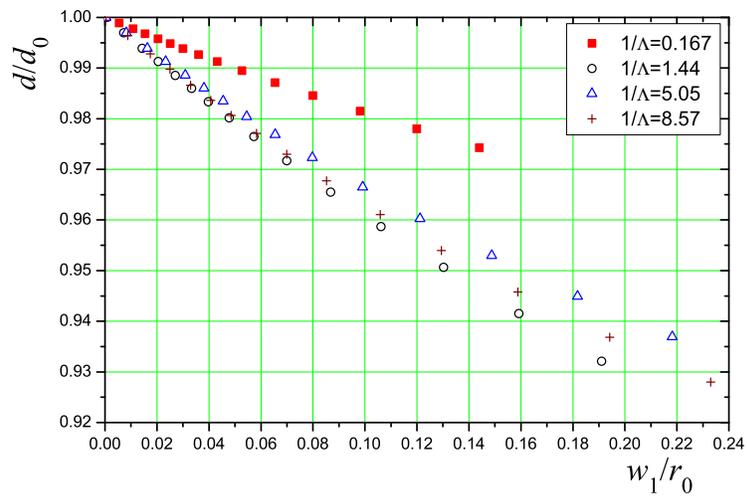}
\caption[Dependence of flame speed $d$ on curvature for Model B]
{Dependence $d(w_1/r_0)$ for Model B.
\label{bdr}}
\end{centering}\end{figure}
 
\section{Markstein numbers from direct numerical 2D simulations}
In this section we estimate how observed numerical flame speed depends on flame curvature, whether one can consistently get Markstein
length parameter $Ma$ that would describe this dependence via \p{mark} for a range of $r_0$\rq s as the flame grows in radius. 
If this is the case,
one may further try to correct for this curvature effect on flame speed in multidimensional simulations
(by adjusting diffusivity and burning rate parameters based on local flame curvature, 
so that to correct for the discrepancy in flame speed between plane and curved flames). Comparison between analytical
and numerical results clarifies whether real physical Markstein effect dominates unrelated effects
(say, growing asphericity of flame surface studied in the previous section leads to apparent growth of $D$ estimated 
assuming spherical flame shape). 

The dependence on $r_0$ we observe
deviates from linear form (\p{mark} with $Ma_2/R^2$ and further terms in expansion omitted). Contrary to chemical flames,
with widths much smaller than radii of curvature studied, our artificial \fl\ widths are much wider: $R_0=30\:\mathrm{km}$ we 
typically use as initial flame radius is just about 15 grid cells, and \fl\ width $W_1$ is about 4 cells. This difference makes such 
deviations expectable, after all Markstein \lq\lq law\rq\rq\ is just an expansion over small parameter, which is a ratio of flame width 
and flame radius of curvature (it is this ratio that determines how strong a stretch on the flame scale is). To make 
comparisons more consistent we use fits of observed flame speed with \p{mark} with 3 terms left in parentheses (that is, \p{mark}
with \lq\lq $\ldots$\rq\rq\ omitted), with $Ma,\,Ma_2$ considered free parameters. We use the same fits to get $Ma,\,Ma_2$ for 
$D_{r_0}$ found analytically. Markstein numbers \[ M=Ma/W_1\] are reported below (as we saw in Sec.~5.1 analytically, Markstein length
$Ma$ scales proportionally to flame width if one changes the latter by varying scale parameters $\tilde{K},R$ of the model while 
keeping diffusivity and burning rate form-factors $K_0(f)$, $\Phi_0(f)$ invariant), see Tab.~\ref{tab51}.

For each model and expansion parameter we present Markstein number $M_\mathrm{st}$ found analytically, together with that
found numerically, for several timesteps and initial setups used. 
Analytical results shown were obtained by fitting $D_{r_0}$ found through quasi-steady technique presented in 
Sec.~\ref{markstein:intro}--\ref{markstein:tech}
with 3-term expansion of the form \p{mark} for a set of approximately equidistant flame curvatures $1/r_0$, ranging from $1/400$
to $1/50$ 
(dimensionless, $r_0$ is scaled with the same factor $\sqrt{\tilde{K}/R}$ as dimensionless coordinate $\chi$ and width $w_1$ in Sec.~5.1).
As a reminder for the sense of scale, flame width $w_1$ is $\sim 2$ for Model B, and $\sim 8$ for sKPP for small expansion, see
Fig.~\ref{w1n_akb} and its caption. This way, when fits are used with at least quadratic in curvature terms left in expansion, exact 
value found
for Markstein number $M$ is not sensitive to precise set of radii used for the fit. Say, if more curvatures are added to cover 
a range of 0 to $1/15$, the $M$ found for Model B at $1/\l=0.167$ changes from 0.2145 to 0.2117 (1.3\%). Besides, when more terms 
are used in fit (higher powers in curvature left in expansion \p{mark}) results for $M$ do not change drastically as well. For
the example above leaving terms up to cubic leads to $M=0.2148$ (for curvature range 0 to $1/15$; even smaller difference for 
reduced range $1/r_0\in[1/400;1/50]$, as for $M$\rq s presented in Tab.~\ref{tab51}); fit with terms up to quartic in curvature yields 
$M=0.2150$.

Numerical Markstein numbers were estimated from 2D runs with central and corner ignition (same setup as in the previous sections). 
For this, flame speeds and radii were
recorded for each timestep; then $D_{r_0}$ was fitted with expansion of the form \p{mark} with three terms (up to $r_0^{-2}$) for 
timesteps from 2000 to the step when flame radius reached the value of $R$ recorded in the column header in Tab.~\ref{tab51}. Flame speed 
was estimated based on integral burning rate, $D=-\frac{1}{\rho_\mathrm{fuel}A}\frac{dm_\mathrm{fuel}}{dt}$; $A$ here denotes flame 
surface, estimated based on burned volume assuming spherical flame shape.

\begin{table}[!hbpt]
{\center \begin{tabular}{l@{\ }r|rrrr|r}\hline 
Model    & $1/\l\;$   &$\!M,\,R_2\:$ &$\!M,\,R_3\:$ &$\!M,\,R_4\:$ &$\!M,\,R_5\:$ & $\!M_\mathrm{st}\:$ \\ \hline
B        & 0.167      &  0.233       & 0.232        &    0.232     &  0.232       &   0.2145            \\
sKPPq$,\,N=512$& 0.167&  0.0360      & 0.1506       &   0.1106     &  0.1107      &   0.0733            \\ 
Bq       & 0.398      &  0.249       & 0.241        &   0.241      &  0.242       &   0.2163            \\
sKPPq    & 0.398      &  0.169       & 0.159        &   0.159      &  0.158       &   0.0680            \\
Bq$,\,N=512$   & 0.734&  0.325       & 0.309        &   0.313      &  0.326       &   0.3246            \\
sKPP     & 0.734      &  0.233       & 0.280        &   0.409      &  0.587       &   0.0588            \\
sKPPq$,\,N=512$& 0.734&  0.251       & 0.294        &   0.416      &  0.584       &   0.0588            \\
B        & 1.44       &  0.384       & 0.309        &   0.305      &  0.325       &   0.4409            \\
B,$\,R_0=15$   & 1.44 &  0.202       & 0.255        &   0.258      &  0.286       &   0.4409            \\
sKPP     & 1.44       &  0.705       & 2.113        &   3.633      &  6.843       &   0.0517            \\
B,$\,R_0=90,\,N=2^{11}$& 2.33& --    & 0.629        &   1.094      &  2.28        &   0.3432            \\
B,$\,R_0=90$& 2.33    & -1.69        & 0.760        &   1.134      &  2.48        &   0.3432            \\
B,$\,R_0=90,\,W_1=4.5$& 2.33 & -1.13 & 0.512        &   0.364      &  0.430       &   0.3432            \\
B,$\,R_0=90,\,W_1=5$  & 2.33 & -1.64 & 0.517        &   0.325      &  0.339       &   0.3432            \\

\hline

\end{tabular}
\caption[Markstein numbers computed numerically and through quasi-steady technique for models B and sKPP at different expansions]
{Markstein numbers computed numerically for models B and sKPP at timesteps, computed
when flame 
radius reached $R_2=R_0+90$ (km), $R_3=R_0+270$, $R_4=R_0+420$ and $R_5=R_0+600$ (the radius is shown as an index 
in the table header). Last column shows Markstein number found via quasi-steady state technique.
\label{tab51}}}\smallskip
\end{table}

As follows from the Table, \p{mark} consistently describes dependence of flame speed on its radius at small expansions $1/\l$, 
one gets almost the same Markstein numbers while fitting $D_R$ over different ensembles of $R$\rq s, meaning that \p{mark}
is a good approximation, omitted terms are not significant for the curvatures studied. As flame remains circular at these 
expansions for small enough radii, one would not expect explicit dependence of flame speed on time, all time dependence is
contained in $D_{R(t)}$ dependence, through Markstein effect and radius growth with time. Deviations from this are observed 
at larger expansions, again expectedly, as corrections to flame speed due to asphericity development (both large-scale, and
LD-like) depend explicitly on time provided for the asphericity to develop. For sKPP these deviations are pronounced at 
$1/\l=0.734$ and above, in accord with our direct observations of flame distortion development in the previous section.
Model B shows similar violation of \p{mark} in simulations at $1/\l=2.33$ ($\rho_\mathrm{fuel}=10^7\:\gcm$ in SN~Ia problem).
This again is in agreement with worse model behavior (faster LD-instability development) at such expansion.
Notice that precise $M$ found numerically differs from that found through steady-state technique increasingly
as expansion increases. Same increasing deviation was observed above for $D$ in 1D simulations, thus it can be expected that
similar discrepancy should be found for curvature corrections to $D$. Increasing flame width damps LD instability development, 
as well as decreases discrepancy (due to discretization) between numerically found Markstein number, and steady-state one (in
continuous calculation). This can be seen in the Table for Model B entries at $1/\l=2.33$ and different flame width. Notice
also, that for $R=R_2$ at $1/\l=2.33$ one gets negative Markstein numbers. Burning is not quasi-steady yet at that time
(timestep 1300) for larger expansions. Studying flame speed as a function of time (what we effectively do through studying
Markstein number evolution) provides another way to characterize flame asphericity development.

For smaller expansions agreement with steady-state results for Model B is encouraging. Similar agreement, 
with further deviation in the same direction due to discretization effects, is observed 
for 3D simulations; say, we get $M=0.280$ for Model B $256^3$ cubic run at $1/\l=0.398$ at $R=R_2=120$~km (curvature of a spherical
flame in 3D being $2/R$).

\chapter{Conclusions}
\label{cha:conclude}
We analyzed a reaction-diffusion system \p{I1} with hydrodynamically determined advection velocity $\mathbf{u}$. 
This model describes premixed physical flames with one dominating reaction, the rate of which can be effectively determined by one 
reaction progress variable $f$, which is the case, in part, for systems with unit Lewis number (ideal gases in part); this system is used
as a tool for tracking unresolved flames in deflagration simulations (Flame Capturing technique, FC, \cite{X95}),
in part in astrophysical simulations of nuclear deflagration in a White Dwarf during an SN~Ia explosion.

One part of the study 
was performed in continuous steady-state approximation (steadily propagating flame solutions) for planar flames (spatially one-dimensional 
problem; Chap.~1 and 2) with the purpose of finding the flame propagation speed and width, as well as for spherical flames in 2D and 3D, 
with the goal to quantify Markstein effect, how the flame front speed depends on the its curvature (Chap.~5). For this part of the study
we assumed heat release proportional to $f$ increase, $\delta Q = q\,df$ ($q$ meaning total heat release in the reaction, per unit mass), 
negligible pressure jump across the flame (thus, in part, thermal enthalpy density increasing linearly with $f$, $dH=\delta Q$), 
$\rho H=\mathrm{const}$ in the process of such isobaric burning. The problem
of finding steady flame profiles $f(x)$ ($f(r)$ for spherical flames) was analytically reduced to finding eigenfunctions of boundary-value
problems, \p{master1} with $p(0)=p(1)=0$ for planar flames, (\ref{Markn}--\ref{Markr}) for spherical flames. These eigenfunctions yield 
flame profiles in dimensionless spatial coordinate: all the quantities were made dimensionless by scaling with appropriate powers of 
characteristic values of reaction rate $R$ and $f$-diffusivity $\tilde{K}$, Eq.~\p{dimless}. 
Dimensionless flame speed $d$, Eq.~\p{dimlessd} is given by an eigenvalue of the corresponding boundary problem in 1D or nD; for the
case of spherical flames flame radius $r_0$ enters explicitly the boundary conditions, thus making flame speed and profile dependent
on flame curvature.

Another part of the study was performed numerically, using full hydrodynamical codes {\tt{ALLA}} (\cite{X98}) and {\tt FLASH} 
(\cite{fryxell}). This part had two major goals. The first one was to check how based on steady analysis calibration of different flame 
models studied was affected by discretization effects in direct simulations. For flame widths used/proposed to be used in FC (3.2--4 cells
between $f=0.1$ and $f=0.9$) flame velocities observed in direct simulations are somewhat smaller than their prescribed values
based on steady-state analysis, the difference increases with parameter $1/\l$ characterizing matter expansion, Eq.~\p{lam}; this difference
remains within 8\% for the models studied at $1/\l<1.44$ (interval of most interest to SN~Ia problem, corresponds to fuel (with composition 
$0.5^{12}C+0.5^{16}O$) density of $3\times 10^7\:\gcm$ and above). 
Flame widths observed in these simulations are larger than their prescribed values; Cf. Tab.~\ref{tab2} and \ref{tab31} for comparison of
continuous and discretized speeds and widths for some models studied. The discrepancy in width was clearly related to
numerical diffusion tails (Fig.~\ref{stepKnum}) by increasing number of computation grid cells within the flame width, Tab.~\ref{tab3}; 
for such wider flames front speed also tends to its value prescribed through continuous steady-state analysis, discrepancy reduced to
about 1\% for 16 cells within flame width $W_1$. 

For thin flames used for FC it is desirable to correct for this discretization discrepancy; this is accomplished
by defining dimensionless flame speed $d$ and width $w$ by direct simulation in 1D, using flame parameters ensuring physical 
flame width $W$ the same as the one to be used for actual simulations (with gravity, arbitrary flame geometry, etc.), see 
Sec.~\ref{nstdy:calibr} for more detail. 

For values of $d$ and $w$ found by any of the methods (for any flame model, specified by dimensionless form-factors of reaction rate and 
diffusivity $\Phi_0$ and $K_0$, \p{dimless}) the procedure to obtain proper normalization factors $R$ and $\tilde{K}$ so that Eq.~\p{I1}
coupled to the rest of hydrodynamic equations yield a \fl\ having prescribed width $W$ and speed $D$ is given by Eq.~\p{KR} as described 
in Sec.~\ref{prescr}. For Model B, in particular, $d$ and $w$ are given by fits \p{Bcalib}, thus yielding efficient procedure for prescribing
$K$ and $\Phi$ based on local $\l$.
\smallskip

Any model with unique propagation speed may be calibrated this way to yield a flame with required properties, in idealized conditions at 
least, in 1D with simple hydrodynamics. Some models are however better than others for use in flame capturing. Certain features 
of $\Phi_0$ and $K_0$ of the model used produce flames with undesirable features. Some of these features can be identified in steady
state study: for flame models with $K_0=1$, the original one (\cite{X95}, Eq.~\p{Phistep}) and KPP \p{Phikolm} $\Phi_0$ goes to zero
either at $f=0$ or 1, thus producing flames with infinite tails. For artificially thickened \fl\ used in FC to deviate least in 
behavior from much thinner physical flame it is imperative to have the flame completely localized in the narrowest possible region.
Central region of the \fl, characterized by large gradients of $f$ must be adequately resolved in simulations, thus have physical
width of about 3 or more grid spacings; therefore flames without long tails, with profile close to linear are to be preferred. 
By small perturbation of the burning rate model KPP may be transformed into a model with finite tails (and unique eigenvalue for 
propagation speed), sKPP, Eq.~\p{skpp}. However unless the perturbation (given by shift parameters $\e_a, \e_f$) is large enough (leading to 
significant numerical noises) model sKPP retains long tails in the profile (Fig.~\ref{fkb}), thus having a disadvantage
compared to other models proposed. 

Using non-constant diffusivity, $K_0=f^r$ allows one to make a model with step-function burning rate have finite tails; choice of parameters
$(f_0;r)=(0.2;0.75)$ makes flame profiles insensitive to flame expansion, another desirable feature to make \fl\ response to hydrodynamics
similar at different densities (in SN~Ia problem), consistent throughout the simulation as the flame propagates into less dense regions
further from the center of the WD. However this model, Model A, as well as other models with large discontinuities in burning rate produces 
significant numerical noises in simulations, about 10\% fluctuations in instant flame speed, Fig.~\ref{st}. Numerical noises, as well as flame
propagation anisotropy, small-scale hydrodynamical instability of the flame front in 2D and 3D are only observed in
actual simulations of discretized on computation grid set of hydrodynamic equations, not in continuous steady-state study. Studying these
intrinsically non-steady features of the flames was another major goal of numerical part of our study, presented mostly in Chap.~3 and 4.

Numerical study showed that different models differ one from another in amount of noise they produce (which is readily observed in 1D),
in flame surface distortion in 2D and 3D. 1D noises and flame speed anisotropy with respect to the grid are numerical artifacts and
must be minimized by choosing appropriate flame model for use in FC. Small scale instability we observed on spherical, cylindrical and 
perturbed planar flames is of physical nature, characteristic for any flame propagation when fuel and ash densities differ.
This instability is of the type studied by Landau and Darrieus in simplified setting, as was demonstrated by quantitative agreement
between characteristic instability lengthscales in our direct simulations, and analytic estimates in Markstein approximation, 
Sec.~\ref{Mark1D}. It is physically impossible to invent a flame model immune to this instability. However some models, like sKPP,
become fast perturbed on very small scales, of order 10 cells at $1/\l\ge 0.734$, whereas another model proposed, Model B, 
Eq.~(\ref{syntf}--\ref{clam}), only starts showing signs of this instability development (at $1/\l\ge 1.44$) when getting close to 
boundaries in $1024\times 1024$ box, in cylindrical flame simulations; flame radius increasing by a factor of 20 by that time, 
Fig.~\ref{2d3mods}.
Such behavior is adequate for SN~Ia simulations. Numerical effects, propagation anisotropy and noises, are also insignificant for
Model B, whereas sKPP shows significant anisotropy at $1/\l\gtrsim 0.734$, which strongly distorts the pattern of LD-type instability.

Flames of Model B are not perturbed significantly by reflecting boundaries, in contrast with sKPP, for which octant simulations are thus 
even more 
questionable. Markstein numbers computed for Model B using quasi-steady state technique, Sec.~\ref{Mark:theory} and estimated based on 
radius dependence of cylindrical flame speed $D(r_0)$ in simulations demonstrate close agreement at small expansions, and are close
enough at $1/\l=1.44$ to suggest that physical Markstein effect dominates over contribution to 
$D(r_0)$ dependence due to propagation anisotropy and LD-instability at densities of $\gtrsim 3\times 10^7\:\gcm$ in WD problem. Based 
on all these features observed we recommend using Model B for Flame Capturing.
\medskip

A few directions to develop that are close to our study are the following.
\\ 
\indent\hspace{3mm}$\bullet$ Larger 3D simulations need to be performed
to clearly see LD-instability on spherical flames for Model B. The results might suggest slight changes in the model,
better behavior in 3D may be possible with somewhat larger $c(\l)$ term in burning rate at large expansions. 
\\
\indent\hspace{3mm}$\bullet$ To use the model when $H\rho$ is significantly nonconstant 
(which is the case in SN~Ia problem at smaller densities, $\sim 10^6\:\gcm$,
near quenching of nuclear burning) steady-state model calibration has to be modified; this is automatically corrected
for if the model is calibrated numerically, as is the case for Model B calibration \p{Bcalib}. 
\\
\indent\hspace{3mm}$\bullet$ Also at smaller densities in SN~Ia problem
assumptions of the flame being isobaric becomes increasingly violated. This is due to decreasing sound speed, and increasing speed
of the flame with respect to the grid, equal to $D(1+1/\l)$ for planar flames. As follows from analysis in Sec.~\ref{bndryd} flame
speed depends on exact density distribution within the flame in part, which is affected by pressure jump across the flame and should
be corrected for if this jump is significant. The jump depends on geometry of the flow: in our simulations this non-isobaricity led to 
systematic increase of flame speed in 2D and 3D as compared to the speed in 1D. The deviation reached 1\% at $1/\l=2.33$, and $\sim 20\%$
at $1/\l=8.57$ (at that density, $8\times 10^5\:\gcm$, sound speed is just a factor of 2 larger than $D(1+1/\l)$, for $D=80\:\kms$ used.)
These problems of equation of state and large pressure jump should not be relevant for most terrestrial flames.
\\
\indent\hspace{3mm}$\bullet$ Quasi-steady state technique presented in Sec.~\ref{markstein:intro} in general setting, with arbitrary 
Lewis number and possibility of having several species involved in reactions with comparable rates is immediately applicable 
for quasi-steady estimates of Markstein widths of various terrestrial and astrophysical problems. The same technique, simplified to 
planar flames, may be used to find steady propagation velocities of flames more general than the ones described by Eq.~\p{I1}. This
may be needed even for some FC models that involve more than one reaction progress variable, with intent to more accurately prescribe
heat release distribution when several characteristic reactions with different enough characteristic temporal and spacial scales 
are to be taken into account.

\appendix
\chapter{Summary of implementation of FC Technique with Flame Model B}
\label{app:1}

Here we summarize a step-by-step instructions for using the Flame Capturing Technique based on Model B we recommend based
on analysis in the thesis. 

Assuming one has a hydrodynamic solver for modeling deflagrations, the steps to use the technique for tracking the thickened flame 
region are  the following:\\
\indent 1) Add one new scalar quantity $f$ to physical variables evolved with the code. \\
\indent 2) Use $\delta Q/dt = q\,df/dt$ as a heat release term in original hydrodynamic equations. $q$ is total specific heat release
of nuclear burning, it depends on local pressure and fuel composition.
\\
\indent 3) Evolve $f$ via 
\begin{equation}\label{I1a}
    \frac{\partial f}{\partial t}+\mathbf{u}\cdot\bnabla f
    =\bnabla(K\bnabla f)+\Phi(f),
\end{equation} 
where $\mathbf{u}$ stands for local gas velocity, source term and diffusivity are given by
\begin{equation}\label{dimlessPhia} 
  \Phi(f)=Rf^{s_f}(1-f)^{s_a}(f-c(\l))    ,
\end{equation}
\begin{equation}\label{dimlessKa} 
 K(f)=\tilde{K}f^{r_f}(1-f)^{r_a}.
\end{equation}
respectively,
\begin{equation}\label{expona}
  s_f=1,\: s_a=0.8,\: r_f=1.2,\: r_a=0.8.
\end{equation}
Normalization factors $\tilde{K}$ and $R$ are functions of locally determined expansion parameter 
$\Lambda={\rho_\mathrm{ash}}/({\rho_\mathrm{fuel}-\rho_\mathrm{ash}})$ (the latter depends on pressure and fuel composition
in the cell), and the values for \fl\ speed $D$ and width $W_1$ (between values $f=0.1$ and $f=0.9$) 
one strives to obtain:
\begin{equation}\label{KRa}
\tilde{K}=\frac{D_f}{d(\l)}\cdotp\frac{W_1}{w_1(\l)},\quad  R=\frac{D_f}{d(\l)}\Bigm/\frac{W_1}{w_1(\l)}.
\end{equation}
Dimensionless $d(\l)$, $w(\l)$, and term $c(\l)$ in Eq.~\p{dimlessPhia} are given by the following fits:
{\setlength\arraycolsep{2pt}
\begin{equation}\label{Bcalib}\begin{array}{rl}
  1/\l\in[0;0.515]:&\left\lbrace\begin{array}{ll}
         c=& 0.005\\
         d=&0.328227-0.10051/\l+0.0244596/\l^2\\
         w_1=&2.075422+0.443918/\l-0.097483/\l^2
  \end{array}\right.\\

  1/\l\in[0.515;0.81]:&\left\lbrace\begin{array}{ll}
         c=&1/\l-0.51\\
         d=&0.497578-0.363476/\l+0.1036/\l^2\\
         w_1=&1.553031+1.50381/\l-0.192923/\l^2
  \end{array}\right.\\

  1/\l\in[0.81;1.5]:&\left\lbrace\begin{array}{ll}
         c=& 0.3\\
         d=&0.172133-0.051673/\l+0.0073512/\l^2\\
         w_1=&2.475085+0.232926/\l-0.0323989/\l^2 
  \end{array}\right.\\

  1/\l\in[1.5;1.9]:&\left\lbrace\begin{array}{ll}
         c=&0.675-0.25/\l\\
         d=&-0.0420695+0.129058/\l-0.0177843/\l^2\\
         w_1=&2.764056-0.0061156/\l+0.0025170/\l^2
  \end{array}\right.\\

  1/\l\in[1.9;8.6]:&\left\lbrace\begin{array}{ll}
         c=& 0.2\\
         d=&0.0139649+0.434752\l-0.507838\l^2+0.250103\l^3\\
         w_1=&3.3891706-2.857323\l+5.045909\l^2-3.744233\l^3.
  \end{array}\right.\\
\end{array}
\end{equation}}%
These fits yield errors in flame speed in 1D not exceeding 0.8\% for $W_1=3.2$ cells, the value we recommend to use,
and $1/\l\in[0;8.6]$, covering expansions of interest for SN~Ia problem.
\newpage
\addcontentsline{toc}{chapter}{References}
\begin{singlespace}
\bibliography{bibtex_file1}

\begin{thebibliography}{64}
\expandafter\ifx\csname natexlab\endcsname\relax\def\natexlab#1{#1}\fi

\bibitem[{{Arnett}(1969)}]{ar69}
{Arnett}, W.~D. 1969, \apss, 5, 180

\bibitem[{{Arnett}(1974)}]{ar74}
---. 1974, \apj, 191, 727

\bibitem[{{Arnett}(1982)}]{arnett82}
---. 1982, \apj, 253, 785

\bibitem[{Baade \& Zwicky(1934)}]{baade}
Baade, W., \& Zwicky, F. 1934, Phys. Rev., 46, 76

\bibitem[{{Brown} {et~al.}(2005){Brown}, {Calder}, {Plewa}, {Ricker},
  {Robinson}, \& {Gallagher}}]{brown05}
{Brown}, E.~F., {Calder}, A.~C., {Plewa}, T., {Ricker}, P.~M., {Robinson}, K.,
  \& {Gallagher}, J.~B. 2005, Nuclear Physics A, 758, 451

\bibitem[{{Cappellaro} {et~al.}(1997){Cappellaro}, {Turatto}, {Tsvetkov},
  {Bartunov}, {Pollas}, {Evans}, \& {Hamuy}}]{cappellaro}
{Cappellaro}, E., {Turatto}, M., {Tsvetkov}, D.~Y., {Bartunov}, O.~S.,
  {Pollas}, C., {Evans}, R., \& {Hamuy}, M. 1997, \aap, 322, 431

\bibitem[{{Chandrasekhar}(1931)}]{chandra}
{Chandrasekhar}, S. 1931, \apj, 74, 81

\bibitem[{{Colgate} \& {McKee}(1969)}]{colgate69}
{Colgate}, S.~A., \& {McKee}, C. 1969, \apj, 157, 623

\bibitem[{Damk\"{o}hler(1940)}]{Damkoehler40}
Damk\"{o}hler, G. 1940, Z.~Elektroch., 46, 601

\bibitem[{Darrieus(1938)}]{dar}
Darrieus, G. 1938, Conferences: La Technique Modern. Congres de Mechanique
  Applique, Paris

\bibitem[{{Dursi} {et~al.}(2003){Dursi}, {Zingale}, {Calder}, {Fryxell},
  {Timmes}, {Vladimirova}, {Rosner}, {Caceres}, {Lamb}, {Olson}, {Ricker},
  {Riley}, {Siegel}, \& {Truran}}]{Dursi03}
{Dursi}, L.~J. {et~al.} 2003, \apj, 595, 955

\bibitem[{{Fryxell} {et~al.}(2000){Fryxell}, {Olson}, {Ricker}, {Timmes},
  {Zingale}, {Lamb}, {MacNeice}, {Rosner}, {Truran}, \& {Tufo}}]{fryxell}
{Fryxell}, B. {et~al.} 2000, \apjs, 131, 273

\bibitem[{Gamezo {et~al.}(2005)Gamezo, Khokhlov, \& Oran}]{X05}
Gamezo, V.~N., Khokhlov, A.~M., \& Oran, E.~S. 2005, \apj, 623, 337

\bibitem[{Gamezo {et~al.}(2003)Gamezo, Khokhlov, Oran, Chtchelkanova, \&
  Rosenberg}]{X03}
Gamezo, V.~N., Khokhlov, A.~M., Oran, E.~S., Chtchelkanova, A.~Y., \&
  Rosenberg, R.~O. 2003, Science, 299, 77

\bibitem[{{Gutierrez} {et~al.}(1996){Gutierrez}, {Garcia-Berro}, {Iben},
  {Isern}, {Labay}, \& {Canal}}]{gutierrez96}
{Gutierrez}, J., {Garcia-Berro}, E., {Iben}, I.~J., {Isern}, J., {Labay}, J.,
  \& {Canal}, R. 1996, \apj, 459, 701

\bibitem[{{Hillebrandt} \& {Niemeyer}(2000)}]{hillniem00mech}
{Hillebrandt}, W., \& {Niemeyer}, J.~C. 2000, \araa, 38, 191

\bibitem[{{Hoeflich} \& {Khokhlov}(1996)}]{Hoeflich96}
{Hoeflich}, P., \& {Khokhlov}, A.~M. 1996, \apj, 457, 500

\bibitem[{{Hoyle} \& {Fowler}(1960)}]{HoFo60}
{Hoyle}, F., \& {Fowler}, W.~A. 1960, \apj, 132, 565

\bibitem[{{Iben} \& {Tutukov}(1984)}]{IbenTutukov84}
{Iben}, Jr., I., \& {Tutukov}, A.~V. 1984, in American Institute of Physics
  Conference Series, Vol. 115, American Institute of Physics Conference Series,
  ed. S.~E. {Woosley}, 11

\bibitem[{{Jordan} {et~al.}(2008){Jordan}, {Fisher}, {Townsley}, {Calder},
  {Graziani}, {Asida}, {Lamb}, \& {Truran}}]{Jordan08}
{Jordan}, IV, G.~C., {Fisher}, R.~T., {Townsley}, D.~M., {Calder}, A.~C.,
  {Graziani}, C., {Asida}, S., {Lamb}, D.~Q., \& {Truran}, J.~W. 2008, \apj,
  681, 1448

\bibitem[{{Khokhlov}(1991)}]{X91}
{Khokhlov}, A.~M. 1991, \aap, 245, 114

\bibitem[{Khokhlov(1994)}]{X94}
Khokhlov, A.~M. 1994, \apjl, 424, L115

\bibitem[{Khokhlov(1995)}]{X95}
---. 1995, \apj, 449, 695

\bibitem[{Khokhlov(1998)}]{X98}
---. 1998, \jcoph, 143, 519

\bibitem[{Khokhlov(2000)}]{X00}
---. 2000, arXiv:astro-ph/0008463

\bibitem[{Kolmogorov {et~al.}(1937)Kolmogorov, Petrovskii, \& Piskunov}]{KPP}
Kolmogorov, A.~N., Petrovskii, I.~G., \& Piskunov, N.~S. 1937, Bull. Moskov.
  Gos. Univ. Mat. Mekh., 1, 1

\bibitem[{Landau(1944)}]{lan}
Landau, L.~D. 1944, Zh. Eksp. Teor. Fiz., 14, 240

\bibitem[{{Leibundgut}(2001)}]{leib}
{Leibundgut}, B. 2001, \araa, 39, 67

\bibitem[{{Livne} \& {Arnett}(1995)}]{livnearnett95}
{Livne}, E., \& {Arnett}, D. 1995, \apj, 452, 62

\bibitem[{Lundmark(1920)}]{lundmark}
Lundmark, K. 1920, Sveska Vetenskapsakad. Handlingar, 8, 60

\bibitem[{Markstein(1964)}]{markstein}
Markstein, G.~H. 1964, Nonsteady Flame Propagation (New York: Pergamon Press)

\bibitem[{{Matalon} {et~al.}(2003){Matalon}, {Cui}, \&
  {Bechtold}}]{MatCuiBec03}
{Matalon}, M., {Cui}, C., \& {Bechtold}, J.~K. 2003, Journal of Fluid
  Mechanics, 487, 179

\bibitem[{Matalon \& Matkowsky(1982)}]{matmat}
Matalon, M., \& Matkowsky, B.~J. 1982, J. Fluid Mech., 124, 239

\bibitem[{Michelson \& Sivashinsky(1977)}]{MS}
Michelson, D.~M., \& Sivashinsky, G.~I. 1977, Acta Astronautica, 4, 1207

\bibitem[{Neumann \& Richtmyer(1950)}]{vn}
Neumann, J.~V., \& Richtmyer, R.~D. 1950, J. of Applied Physics, 21, 232

\bibitem[{Niemeyer {et~al.}(2002)Niemeyer, Reinecke, \&
  Hillebrandt}]{Reinlanl02}
Niemeyer, J.~C., Reinecke, M., \& Hillebrandt, W. 2002

\bibitem[{{Nomoto} {et~al.}(1984){Nomoto}, {Thielemann}, \& {Yokoi}}]{w7}
{Nomoto}, K., {Thielemann}, F.-K., \& {Yokoi}, K. 1984, \apj, 286, 644

\bibitem[{{Nomoto} {et~al.}(2000){Nomoto}, {Umeda}, {Kobayashi}, {Hachisu},
  {Kato}, \& {Tsujimoto}}]{nomoto00}
{Nomoto}, K., {Umeda}, H., {Kobayashi}, C., {Hachisu}, I., {Kato}, M., \&
  {Tsujimoto}, T. 2000, in American Institute of Physics Conference Series,
  Vol. 522, American Institute of Physics Conference Series, ed. S.~S. {Holt}
  \& W.~W. {Zhang}, 35--52

\bibitem[{{Ostriker} {et~al.}(1974){Ostriker}, {Richstone}, \& {Thuan}}]{os74}
{Ostriker}, J.~P., {Richstone}, D.~O., \& {Thuan}, T.~X. 1974, \apjl, 188, L87+

\bibitem[{{Pelce} \& {Clavin}(1982)}]{pelce82}
{Pelce}, P., \& {Clavin}, P. 1982, Journal of Fluid Mechanics, 124, 219

\bibitem[{{Phillips}(1993)}]{phillips}
{Phillips}, M.~M. 1993, \apjl, 413, L105

\bibitem[{{Plewa}(2007)}]{Plewa07}
{Plewa}, T. 2007, \apj, 657, 942

\bibitem[{{Plewa} {et~al.}(2004){Plewa}, {Calder}, \& {Lamb}}]{gcd}
{Plewa}, T., {Calder}, A.~C., \& {Lamb}, D.~Q. 2004, \apjl, 612, L37

\bibitem[{{Press} {et~al.}(1992){Press}, {Teukolsky}, {Vetterling}, \&
  {Flannery}}]{numfor}
{Press}, W.~H., {Teukolsky}, S.~A., {Vetterling}, W.~T., \& {Flannery}, B.~P.
  1992, Numerical recipes in Fortran: the art of scientific computing, 2nd edn.
  (New York: Cambridge University Press)

\bibitem[{{Rastigejev} \& {Matalon}(2006)}]{RastMat06}
{Rastigejev}, Y., \& {Matalon}, M. 2006, Journal of Fluid Mechanics, 554, 371

\bibitem[{Reinecke {et~al.}(2002{\natexlab{a}})Reinecke, Hillebrandt, \&
  Niemeyer}]{Rein02}
Reinecke, M., Hillebrandt, W., \& Niemeyer, J.~C. 2002{\natexlab{a}}, \aap,
  386, 936

\bibitem[{Reinecke {et~al.}(2002{\natexlab{b}})Reinecke, Hillebrandt, \&
  Niemeyer}]{Rein02b}
---. 2002{\natexlab{b}}, \aap, 391, 1167

\bibitem[{R\"{o}pke \& Hillebrandt(2005)}]{Roepke05}
R\"{o}pke, F.~K., \& Hillebrandt, W. 2005, \aap, 429, L29

\bibitem[{{R{\"o}pke} {et~al.}(2007){R{\"o}pke}, {Hillebrandt}, {Schmidt},
  {Niemeyer}, {Blinnikov}, \& {Mazzali}}]{Roepke07}
{R{\"o}pke}, F.~K., {Hillebrandt}, W., {Schmidt}, W., {Niemeyer}, J.~C.,
  {Blinnikov}, S.~I., \& {Mazzali}, P.~A. 2007, \apj, 668, 1132

\bibitem[{Sivashinsky(1977)}]{Siv77}
Sivashinsky, G.~I. 1977, Acta Astronautica, 4, 1177

\bibitem[{Timmes \& Woosley(1992)}]{TimWoo92}
Timmes, F.~X., \& Woosley, S.~E. 1992, \apj, 396, 649

\bibitem[{{Travaglio} {et~al.}(2004){Travaglio}, {Hillebrandt}, {Reinecke}, \&
  {Thielemann}}]{Travaglio04}
{Travaglio}, C., {Hillebrandt}, W., {Reinecke}, M., \& {Thielemann}, F.-K.
  2004, \aap, 425, 1029

\bibitem[{Truran {et~al.}(1967)Truran, Arnet, \& Cameron}]{truran67}
Truran, J.~W., Arnet, W.~D., \& Cameron, A.~G.~W. 1967, Canad. J. Phys., 45,
  2315

\bibitem[{{Webbink}(1984)}]{webbink84}
{Webbink}, R.~F. 1984, \apj, 277, 355

\bibitem[{Williams(1985)}]{williams}
Williams, F.~A. 1985, Combustion theory, 2nd edn. (Reading, MA: Addison-Wesley)

\bibitem[{Woodward \& Colella(1984)}]{cowo}
Woodward, P., \& Colella, P. 1984, \jcoph, 54, 174

\bibitem[{{Woosley} \& {Weaver}(1994)}]{woosley94}
{Woosley}, S.~E., \& {Weaver}, T.~A. 1994, \apj, 423, 371

\bibitem[{Xin(2000)}]{xin}
Xin, J. 2000, SIAM Rev., 42, 161

\bibitem[{Zeldovich {et~al.}(1985)Zeldovich, Barenblatt, Librovich, \&
  Makhviladze}]{zelbook}
Zeldovich, Y.~B., Barenblatt, G.~I., Librovich, V.~B., \& Makhviladze, G.~M.
  1985, The Mathematical Theory of Combustion and Explosions (New York:
  Consultants Bureau)

\bibitem[{Zeldovich \& Frank-Kamenetskii(1938)}]{z-fk}
Zeldovich, Y.~B., \& Frank-Kamenetskii, D.~A. 1938, Zh. Fiz. Khim., 12, 100

\bibitem[{Zhang {et~al.}(2007)Zhang, Messer, Khokhlov, \& Plewa}]{Zhang06}
Zhang, J., Messer, O.~E.~B., Khokhlov, A.~M., \& Plewa, T. 2007, \apj, 656, 347

\bibitem[{Zhao {et~al.}(2006)Zhao, Strom, \& Jiang}]{sn185}
Zhao, F.-Y., Strom, R.~G., \& Jiang, S.-Y. 2006, Chinese J. of Astronomy and
  Astrophysics, 6, 635

\bibitem[{Zhiglo(2007)}]{zh07}
Zhiglo, A.~V. 2007, \apjs, 169, 386

\bibitem[{{Zwicky}(1938)}]{zwicky38}
{Zwicky}, F. 1938, \apj, 88, 522

\end{thebibliography}
\bibliographystyle{apj}
\end{singlespace}


\end{document}